# Dynamic Stabilization of Water Bottles

Yanwen Gu[1], Yunzhou Bai[1], Yuxi Xin[1], Lintao Xiao[2], Sihui Wang[*,3], Hanchao Sun[**,3]

## Abstract

The motion of water filled bottles is studied when it is thrown into the air and falls back to the floor, including the possibilities of an upright landing or rolling down before it finally reaches static state.

When dealing with the process after throwing a water bottle, the free falling (bottle falls without initial angular velocity) and flipping (bottle falls with initial angular velocity) are considered. In theory, the physical principles behind the motion are analyzed. In addition, the impacts of initial angle, linear velocity, angular velocity and the water amount on the uprightness of the bottle are discussed. In experiment of throwing bottle, we changed the water amount, angular velocity, and releasing height, and examined the impacts of these factors. The results suggest that a certain amount of water and spinning result in higher possibility of upright landing.

When dealing with rolling bottle, theoretically we build the bottle-and-bead model to describe the coupled motion of water and the bottle. Analytical solutions are obtained for small amplitude and the numerical solution can be done in a general situation. In the experiment of rolling bottle, we firstly verified the theoretical model, and then addressed the impact of initial conditions and water amount on the motion patterns of the bottle.

1. Nanjing Foreign Language School
2. IIIS, Tsinghua University
3. School of Physics, Nanjing University

* wangsihui@nju.edu.cn
** sunhanchao1995@163.com

# Contents

# Introduction

There was once a popular school game: throw a bottle into the air with a spin and try to land it on a horizontal surface in an upright position. Many studies have been conducted on this phenomenon, in which some reasonable assumptions, theoretical modeling, and experimental parameters are studied in articles like *Water Bottle Flipping Physics*[1] and *Research on the Stable Landing of Flipped Bottle*[2]. *The complex physics of that viral water bottle trick*[4] only qualitatively described the process of a bottle from being thrown out to falling down. Similar studies on this topic (*Problems for the 31st IYPT 2018*[3]) is also included in the IYPT Tournament.

These works mostly concentrated on static descriptions and the key factors of upright landing are limited to lowering the center of mass and reducing the angular velocity before landing. The major parameter discussed is the amount of water which affects the center of mass and angular velocity.

Nevertheless, they failed to realize that appropriate rotation is not only harmless but also beneficial to a successful upright landing, nor they analyzed the favorable or unfavorable conditions and the physics behind the game. In this paper, we firstly analyze the general process of bottle-throwing; and discuss free falling and flipping respectively. Then we explore the effects of parameters such as the initial angle, the releasing height, as well as the impact of water amount on the bottle flipping game. We also design experiments on control variables, including coefficient of restitution, height, water amount, and angular velocity.

When dealing with a rolling water bottle, theoretically we will build the bottle-and-bead model to describe the coupled motion of water and the bottle. Analytical solutions describing the forward and oscillating motion patterns are obtained for small amplitudes and will be verified in our experiment. We then address the impact of initial conditions and water on the motion patterns of the bottle. Theoretical analysis and control variable experiments are conducted to reveal **how a moving bottle reaches stabilization.**

# 1. Theoretical analysis

## 1.1 Throwing

### 1.1.1 Mass center and critical tilt angle

First, we calculate the position of the center of mass in the system. We suppose the container shape is a standard cylinder. The base radius is $r$, the height is $H$, and the mass is $M$; The density of the internal liquid is $\rho$, and the height is $h$(see Figure 1.1).

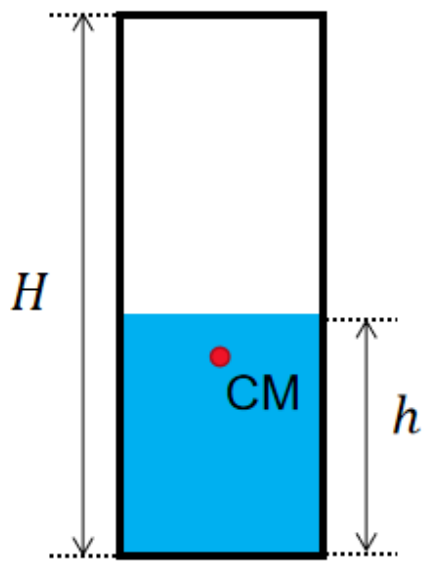

**Figure 1.1 The position of the center of mass**

Then, the position of the center of mass in the system relative to the height of the bottom of the container $h_{cm}$ can be expressed as:

$$h_{cm} = \frac{\rho \pi r^2 h^2 + MH}{2(\rho \pi r^2 h + M)}$$

Then we calculate the maximum tilt angle (critical angle) at which the bottle can stand upright under static conditions. The critical angle corresponds to the case where the gravity line just exceeds the contacting point (see Figure 1.2).

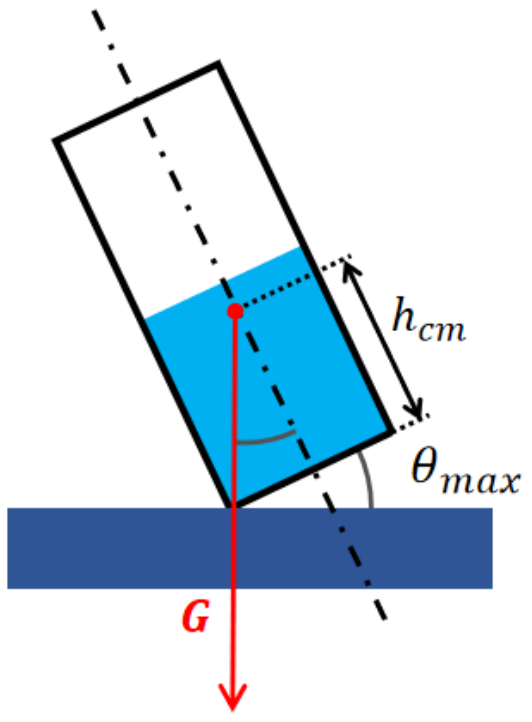

**Figure 1.2 The critical tilt angle**

We can easily find: $\theta_{\max} = \arctan(\frac{r}{2h_{cm}}) = \arctan(\frac{r(\rho \pi r^2 h + M)}{\rho \pi r^2 h^2 + MH})$.

$h_{cm}$ and $\theta_{max}$ changes as the water level h increases (see Figure 1.3 and Figure 1.4) .

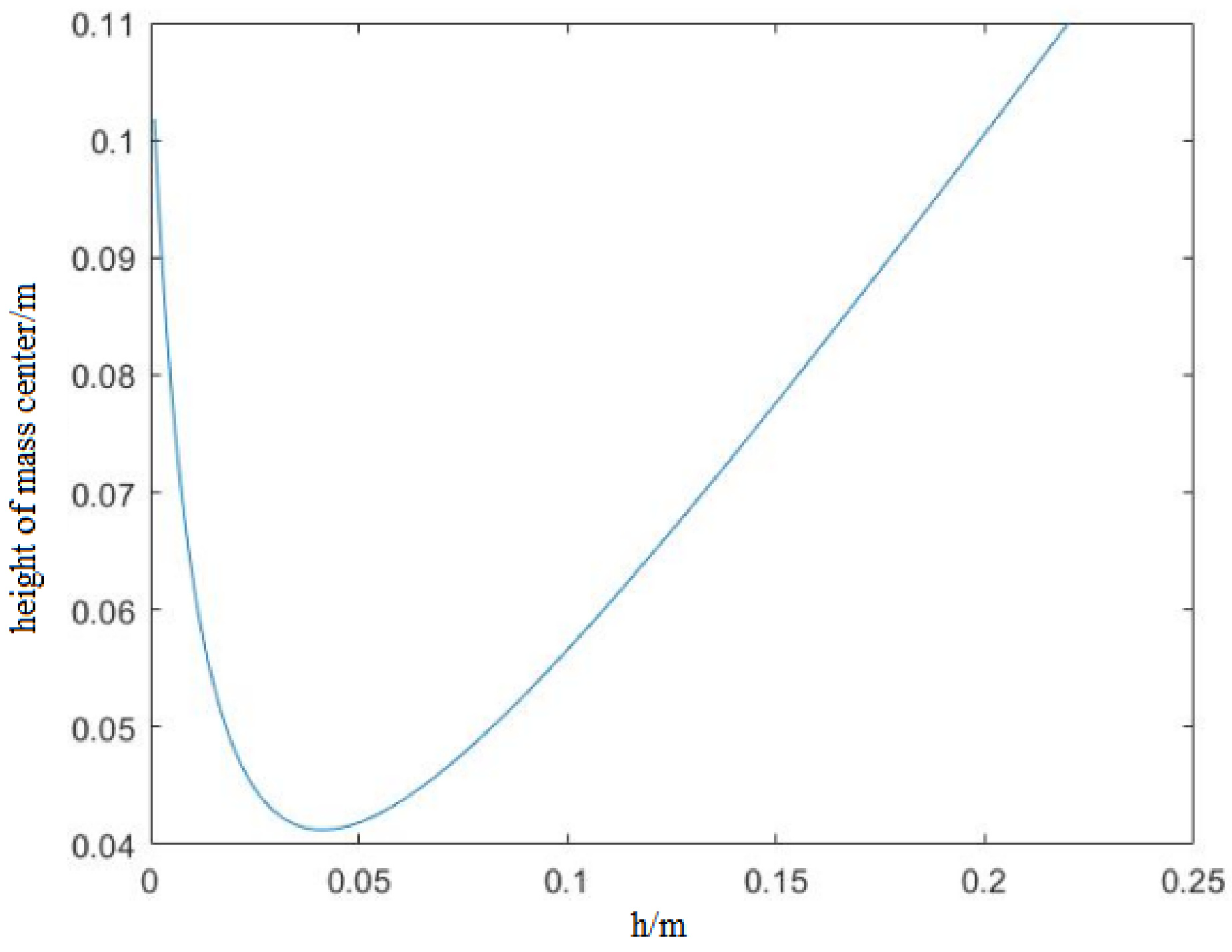

**Figure 1.3 The relationship between the height of center of mass and the water level**

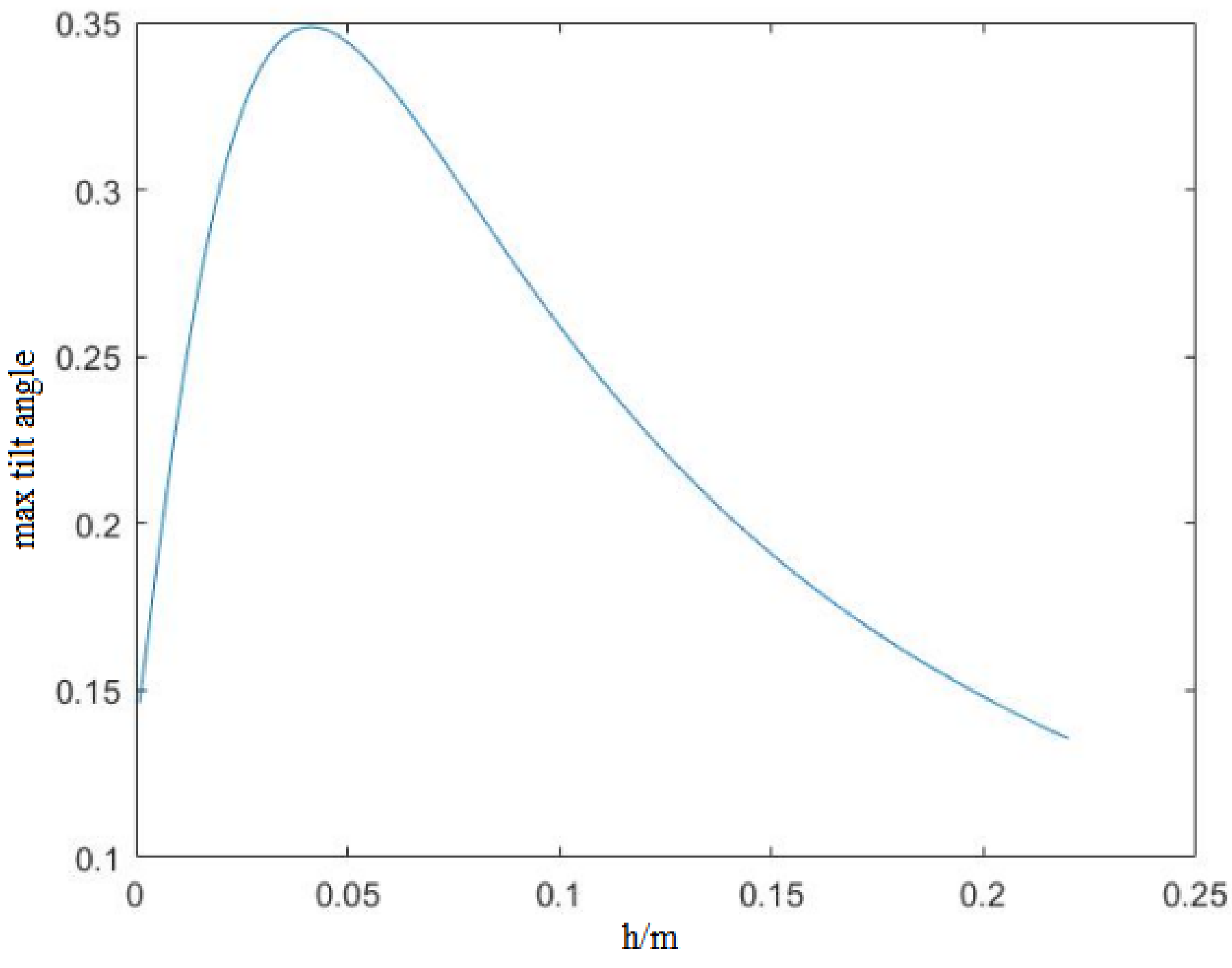

**Figure 1.4 the relationship between maximum tilt angle and water surface height**

As we see from the figure 1.3 and 1.4, when the amount of water increases, the height of center mass goes down and then up. Also, lower center mass directly corresponds to larger maximum tilt angle, which enhances the bottle's standing possibility when throwing.

### 1.1.2 Collision

Next, we create the basic collision model.

Suppose when the bottle touches the ground, the angle between the bottle and the ground is $\theta$, the horizontal velocity is $V_{x0}$, the vertical velocity is $V_{y0}$, and the angular velocity is $\omega_0$; after the collision, the velocities are $V_{x1}$ and $V_{y1}$, the angular velocity is $\omega_1$, the coefficient of restitution between the bottle and the ground is $e$, the static friction coefficient is $\mu$, the supporting force provided by the ground to the container when collision is $N$, and the frictional force is $f$ (see Figure 1.5).

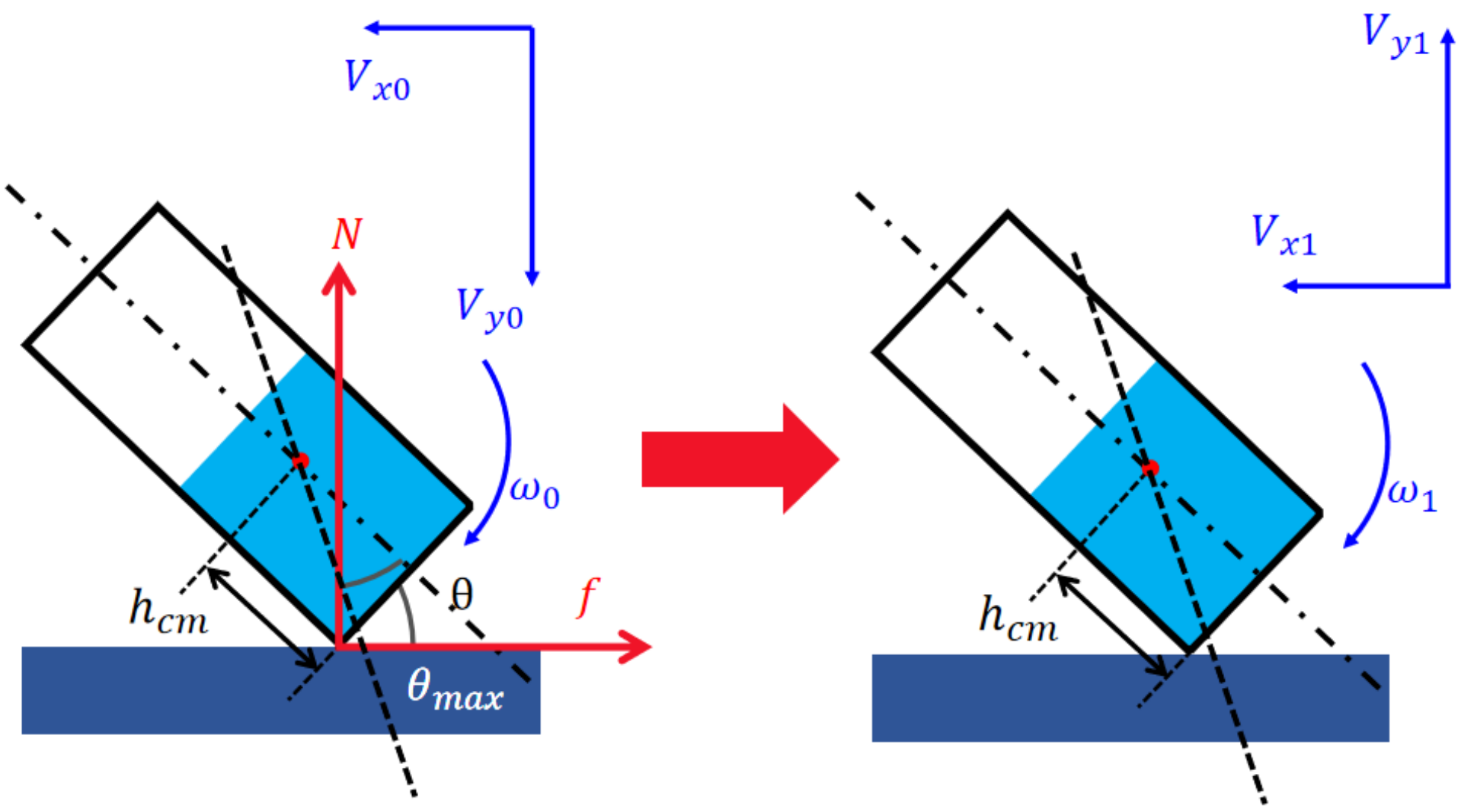

**Figure 1.5 The collision model (left before collision, right after collision)**

First, the equations based on the momentum theorem is as follows:

$$\int f dt = (M+m)(V_{x0} - V_{x1}) \ldots\ldots①$$

$$\int N dt = (M+m)(V_{y0} + V_{y1}) \ldots\ldots②$$

$$f = \mu N \ldots\ldots③$$

Second, the expression based on the angular momentum theorem is as follows:

$$J(\omega_0 - \omega_1) = (M+m)\sqrt{r^2 + h_{cm}{}^2}[(V_{x0} - V_{x1})\cos(\theta - \theta_{max}) + (V_{y0} + V_{y1})\sin(\theta - \theta_{max})] \ldots\ldots④$$

Third, the expression based on the definition of restitution coefficient $e$ is as follows:

$$e = \frac{V_{y1} - \omega_1\sqrt{r^2 + h_{cm}{}^2}\sin(\theta - \theta_{\max})}{V_{y0} + \omega_0\sqrt{r^2 + h_{cm}{}^2}\sin(\theta - \theta_{\max})} \text{......⑤}$$

Last, the solution based on the above expressions ①②③④⑤ is as follows:

$$\omega_1 = \{\Big[\Big(eV_{y0}J + e\sin(\theta - \theta_{max})J\,\omega_0\sqrt{r^2 + h_{cm}^2} - V_{y0}\sin^2(\theta - \theta_{max})m(r^2 + h_{cm}^2\Big) - V_{y0}\sin^2(\theta - \theta_{max})M(r^2 + h_{cm}^2\Big) - V_{y0}\sin(\theta - \theta_{max})\,m(r^2 + h_{cm}^2)\,\mu\cos(\theta - \theta_{max}) - V_{y0}\sin(\theta - \theta_{max})M(r^2 + h_{cm}^2)u\cos(\theta - \theta_{max}) + \sin(\theta - \theta_{max})\,J\omega_0\sqrt{r^2 + h_{cm}^2}\Big]/[\sin^2(\theta - \theta_{max})m(r^2 + h_{cm}^2) + \sin^2(\theta - \theta_{max})M(r^2 + h_{cm}^2) + \sin(\theta - \theta_{max})m(r^2 + h_{cm}^2)\mu\cos(\theta - \theta_{max}) + \sin(\theta - \theta_{max})M(r^2 + h_{cm}^2)\mu\cos(\theta - \theta_{max}) + J)] - eV_{y0} - e\sin(\theta - \theta_{max})\omega_0\sqrt{r^2 + h_{cm}^2}\}/[\sin(\theta - \theta_{max})\sqrt{r^2 + h_{cm}^2}$$

$$V_{x1} = V_{x0} - V_{y0}\mu - \mu[eV_{y0}J + gJ\omega_0\sqrt{r^2 + h_{cm}^2} + e\sin(\theta - \theta_{max})J\omega_0\sqrt{r^2 + h_{cm}^2} - V_{y0}\sin^2(\theta - \theta_{max})(M + m)(r^2 + h_{cm}^2) - V_{y0}\sin(\theta - \theta_{max})(M + m)(r^2 + h_{cm}^2)\mu\cos(\theta - \theta_{max})]/[J + \sin^2(\theta - \theta_{max})(M + m)(r^2 + h_{cm}^2) + \sin(\theta - \theta_{max})(M + m)(r^2 + h_{cm}^2)\mu\cos(\theta - \theta_{max})]$$

$$V_{y1} = [eV_{y0}J + \sin(\theta - \theta_{max})J\omega_0\sqrt{r^2 + h_{cm}^2} + e\sin(\theta - \theta_{max})J\omega_0\sqrt{r^2 + h_{cm}^2} - V_{y0}\sin^2(\theta - \theta_{max})(M + m)(r^2 + h_{cm}^2) - V_{y0}\sin(\theta - \theta_{max})(M + m)(r^2 + h_{cm}^2)\mu\cos(\theta - \theta_{max})]/[J + \sin^2(\theta - \theta_{max})(M + m)(r^2 + h_{cm}^2) + \sin(\theta - \theta_{max})(M + m)(r^2 + h_{cm}^2)\mu\cos(\theta - \theta_{max})]$$

After releasing, a bottle may collide with the floor many times. Later, we only consider the first and second collisions; that means, when the tilt angle for the second collision does not exceed $\theta_{max}$, the bottle stands upright, otherwise it falls down. Then, according to the different tilt angle, post-collision angular velocity direction and contacting point, the collision is classified as the following 6 cases (see Figure 1.6). In each figure, there is a different situation in the matching of the angular momentum of the system about its center of mass and the moment of impulse given to the system by the ground.

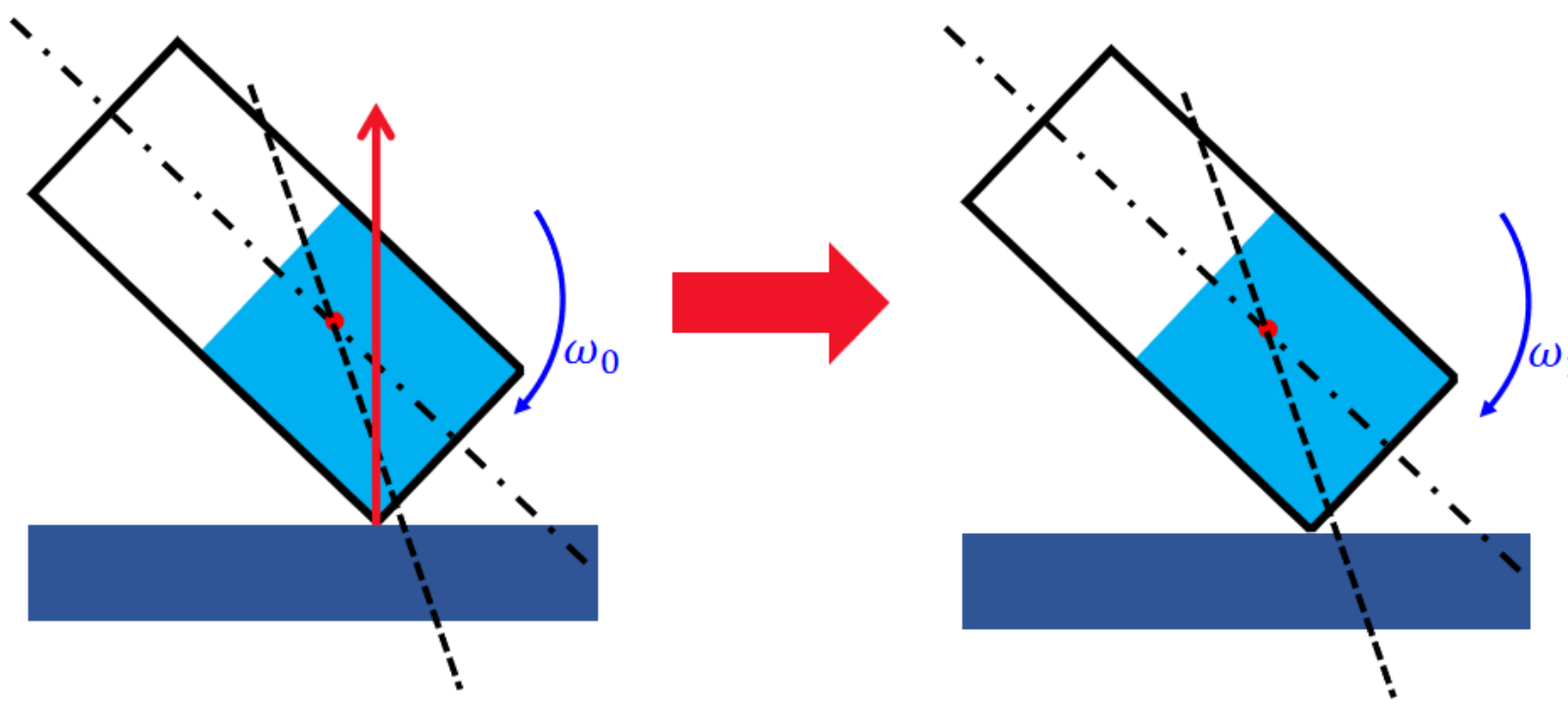

a. Before collision, the angular velocity is inward, the supporting force line exceeds the center of mass, and the angular velocity does not reverse

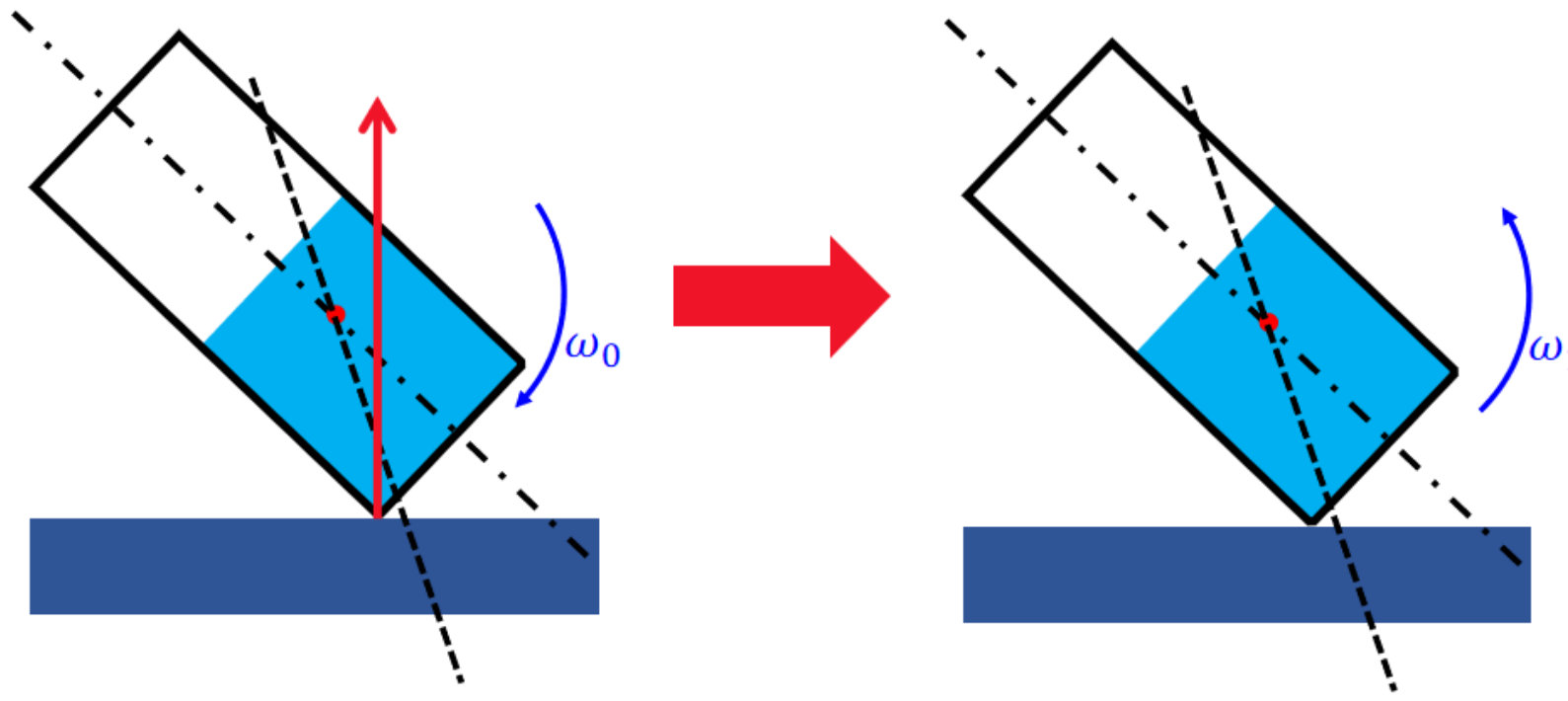

b. Before collision, the angular velocity is inward, the supporting force line is on the right of the center of mass, and the angular velocity reverses

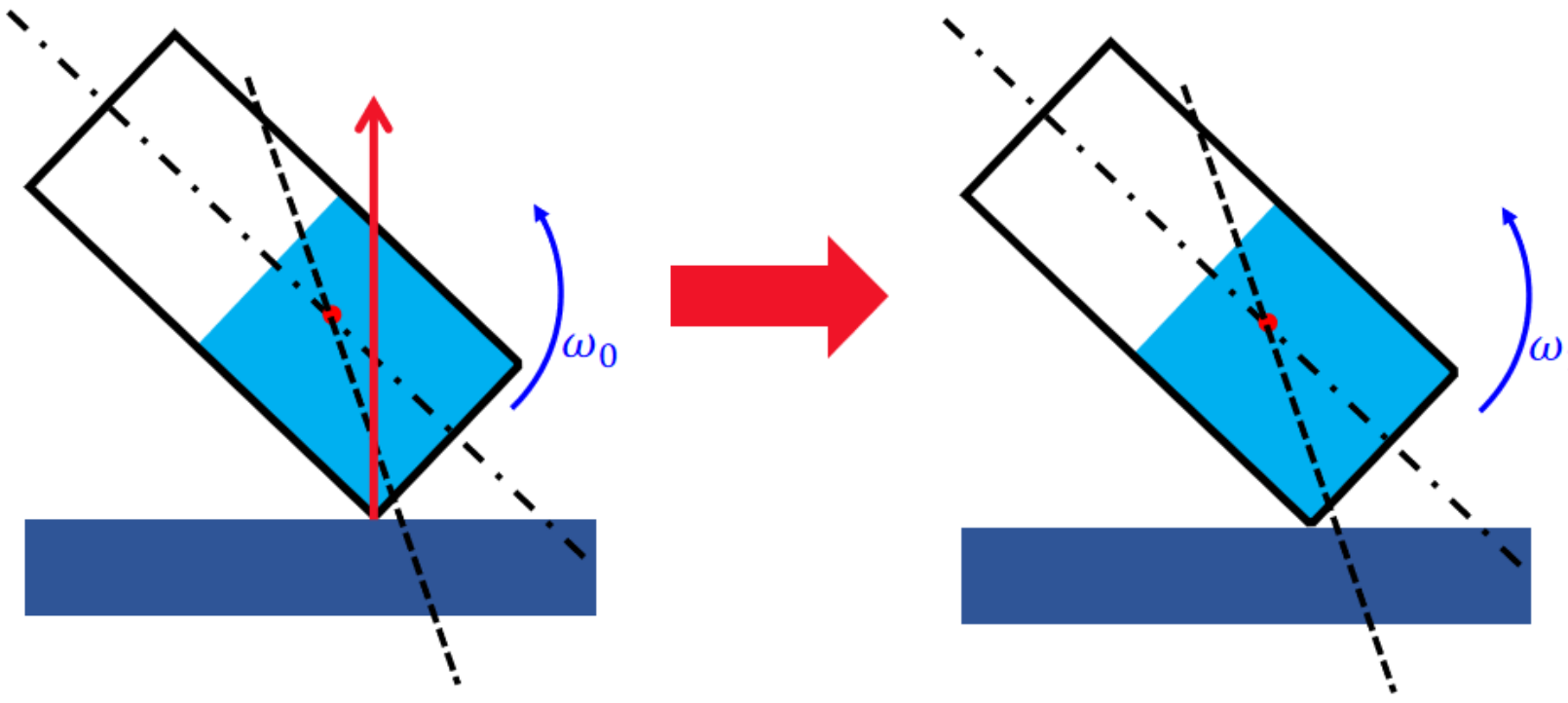

c. Before collision, the angular velocity is outward, the supporting force line is on the right of the center of mass, and the angular velocity does not reverse

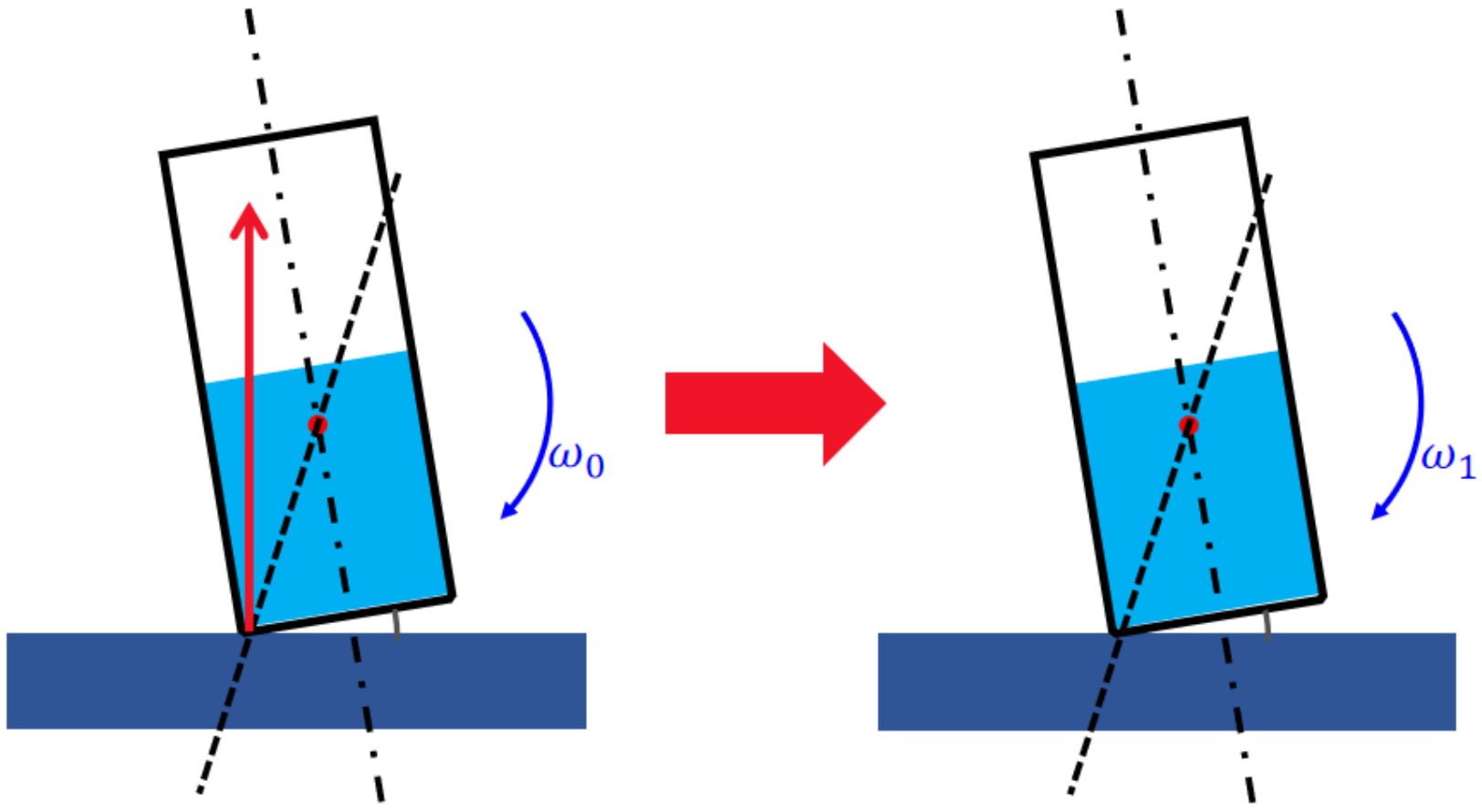

d. Before collision, the angular velocity is inward, the supporting force line is on the left of the center of mass, and the angular velocity does not reverse

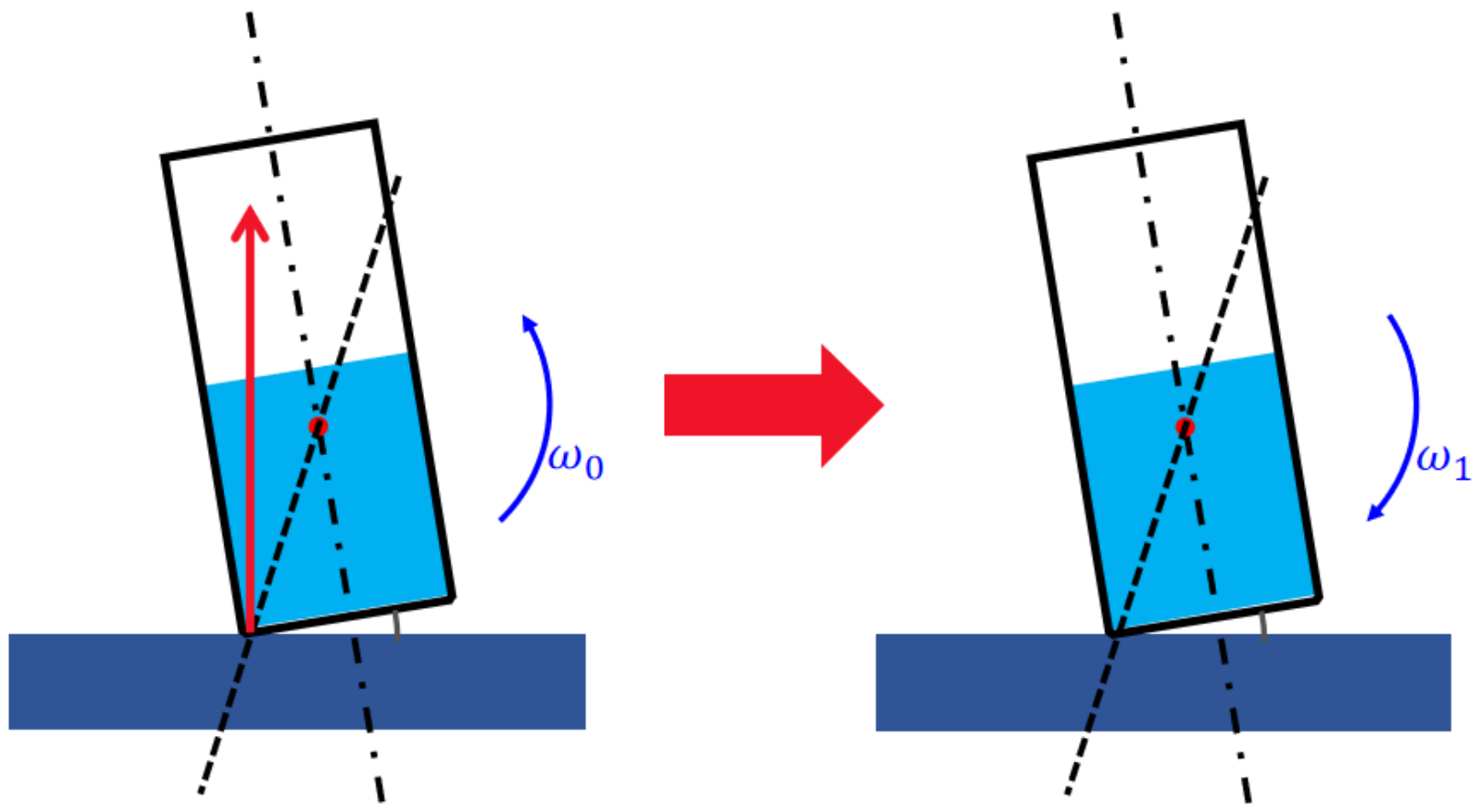

e. Before collision, the angular velocity is outward, the supporting force line is on the left of the center of mass, and the angular velocity reverses

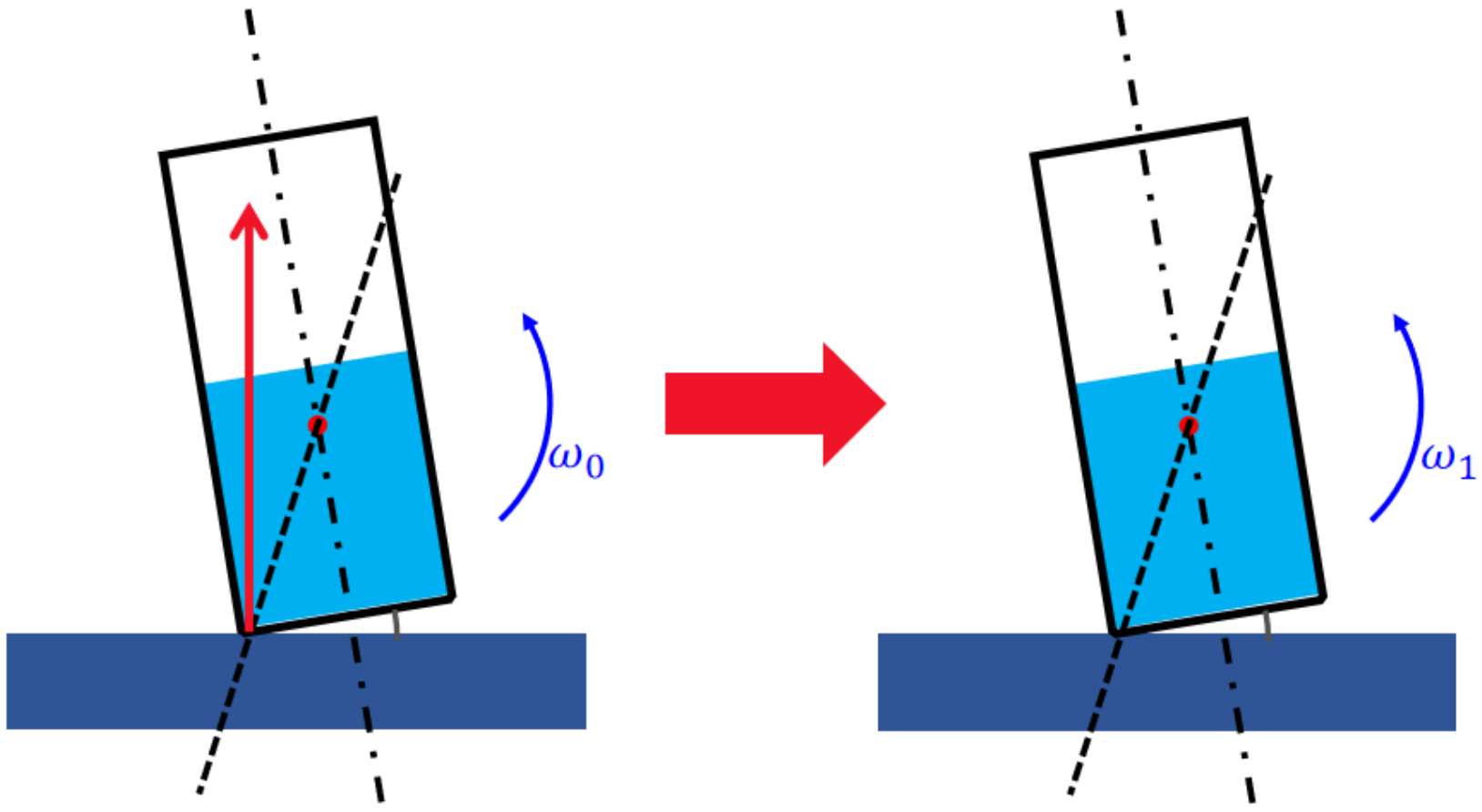

f. Before collision, the angular velocity is outward, the supporting force line is on the left of the center of mass, and the angular velocity does not reverse

**Figure 1.6 Description before and after collision**

From the direction of the angular velocity after collision and the first tilt angle, we can easily exclude situation b and c in that they can never let the bottle stand upright. In situation a, the standing condition needs the extremely accurate matching between the angular momentum and the moment of impulse, and we have only see few of this situation happening in experiments. Therefore, when considering the possibility of bottles standing upright in experiment condition, we think situation a does not play a vital role.

The noteworthy situations are d, e, and f. In these cases, the bottles all have possibilities to stand upright. In situation e and f, particularly, the direction of the angular momentum and the moment of impulse cancel out, thus have larger standing probability.

### 1.1.3 Free Falling

In this process, the bottle does not have initial angular velocity and horizontal linear velocity, it. Then we can solve the angular velocity and linear velocity after collision.

$$\omega_1 = \{[(eV_{y0}J - V_{y0}\sin^2(\theta-\theta_{max})\, m(r^2+h_{cm}^2) - V_{y0}\sin^2(\theta-\theta_{max})\, M(r^2+h_{cm}^2) - V_{y0}\sin(\theta-\theta_{max}) m(r^2+h_{cm}^2)\mu\cos(\theta-\theta_{max}) - V_{y0}\sin(\theta-\theta_{max}) M(r^2+h_{cm}^2)u\cos(\theta-\theta_{max})]/[\sin^2(\theta-\theta_{max}) m(r^2+h_{cm}^2) + \sin^2(\theta-\theta_{max}) M(r^2+h_{cm}^2) + \sin(\theta-\theta_{max}) m(r^2+h_{cm}^2)\mu\cos(\theta-\theta_{max}) + \sin(\theta-\theta_{max}) M(r^2+h_{cm}^2)\mu\cos(\theta-\theta_{max}) + J)] - eV_{y0}\}/[\sin(\theta-\theta_{max})\sqrt{r^2+h_{cm}^2}]$$

$$V_{x1} = -V_{y0}\mu - \mu[eV_{y0}J - V_{y0}\sin^2(\theta-\theta_{max})(M+m)(r^2+h_{cm}^2) - V_{y0}\sin(\theta-\theta_{max})(M+m)(r^2+h_{cm}^2)\mu\cos(\theta-\theta_{max})]/[J + \sin^2(\theta-\theta_{max})(M+m)(r^2+h_{cm}^2) + \sin(\theta-\theta_{max})(M+m)(r^2+h_{cm}^2)\mu\cos(\theta-\theta_{max})]$$

$$V_{y1} = [eV_{y0}J + e\sin(\theta-\theta_{max})J\omega_0\sqrt{r^2+h_{cm}^2} - V_{y0}\sin^2(\theta-\theta_{max})(M+m)(r^2+h_{cm}^2) - V_{y0}\sin(\theta-\theta_{max})(M+m)(r^2+h_{cm}^2)\mu\cos(\theta-\theta_{max})]/[J + \sin^2(\theta-\theta_{max})(M+m)(r^2+h_{cm}^2) + \sin(\theta-\theta_{max})(M+m)(r^2+h_{cm}^2)\mu\cos(\theta-\theta_{max})]$$

The solutions are complex. In order to find out the maximum height in which the bottle will still stand upright in free falling, we simplify the motion of the system between the first and the second collision as a combination of a vertical upward motion starting from the ground and a uniform circular motion. Since the overall rotation angle of the bottles are not large and the maximum tilt angle is not large in the case of standing up, we ignore the error of the distance between the lowest point of the bottle and the ground caused by the rotation of the bottle.

Therefore, the time for the bottle to rotate from the initial state to maximum tilt angle can be expressed as:

$$t_1 = \frac{(\theta_{max} + \theta)}{\omega_1}$$

And the time the bottles staying in air can be expressed as:

$$t_2 = \frac{2V_{y1}}{g}$$

When $t_1 > t_2$, it means the bottle does not have enough time to rotate to its maximum tilt angle when falling to the ground. Thus, in this case, the bottle will stand upright.

So we draw two graphs. Figure 1.7 accounts for the relationship of angular velocity and linear velocity after collision with linear velocity before collision; figure 1.8 stands for the relationship of two times with linear velocity before collision.

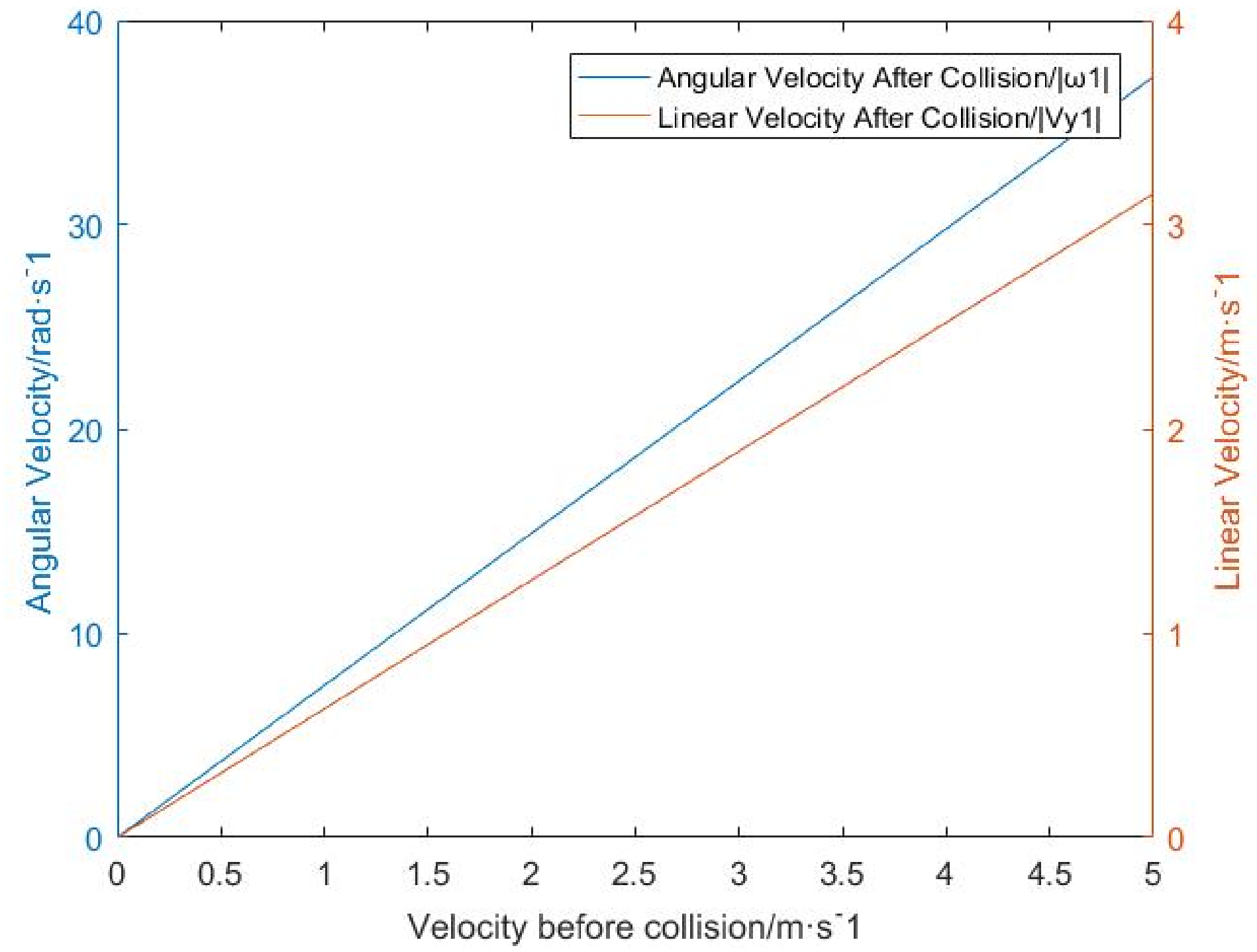

**Figure 1.7 Angular & Linear velocity in free falling**

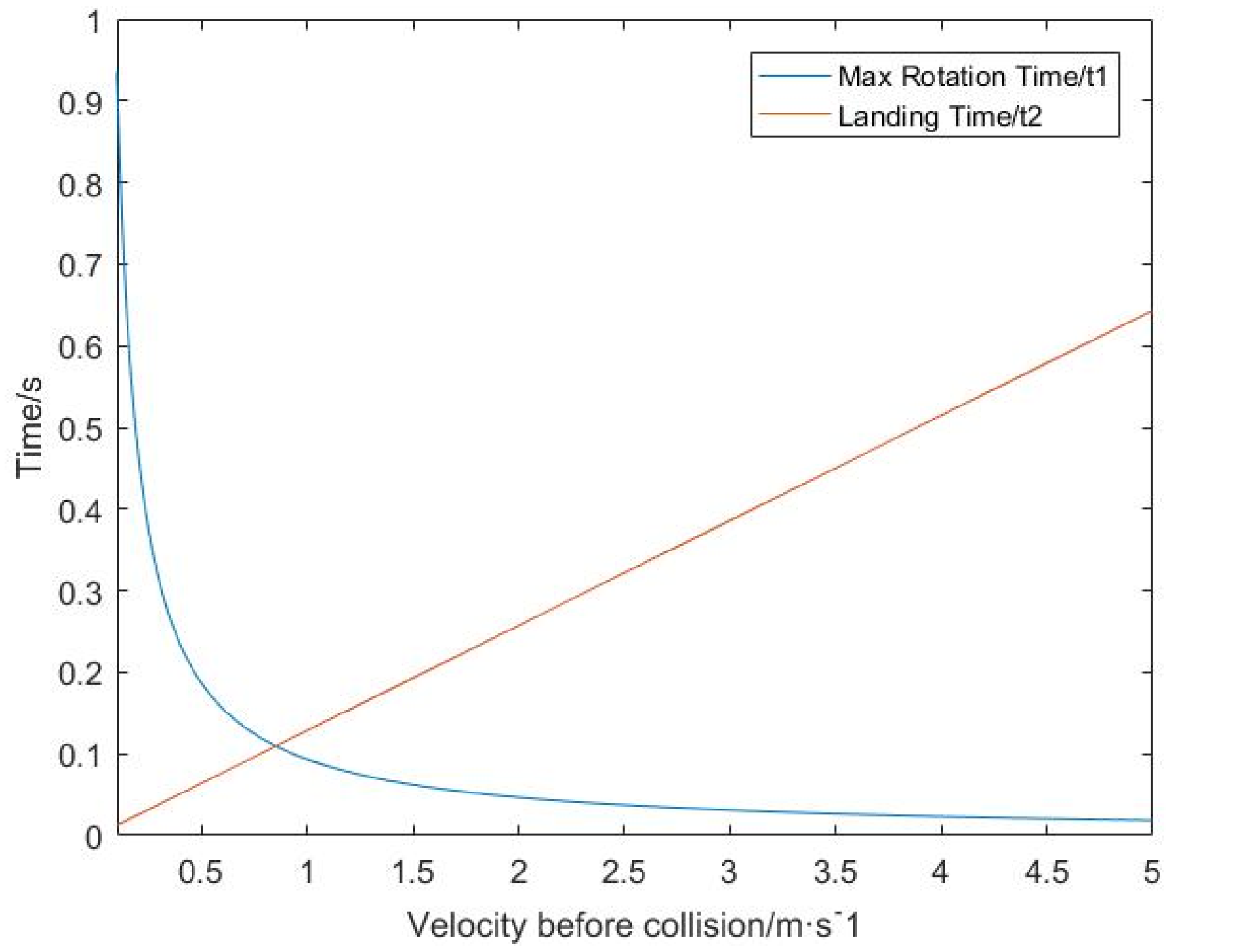

**Figure 1.8 Max rotation time & Landing time in free falling**

Figure 1.7 and Figure 1.8 are drawn using one group of experimental data. It can be seen that, when $V_{y1}$ increases, $t_1$ dramatically decreases and then tends to be zero, while $t_2$ steadily increases. The limit when $t_1 = t_2$ is approximately $0.8m/s$. In this case, the bottle falls at a height of about $3.2cm$.

In other experiments, the maximum free falling height are on the order of single digits.

Therefore we can conclude: Free falling is to the disadvantage of making bottles stand upright. In other words, if we want to make bottles stand, certain amount of angular speed is always necessary, which is opposite to all references we have learned.

### 1.1.4 Flipping

When flipping, the bottle has an initial angular velocity after releasing. Figure 1.9 and Figure 1.10 show the angular velocity and linear velocity after collision with respect to the velocity before collision and the graph of maximum rotation time and landing time.

In these figures, the angular velocity before collision is changing in relation with linear velocity before collision(or throwing time). This is because the angular velocity must make the bottle rotate approximately $180^{\circ}$ in the air. Therefore, when the linear velocity before collision is increasing(so does the falling height), the relevant angular velocity is decreasing since the bottle stays longer in the air. When these two velocities interact with each other, causing the angular momentum and the angular impulse to compensate, the angular velocity after collision will be approximately 0, causing a larger possibility in standing upright. Therefore, we further confirm our guess that angular velocity is helpful to the bottles' standing upright.

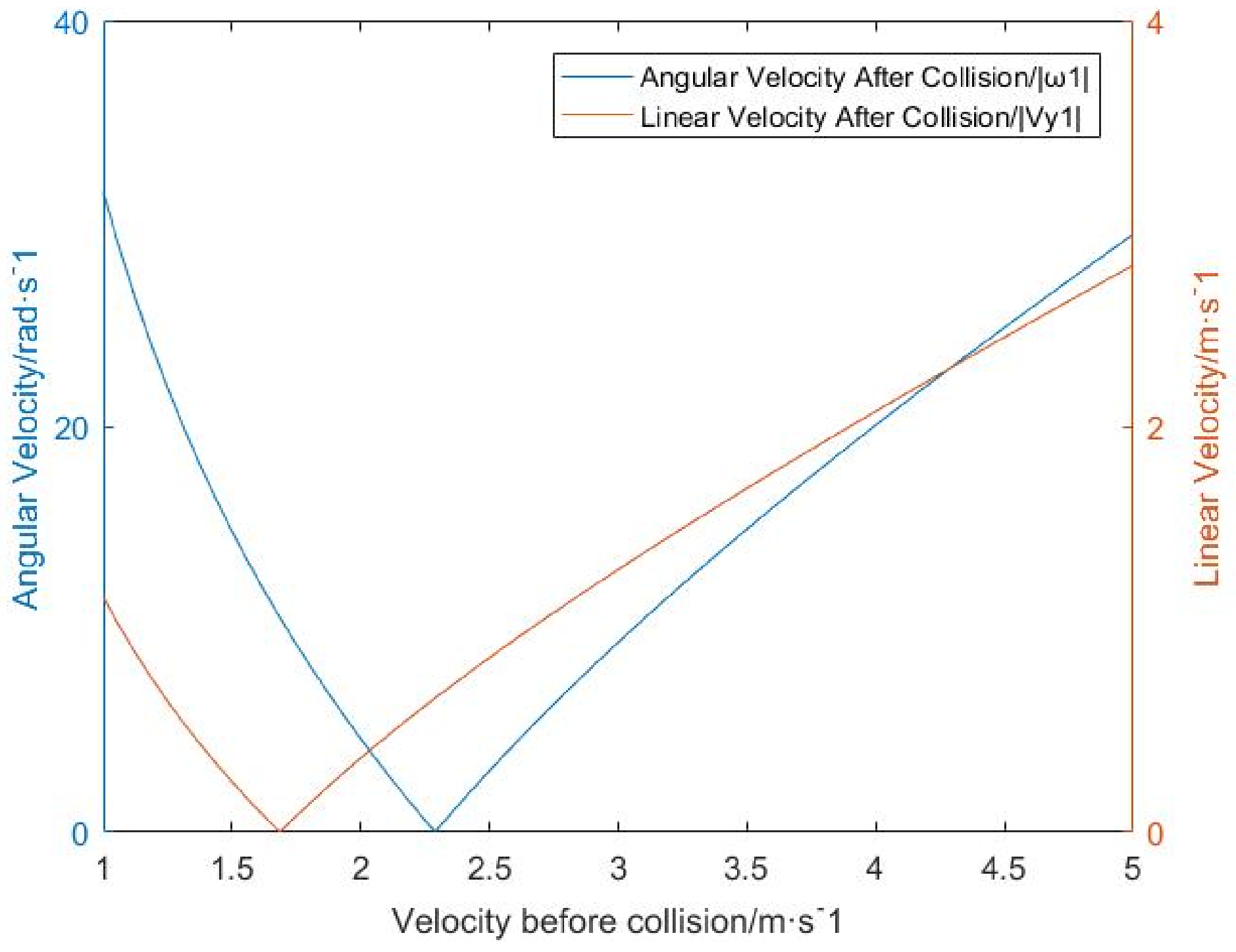

**Figure 1.9 Angular & Linear velocity in rotation**

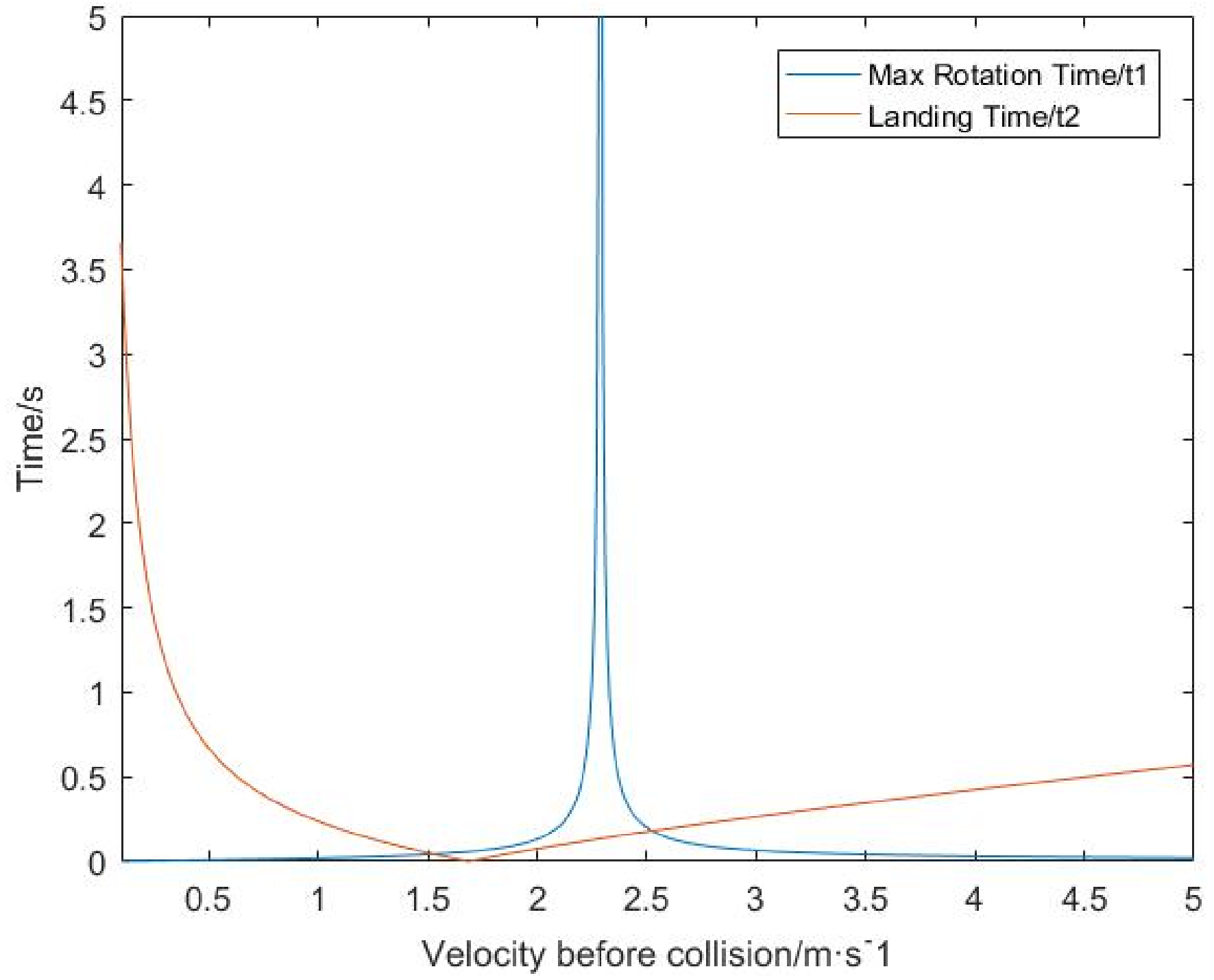

**Figure 1.10 Max rotation time & Landing time in rotation**

### 1.1.5 Discussion: the mobility of water

In analysis before, we consider water as a rigid body. However, in all situations, water is liquid and will flow. Therefore, we here discuss several influence of the stability of the system caused by the mobility of water.

1. Water absorbs colliding energy

When colliding, part of the colliding energy will be transformed into the kinetic energy and the gravitational potential energy of water. Therefore, when colliding, the recovery coefficient will be less when considering water as rigid.

2. The flow and separation of water from the bottle causes the velocity of the bottle to decrease

When colliding, water tends to distribute along the edge of the bottle, causing the rotational inertia of the system to be larger than previously considered. Therefore, according to angular momentum conservation, the angular velocity of the speed will be less. Such separation will also decrease the linear velocity of the bottle, causing a larger possibility to stand upright.

3. The splashing water falls back to help stabilizing.

When the splashing water falls back, its velocity counteracts with that of the bottle, causing its velocity to decrease or to prevent it from shaking on the ground.

4. The water sloshing after landing has a damping effect

After the bottle successfully stands on the ground, it vibrates. The water will stop the bottle more quickly, acting like a damping damper in some high buildings(see figure 1.11).

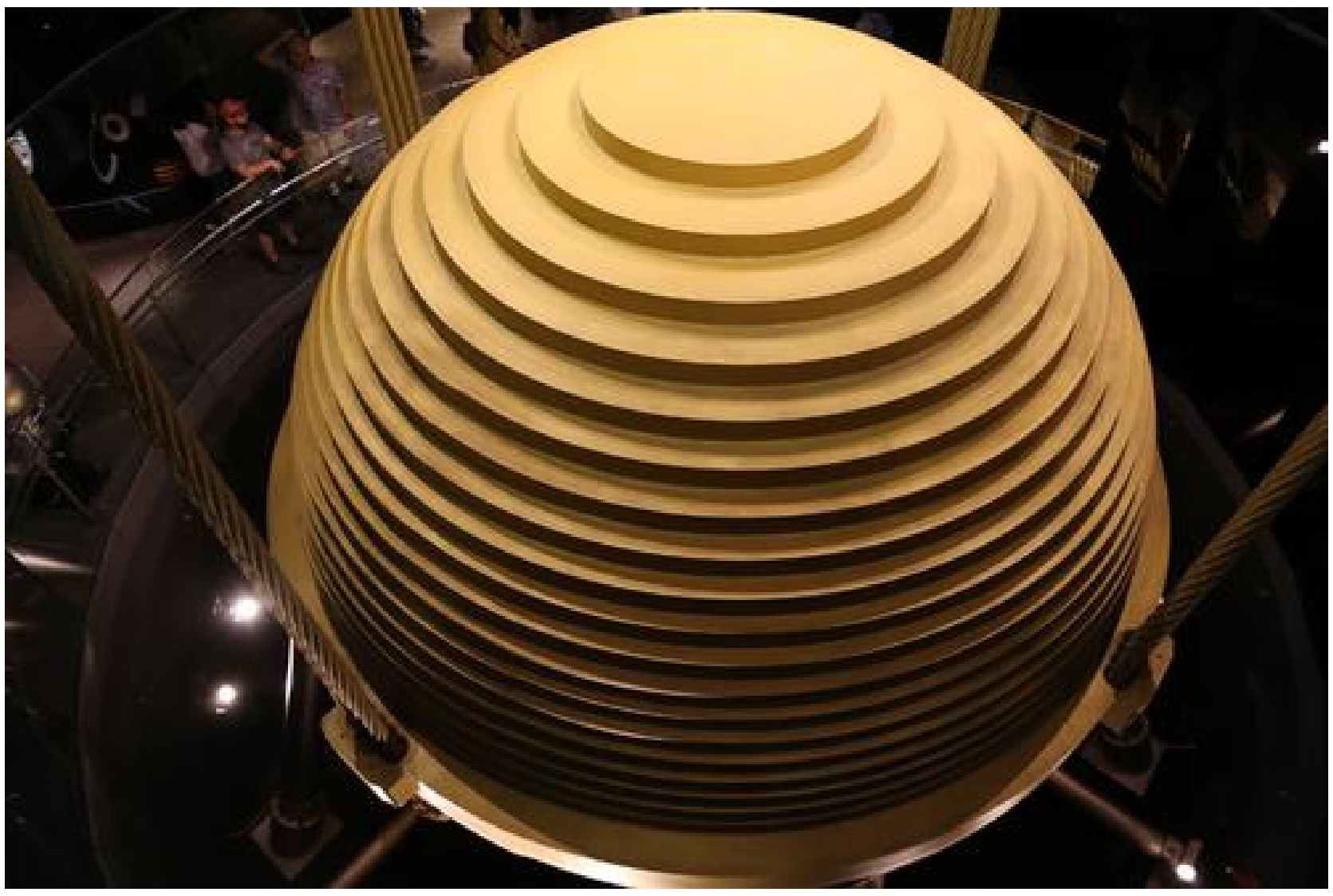

**Figure 1.11 A damping damper**

## 1.2 Rolling

We use the dynamic model of a bottle filled with a steel bead to simulate the movement of a water filled bottle. When comparing with the preliminary experiment, we find that the two systems, meaning the water bottle and the bottle with bead, have similar motion modes. Theoretically, the motion mode of the bottle with bead can be solved by Newton mechanics under ideal conditions, thus helping to understand the motion mode of the water bottle. Studying the influence of various physical parameters on the bottle with bead is also helpful to analyze and predict the motion mode of the water bottle.

### 1.2.1 Physics Model: Bottle and Bead

We ignore all the damping of the bottle in motion and simplify the parameters of actual situation, and only study the following simplified model.

A homogeneous solid steel bead with a radius of r and a mass of m is placed in a thin cylinder with radius R and mass M. The central axis of the cylinder and the central axis C of the bead stay horizontal. And their cross sections are shown in Figure 1.12. The cylinder is placed on a surface so that it can only roll on the ground without slipping, and the bead can only slide without friction (to simulate the motion of water, we assume that the bead is sliding instead of rolling).

The horizontal position of the central axis is marked as X, and the angle between the vertical line and the line of the bead and the cylinder center is $\theta$, the gravitational acceleration is marked as g (see Figure 1.12).

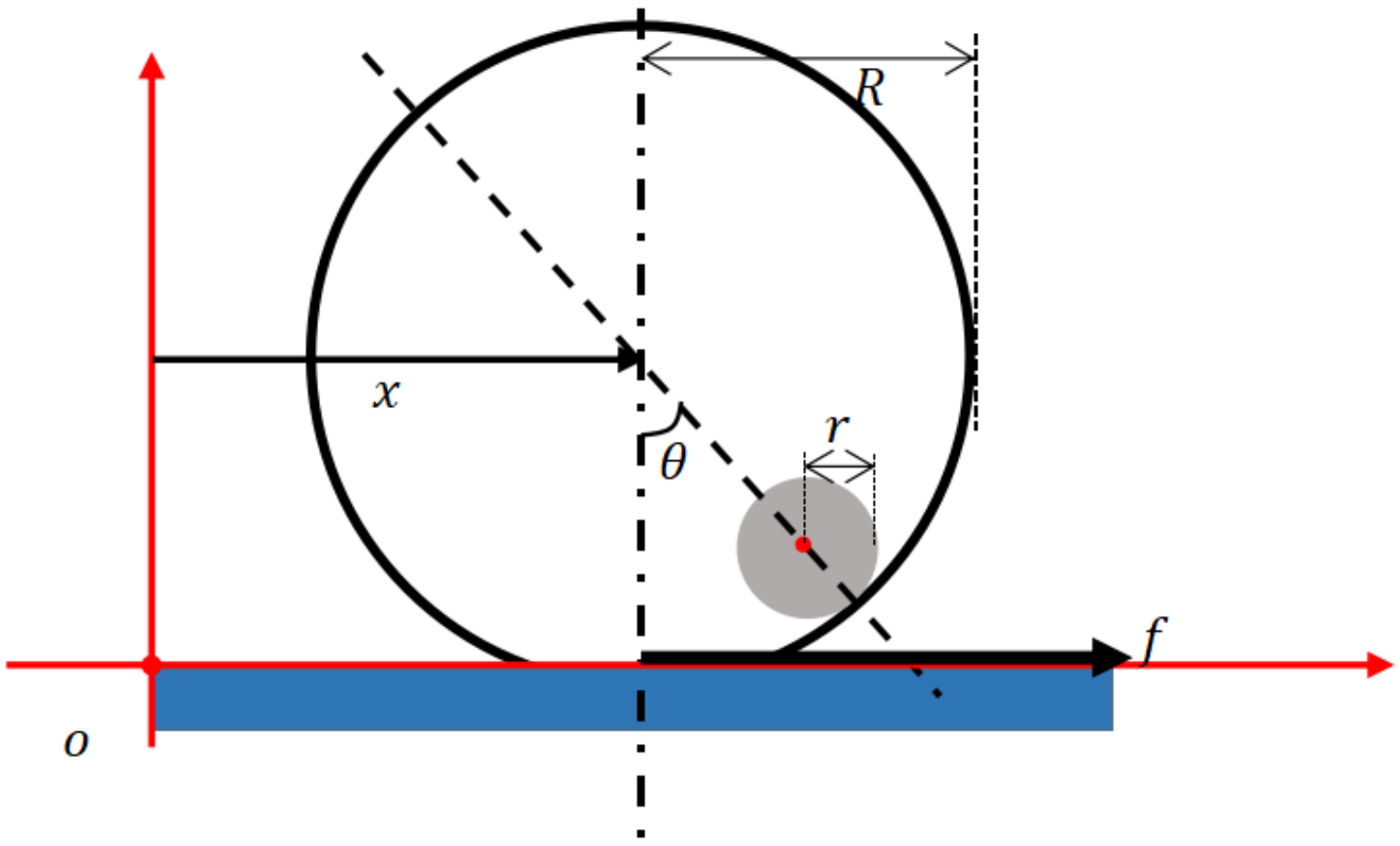

**Figure 1.12 Physical diagram**

Assume that the friction on the bottle is $f$, and take the whole bottle and bead as the system, according to the momentum theorem

$$f = \frac{d}{dt}\{M\frac{dx}{dt} + m\left[\frac{dx}{dt} + (R-r)cos\theta\frac{d\theta}{dt}\right]\}\ldots\ldots①$$

With the bottle as the object, we obtain the following expression according to the angular momentum theorem:

$$-fR = \frac{d}{dt}(MR^2\frac{d(\frac{x}{R})}{dt})\ldots\ldots②$$

Applying Newton's Second Law to the bead in the non-inertial system where the bottle is located

$$m(R-r)\frac{d^2\theta}{dt^2} = -(mgsin\theta + mcos\theta\frac{d^2x}{dt^2})\ldots\ldots③$$

When the angle $\theta$ is small enough, the expressions ① ② ③ can be simplified using $sin\theta \approx \theta, cos\theta \approx 1$:

$$f = (M+m)\frac{d^2x}{dt^2} + m(R-r)\frac{d^2\theta}{dt^2}\ldots\ldots④$$

$$-f = M\frac{d^2x}{dt^2}\ldots\ldots⑤$$

$$(R-r)\frac{d^2\theta}{dt^2} = -(g\theta + \frac{d^2x}{dt^2})\ldots\ldots⑥$$

Add expression ④ and ⑤, and we'll get:

$$(2M+m)\frac{d^2x}{dt^2}+m(R-r)\frac{d^2\theta}{dt^2}=0\ldots\ldots ⑦$$

Set $x=Acos(\omega t+\varphi);\ \theta=Bcos(\omega t+\varphi)$, and we'll get:

$$-A(2M+m)-m(R-r)B=0\ldots\ldots ⑧$$

$$\omega^2A+\omega^2B(R-r)=gB\ldots\ldots ⑨$$

And apply expression ⑧ as follows:

$$A=\frac{-m(R-r)}{2M+m}B\ldots ⑩$$

Here B is a constant that only comes from the initial conditions.

From ⑨ and ⑩,

$$\omega^2\frac{-m(R-r)}{2M+m}+\omega^2(R-r)=g\ldots\ldots ⑪,$$

$$\omega^2=\frac{2M+m}{2M}\frac{g}{R}\ldots\ldots ⑫$$

Obviously, there is another solution, i.e. $A=B=0;\ \omega^2=0;\ x=vt$.

Conclusion:

Solution 1: $x=Acos(\omega t+\varphi);\ \theta=Bcos(\omega t+\varphi)$,

where $A=\frac{-m(R-r)}{2M+m}B$; $\omega^2=\frac{2M+m}{2M}\frac{g}{R}$ . (14)

Solution 2: $A=B=0;\ \omega^2=0;\quad x=vt$.

$\omega$ is the angular vibration frequency of the bead and the bottle as well. In Solution 2, the bottle kepts in uniform motion while the bead stays stationary at the bottom of the bottle, which is not easy to achieve in experiment.

In practice, the general motion of the steel ball and the bottle is often a superposition of the above two solutions, that is, uniform linear motion and simple harmonic vibration.

If the angle $\theta$ cannot be set as small, reduce f in expressions ① ② and simplify expressions ① ② ③ to get the following equations:

$$\frac{d^2\theta}{dt^2} = \frac{-mgsin\theta - mcos\theta\frac{d^2x}{dt^2}}{m(R-r)} \ldots\ldots ⑬$$

$$\frac{d^2x}{dt^2} = \frac{-m(R-r)cos\theta * \frac{d^2\theta}{dt^2} + m(R-r)sin\theta * (\frac{d\theta}{dt})^2}{2M+m} \ldots\ldots ⑭$$

As there is no analytical solution in this expression, we bring it into MatLab programming to obtain the numerical solution by Runge-Kutta method and draw the diagrams of time-linear velocity, time-displacement, time-angular velocity and time-angular displacement (the constant values comes from a set of experimental data).

### 1.2.2 Motion mode analysis

The following diagrams can be divided into two motion modes. When the bottle has no initial velocity, its motion mode is similar to simple harmonic vibration, as the analytical solution (14) above. The time-displacement and time-angular displacement are shown in Figure 1.13 respectively .

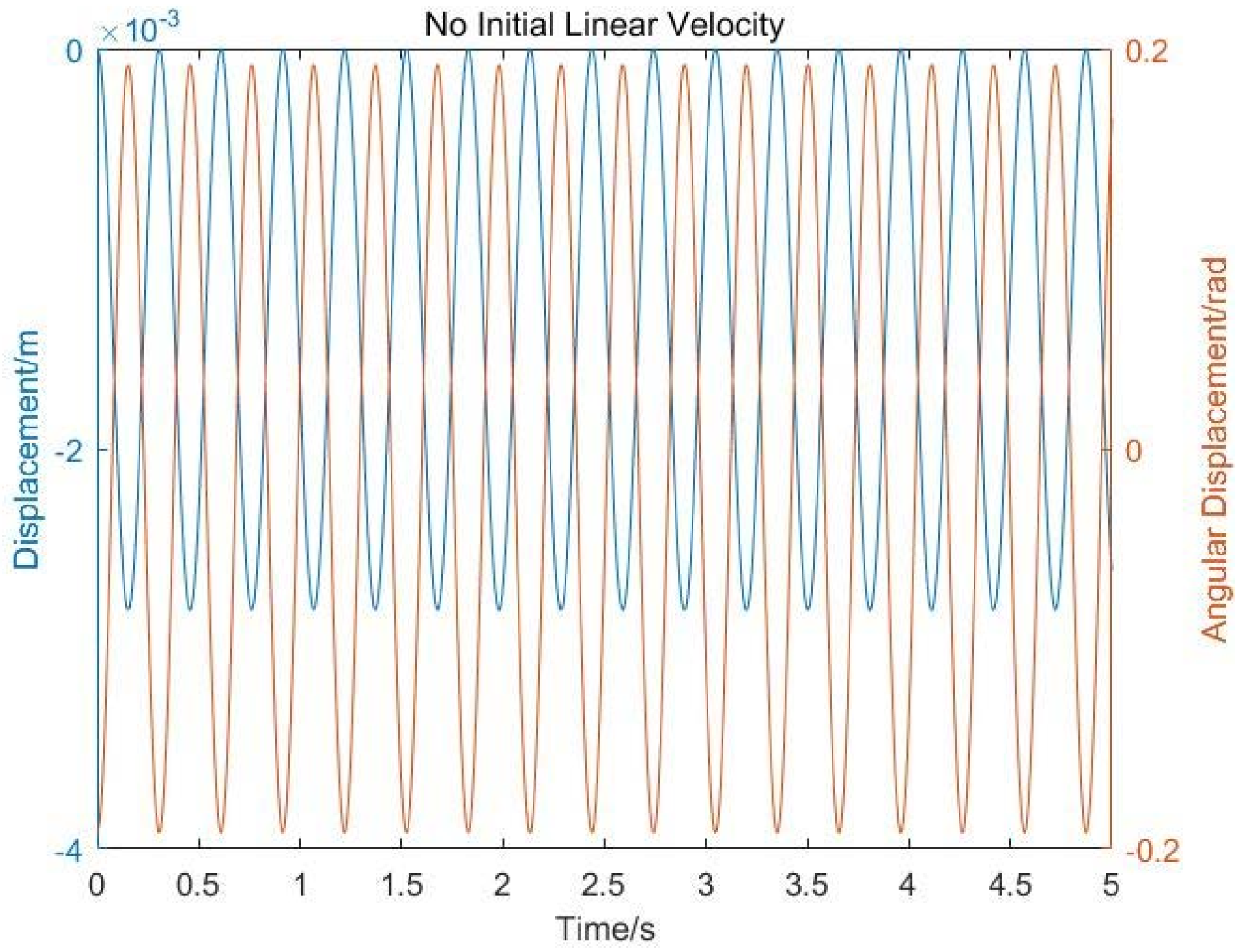

**Figure 1.13 System motion without initial linear velocity**

When the bottle has initial linear velocity, however, the motion mode of the bottle in this case is a superposition of horizontal motion and vibration with angular displacement still vibrating (see Figure 1.14).

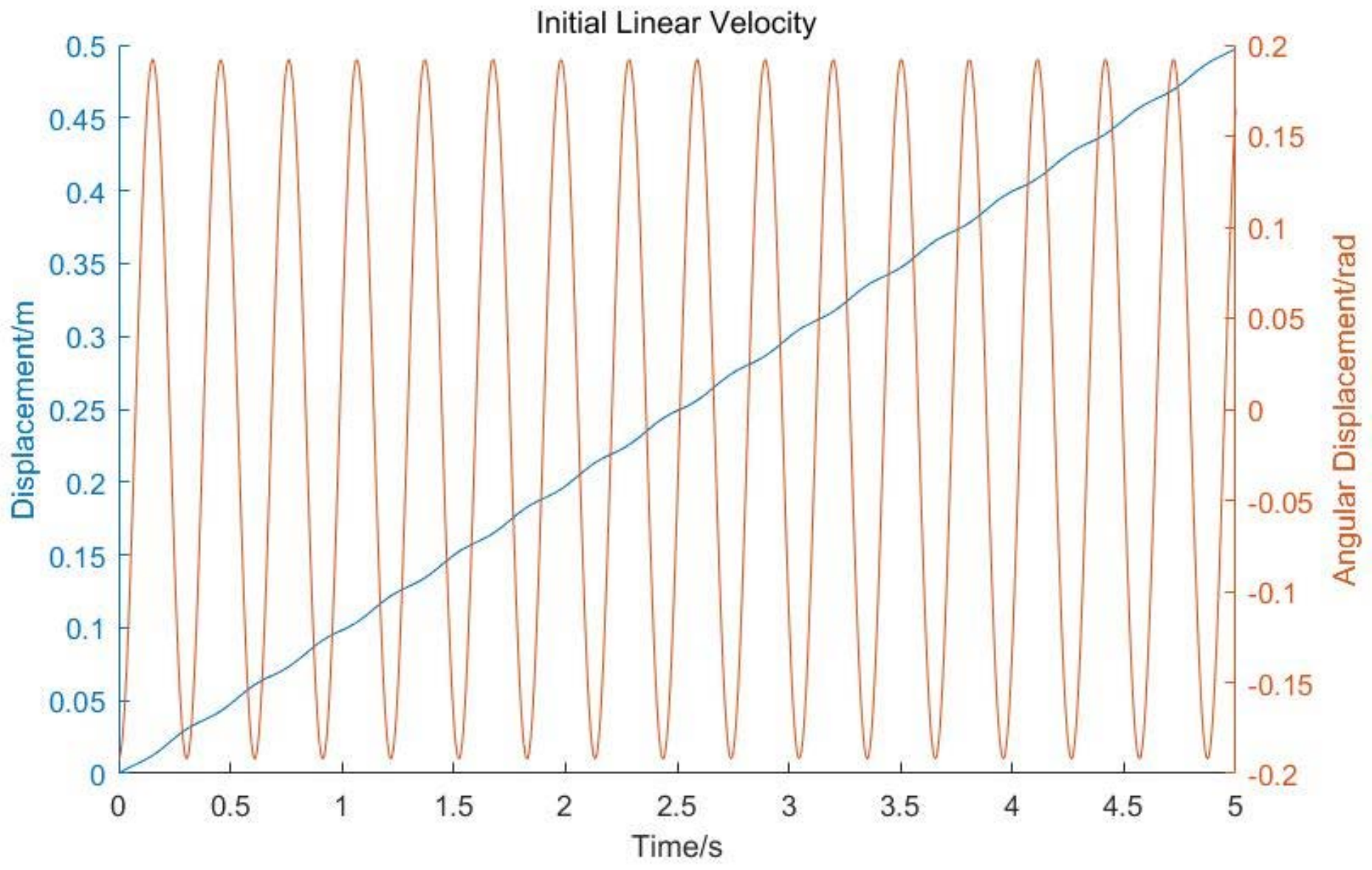

**Figure 1.14 System motion without initial linear velocity**

However, different from the actual experiment, neither the damping of the bead or water in the bottle nor that of the bottle on the ground are given in this diagram. Therefore, damping term is not reflected in the motion expressions above.

# 2. Experiment

## 2.1 Throwing

### 2.1.1 Initial Measurement - Coefficient of Restitution

To verify the buffering effect of water in falling bottles, we measured the falling height and rebound height of free-falling an empty bottle and bottle with different water amounts, and calculated their restitution coefficients.

The experimental devices are shown in Figure 2.1. There is a fixed ruler on the wall. The bottle is released vertically from a stationary state at a certain height and photographed by a camera to measure its rebound height.

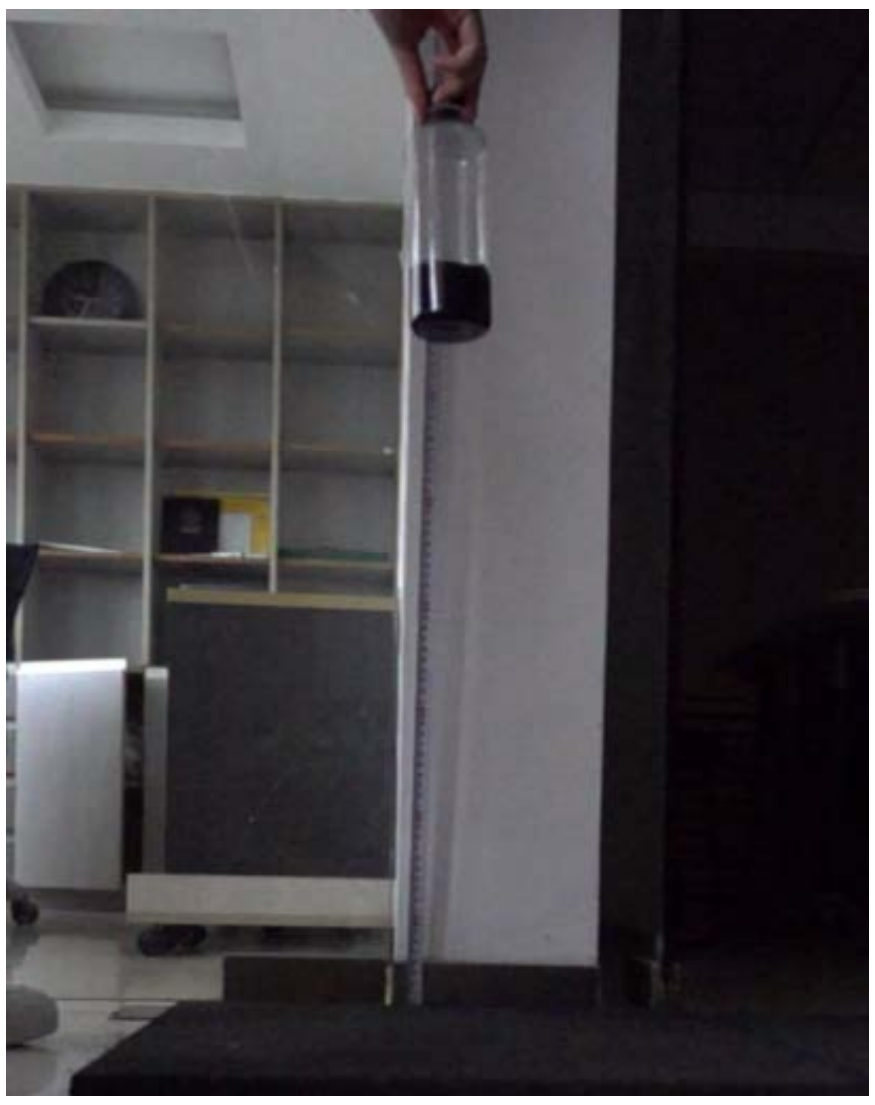

**Figure 2.1 Measure the coefficient of restitution**

We release the bottle with different amount of water (0%, 25%, 55%, 75% and 100%), from different heights ($h_1$) and bring this with the rebound height $h_2$ into $e = \sqrt{\frac{h_2}{h_1}}$ to calculate the coefficient of restitution.

The experimental data are as follows:

| **Water amount 100%** | | | |
|---|---|---|---|
| **Release height/(cm)** | **Rebound height/(cm)** | **Coefficient of restitution** | **Average** |
| 72 | 36 | 0.707 | 0.702 |
| 70 | 34 | 0.697 | |

| **Water amount 75%** | | | |
|---|---|---|---|
| **Release height/(cm)** | **Rebound height/(cm)** | **Coefficient of restitution** | **Average** |
| 92 | 26 | 0.532 | 0.568 |
| 92 | 29 | 0.561 | |
| 48 | 17 | 0.595 | |
| 74 | 23 | 0.558 | |
| 93 | 33 | 0.596 | |

| Water amount 55% | | | |
|---|---|---|---|
| **Release height/(cm)** | **Rebound height/(cm)** | **Coefficient of restitution** | **Average** |
| 78 | 11 | 0.376 | 0.335 |
| 69 | 6 | 0.295 | |

| Water amount 25% | | | |
|---|---|---|---|
| **Release height/(cm)** | **Rebound height/(cm)** | **Coefficient of restitution** | **Average** |
| 64 | 7 | 0.331 | 0.378 |
| 81 | 10 | 0.351 | |
| 68 | 10 | 0.383 | |
| 58 | 9 | 0.394 | |
| 80 | 15 | 0.433 | |

| Water amount 0% | | | |
|---|---|---|---|
| **Release height/(cm)** | **Rebound height/(cm)** | **Coefficient of restitution** | **Average** |
| 72 | 33 | 0.677 | 0.674 |
| 69 | 31 | 0.670 | |

**Table 2.1 Coefficient of restitution**

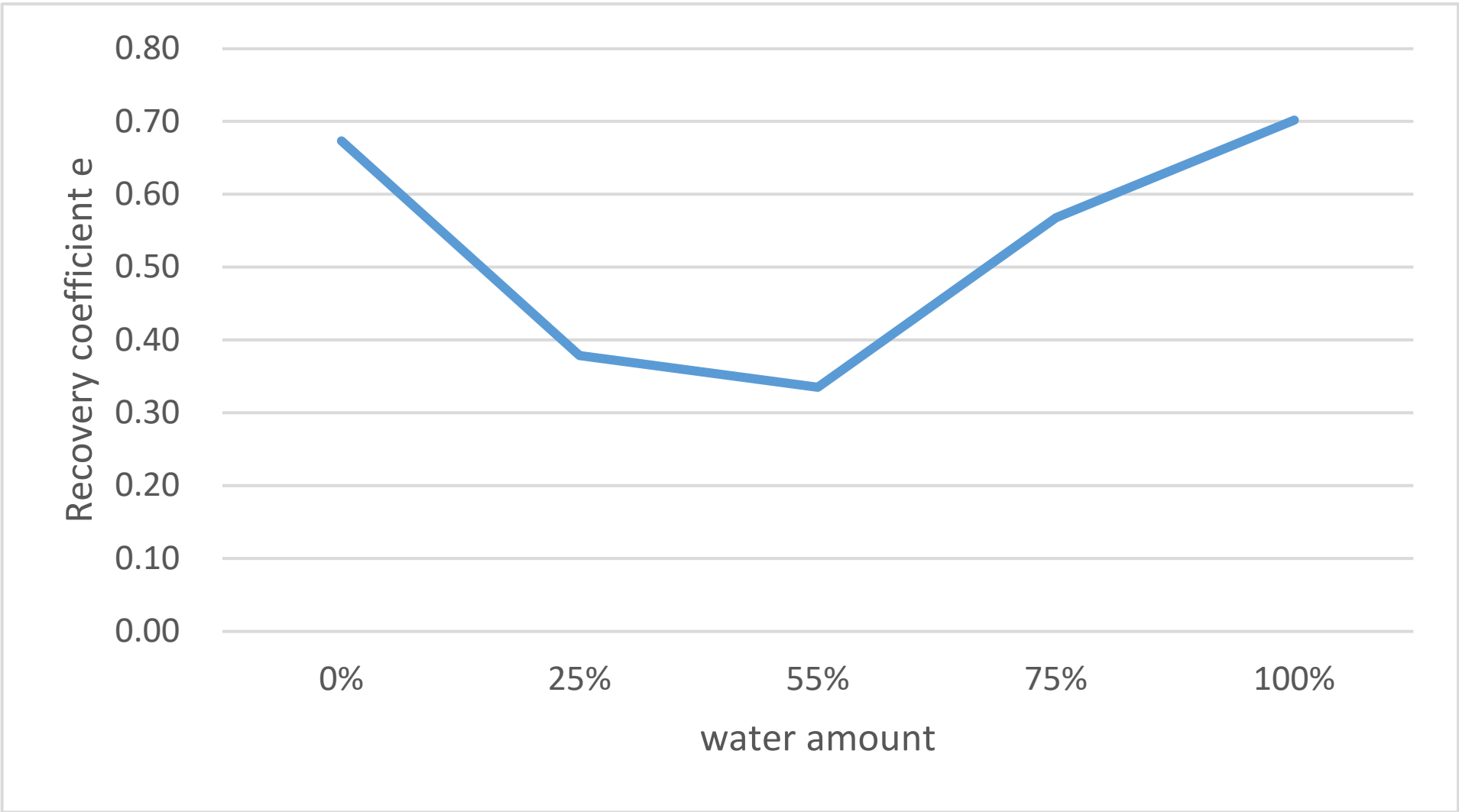

**Figure 2.2 Coefficients of restitution with different water amount**

It can be seen from figure 2.2, the amount of water could largely buffer the rebound of the bottle. When the amount of water is between 25% and 55%, the bottle coefficient of restitution is the smallest and the buffer effect is the largest.

### 2.1.2 Probability of Upright Landing after Free Falling

For all the following experiments, we released the bottle vertically for 100 times at the same condition (height, water amount, angle etc.), and then count the number of upright landings.

Figure 2.3 shows the probability that after released vertically from the height of 20cm and 40cm, a bottle containing 20% water has a probability of no more than 2% to land upright. And none was successful from higher heights.

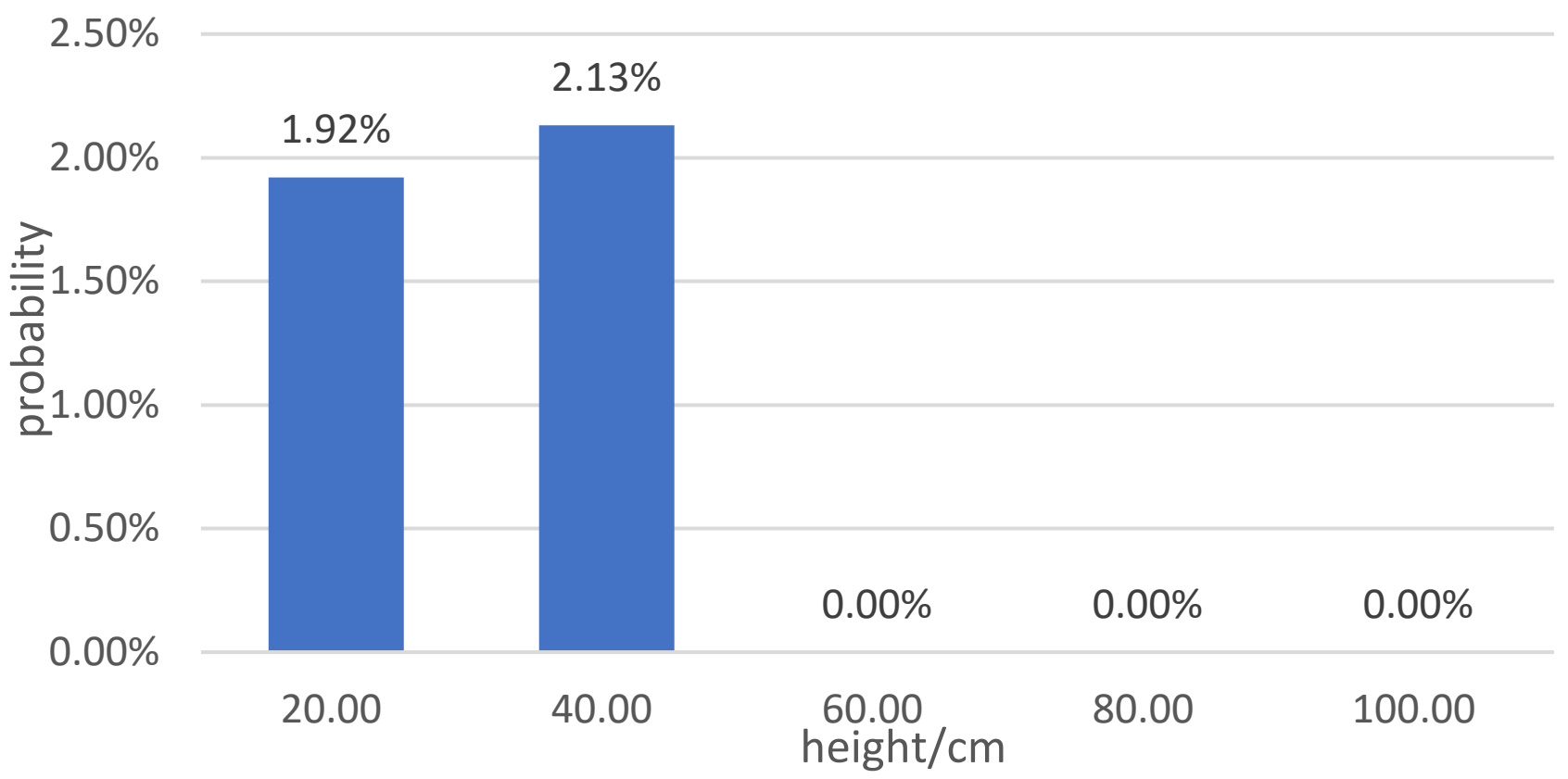

**Figure 2.3 Landing upright probability and release height - 20% water**

The figure2.4 shows the probability that after released vertically from 2cm, a bottle containing 35% water has a probability of 50 % to successfully land upright, and it only has no more than 10% to land upright from higher heights:

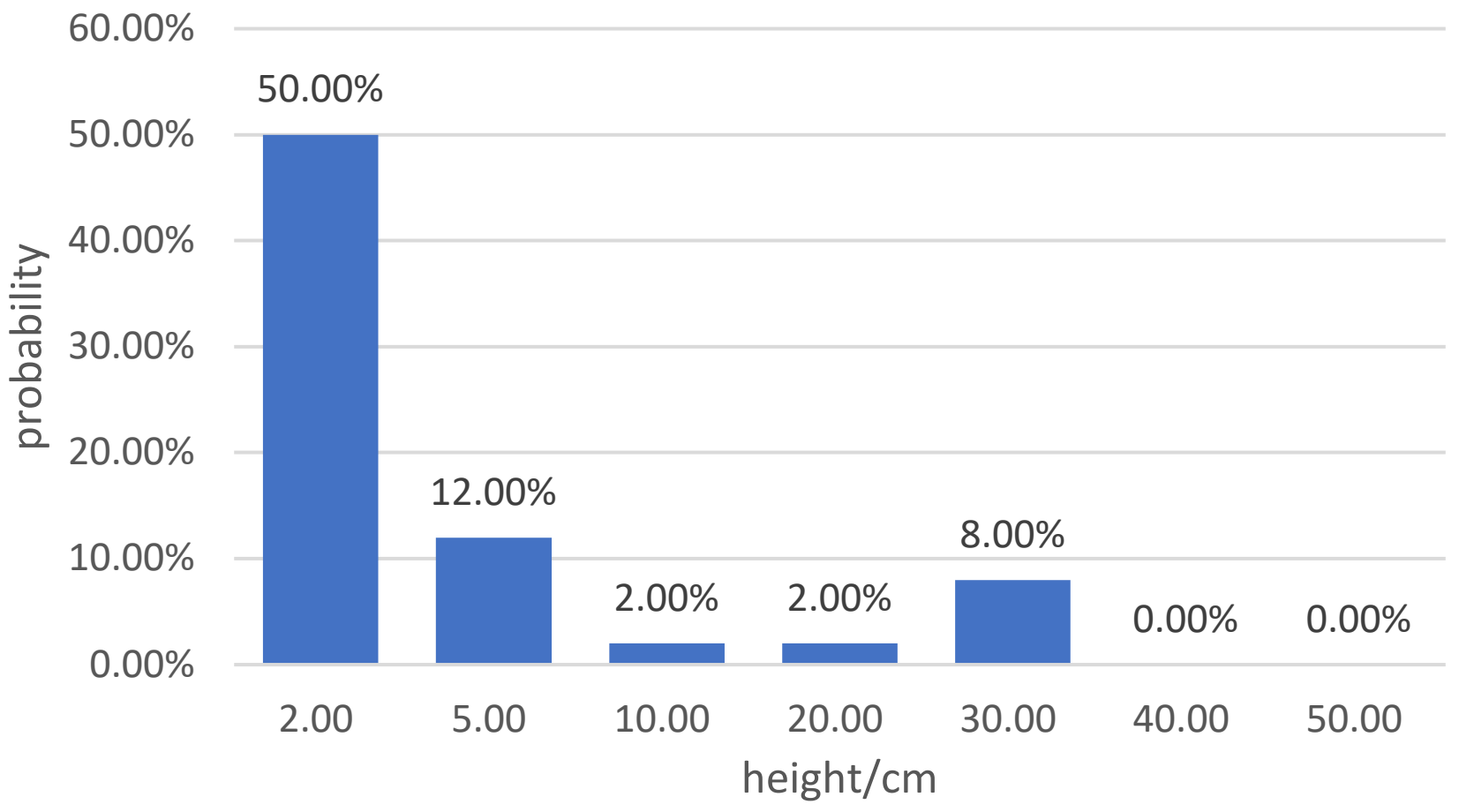

**Figure 2.4 Landing upright probability and release height - 35% water**

Figure2.5 shows the probability that after released vertically from 20cm or 40cm, a bottle containing 40% water has a probability of no more than 5% to successfullyland upright, and none was successful to land upright from other heights.

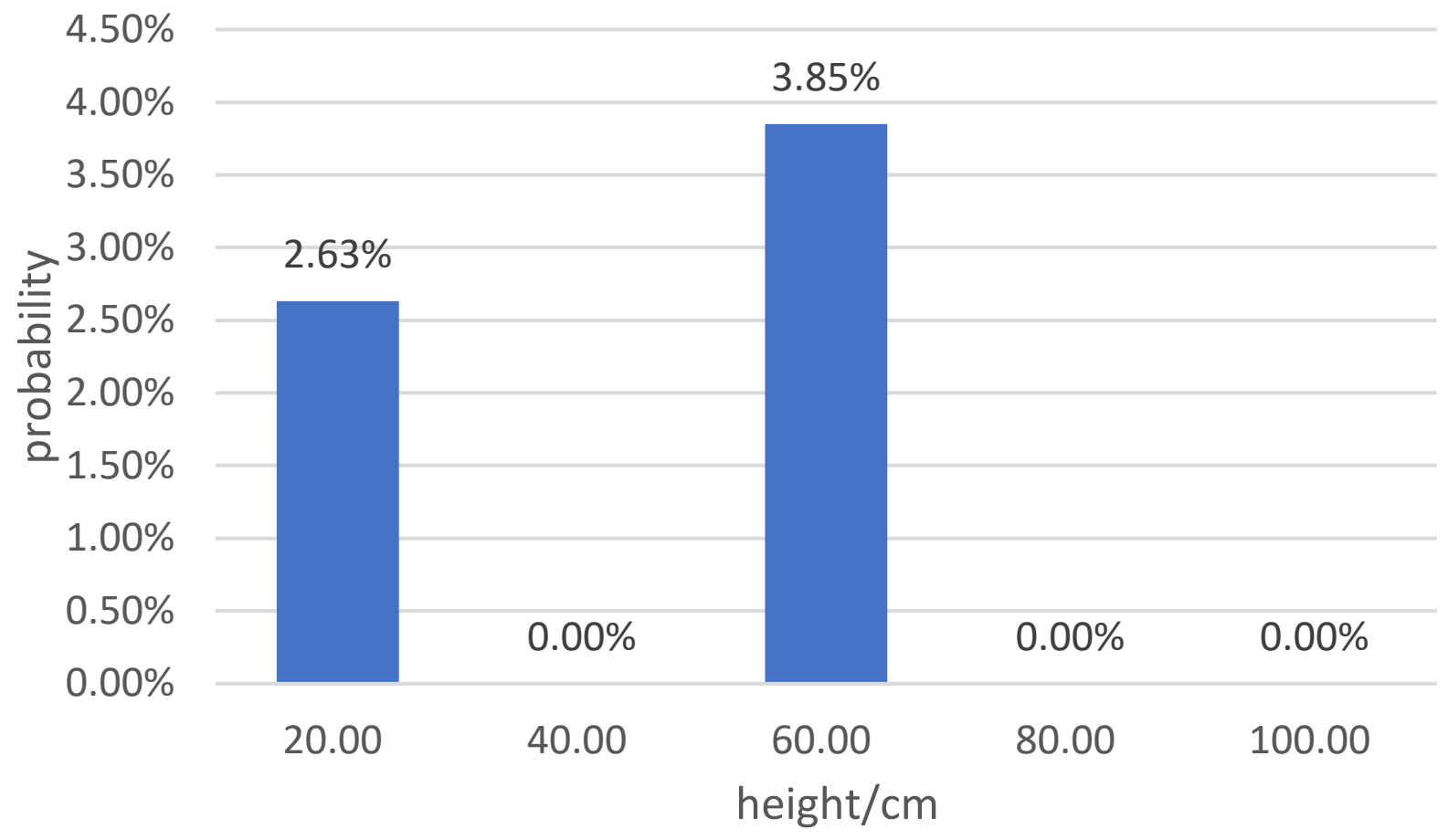

**Figure 2.5 Landing upright probability and release height - 40% water**

Figure2.6 shows the probability that after released vertically from 20cm, a bottle containing 60% water has a probability of about 10 % to successfully land upright. It only has no more than 10% to land upright from higher height, and it cannot land upright when released from 6ocm:

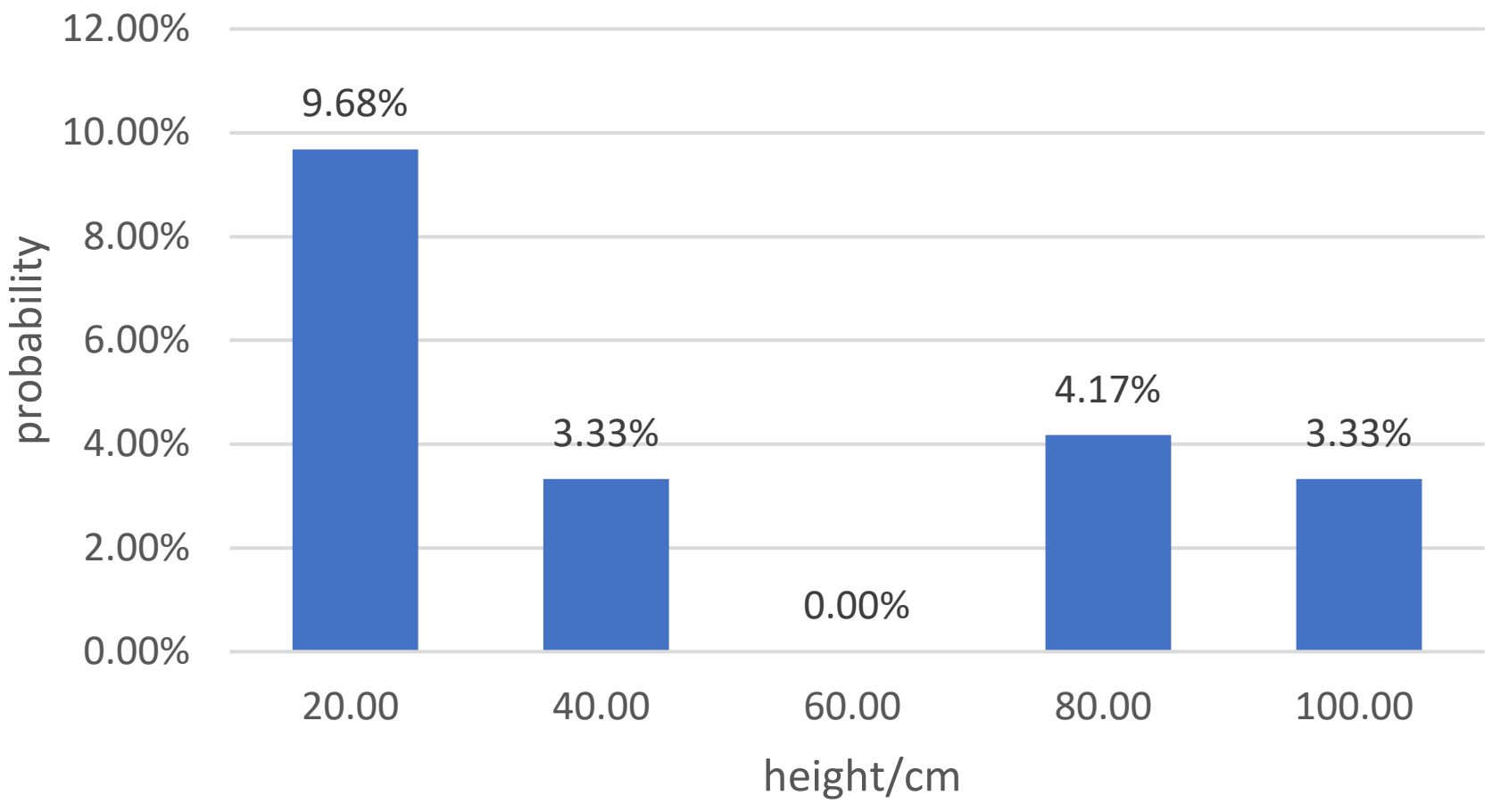

**Figure 2.6 Landing upright probability and release height - 60% water**

Figure2.7 shows the probability that after released vertically from 20cm or 60cm, a bottle containing 80% water has a probability of no more than 5% to successfully land upright, and none was successful to land upright from other heights:

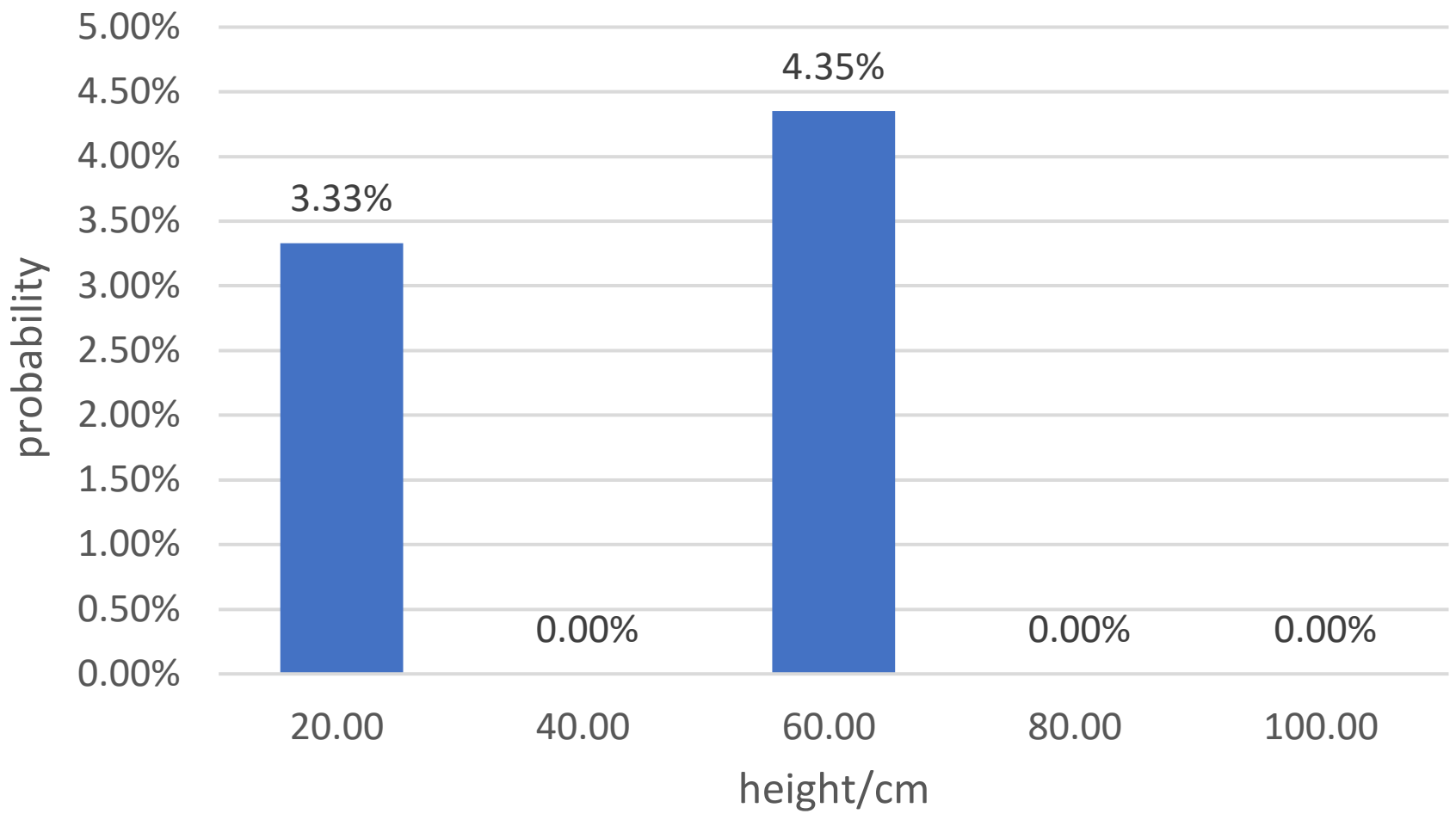

**Figure 2.7 Landing upright probability and release height - 80% water**

Figure2.8 shows the probability that a bottle containing 100% water cannot land upright after it is released vertically from any height:

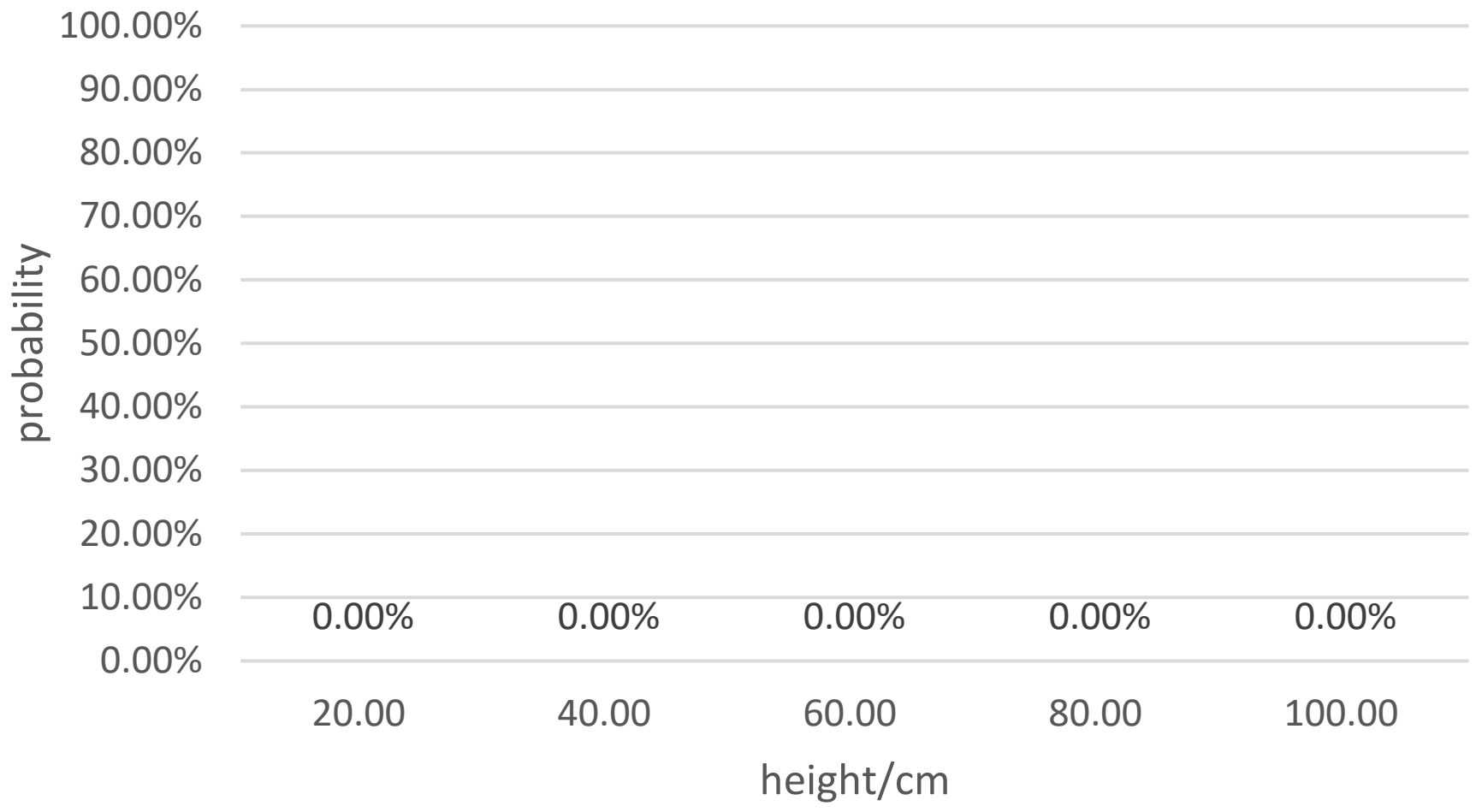

**Figure 2.8 Landing upright probability and release height - 100% water**

In general, the upright landing probability is pretty low for free falling. Before collision, the bottle's angular momentum is nearly zero. When the bottle hits the ground, one side usually hits the ground first, and the impulsive moment during collision may easily cause the bottle to tip over since the bottle has little angular momentum to resist the impulse.

### 2.1.3 Probability of Upright Landing after Flipping

In the flipping experiment, before releasing, the bottle is placed upside down (horizontal) on the platform to realize a 180° (90°) rotation. Then it is released with a slight push.

Figure2.9 shows the probability that a bottle containing 55% water has a highest probability to land upright after it is released from angle 180° at a height of 86.4cm, it has a probability of no more than 6% to land upright from angle 90°, and it cannot land upright from angle 270°:

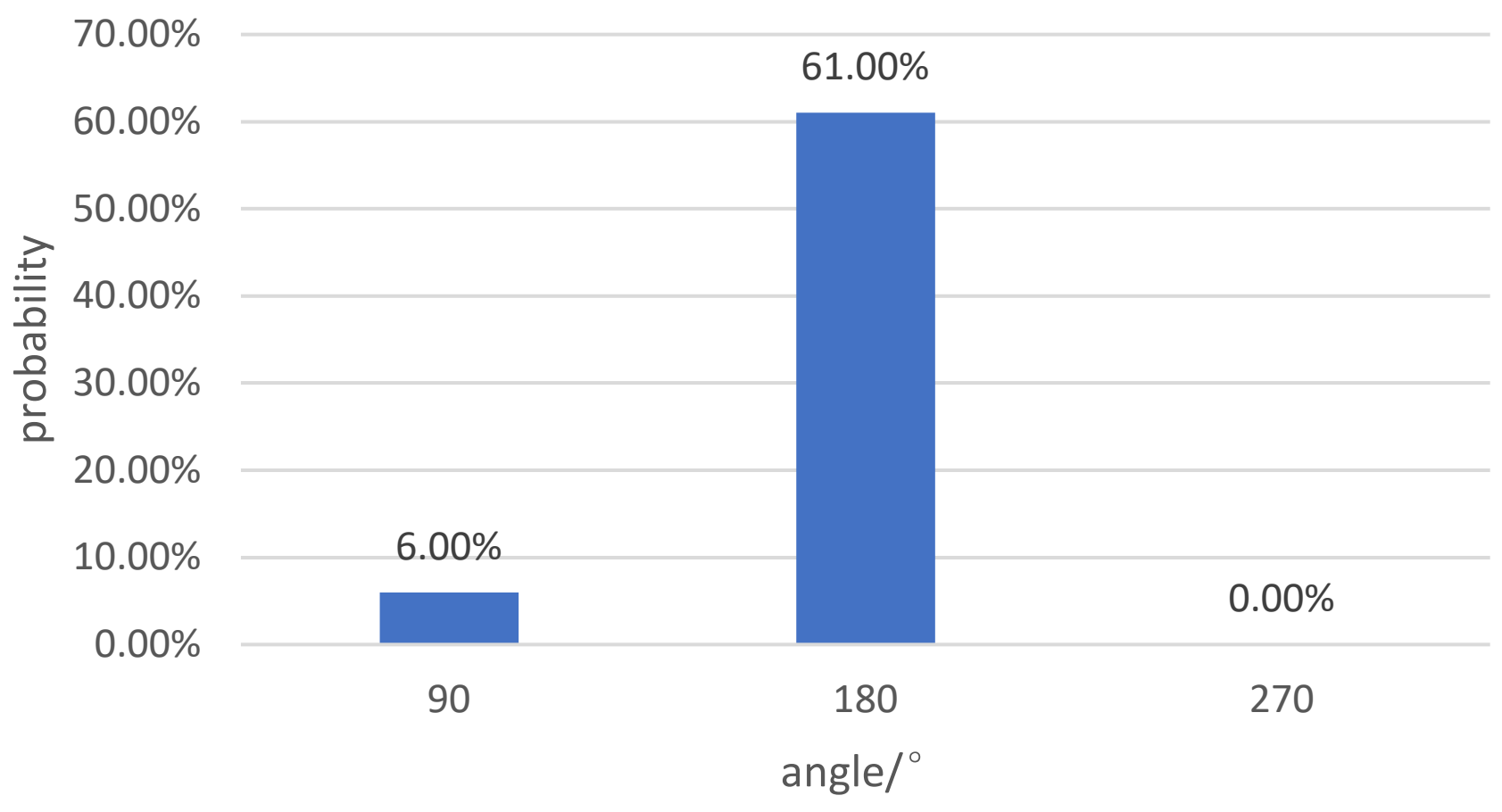

**Figure 2.9 Landing upright probability and release angle - regular bottle**

We also did experiments with irregular shaped bottle. Figure 2.10 shows that the probability of landing upright is still high when the bottle changes to an irregular bottle but maintains the same water amount and release height.

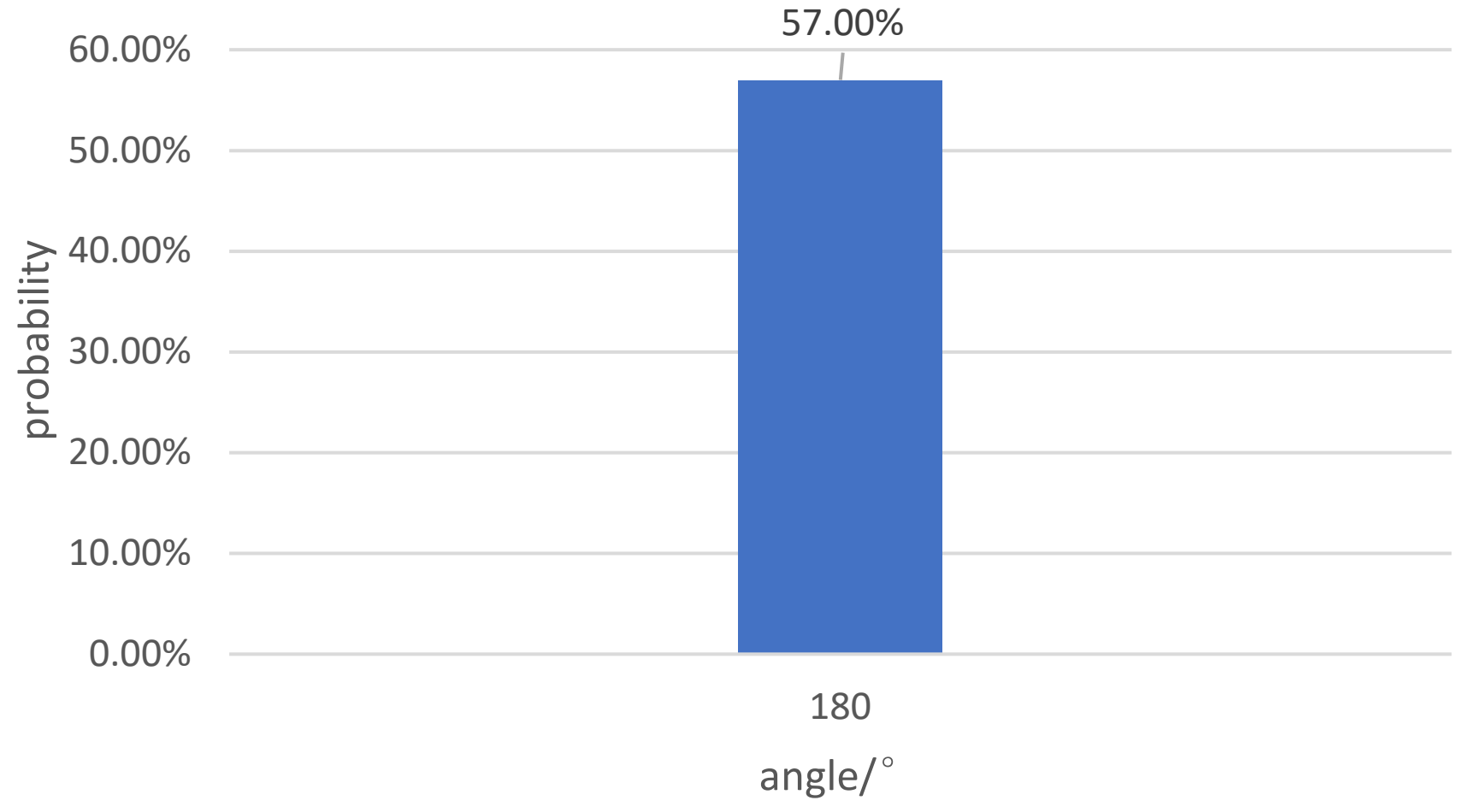

**Figure 2.10 Landing upright probability and release angle - irregular bottle**

Figure2.11 shows the probability that a bottle containing 35% water has highest probability, about 50%, to land upright after it is released from 180° rotating angle at the height of 50- 90cm, it has no more than 40% to land upright from higher or lower heights:

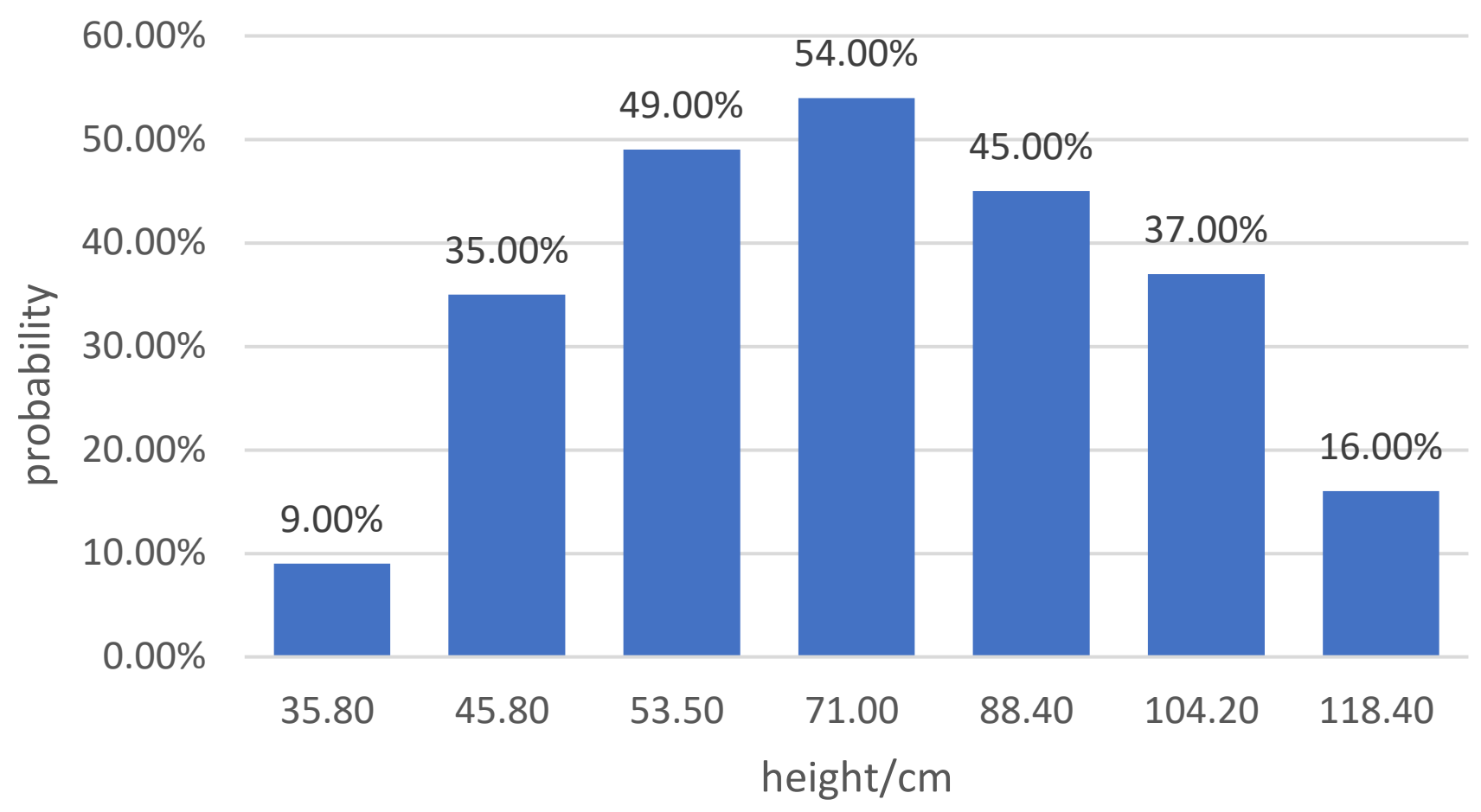

**Figure 2.11 Landing upright probability and release height**

Figure2.12 shows the probability that a bottle containing 35% to 45% amount of water has the highest probability (54% ) to land upright after it is released from 180° rotating angle at a height of 71cm, it has no more than 40% to land upright from higher or lower heights:

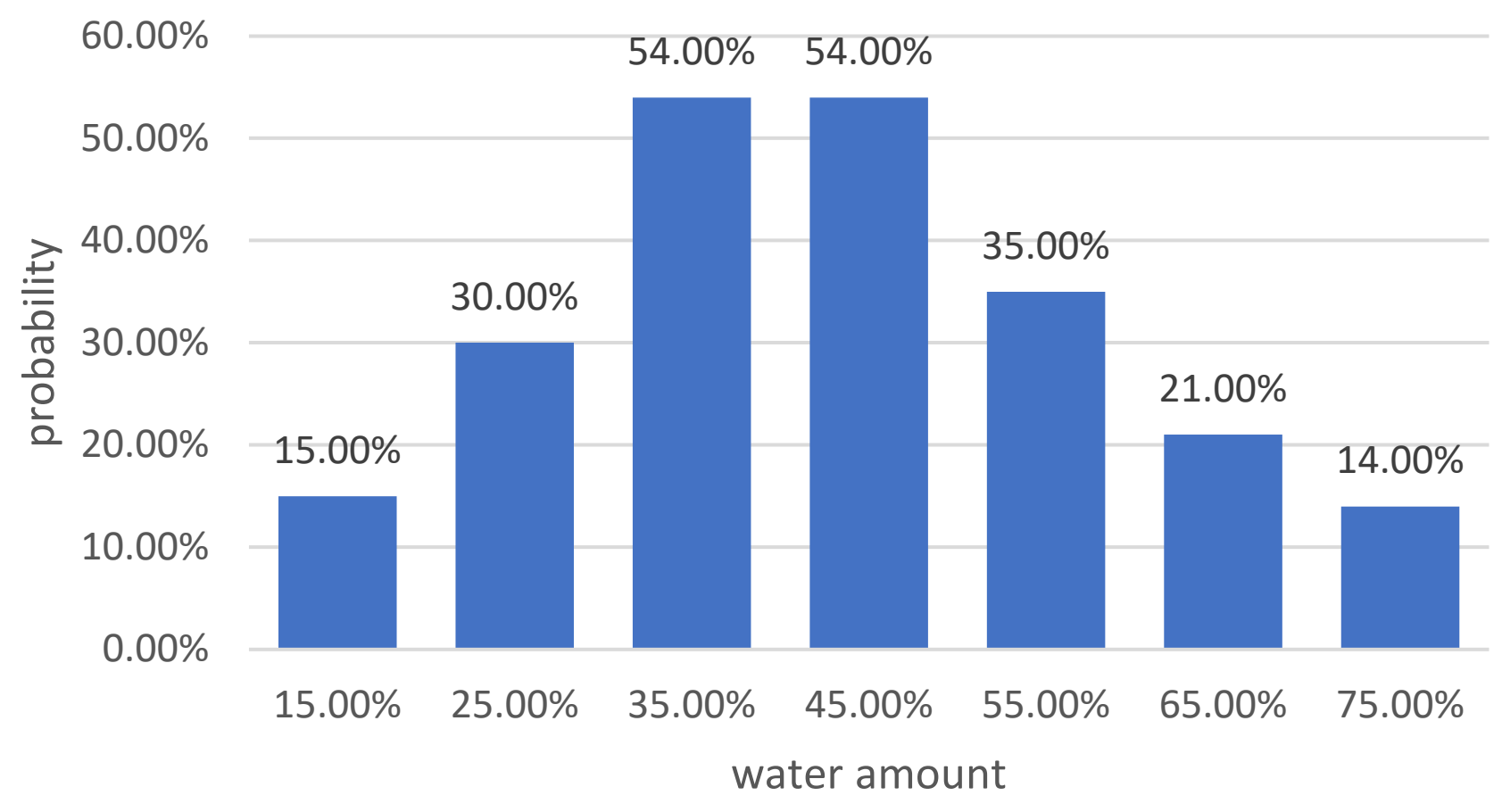

**Figure 2.12 Landed upright probability and water amount**

Conclusion: we find the conditions contributed to the upright landing of a bottle, including a certain rotation angle, water amount and appropriate releasing height.

Figure 2.13 shows that a bottle containing 55% water has a higher probability to land upright on the rough ground after it is released from 180° rotating angle at a height of 86.4cm

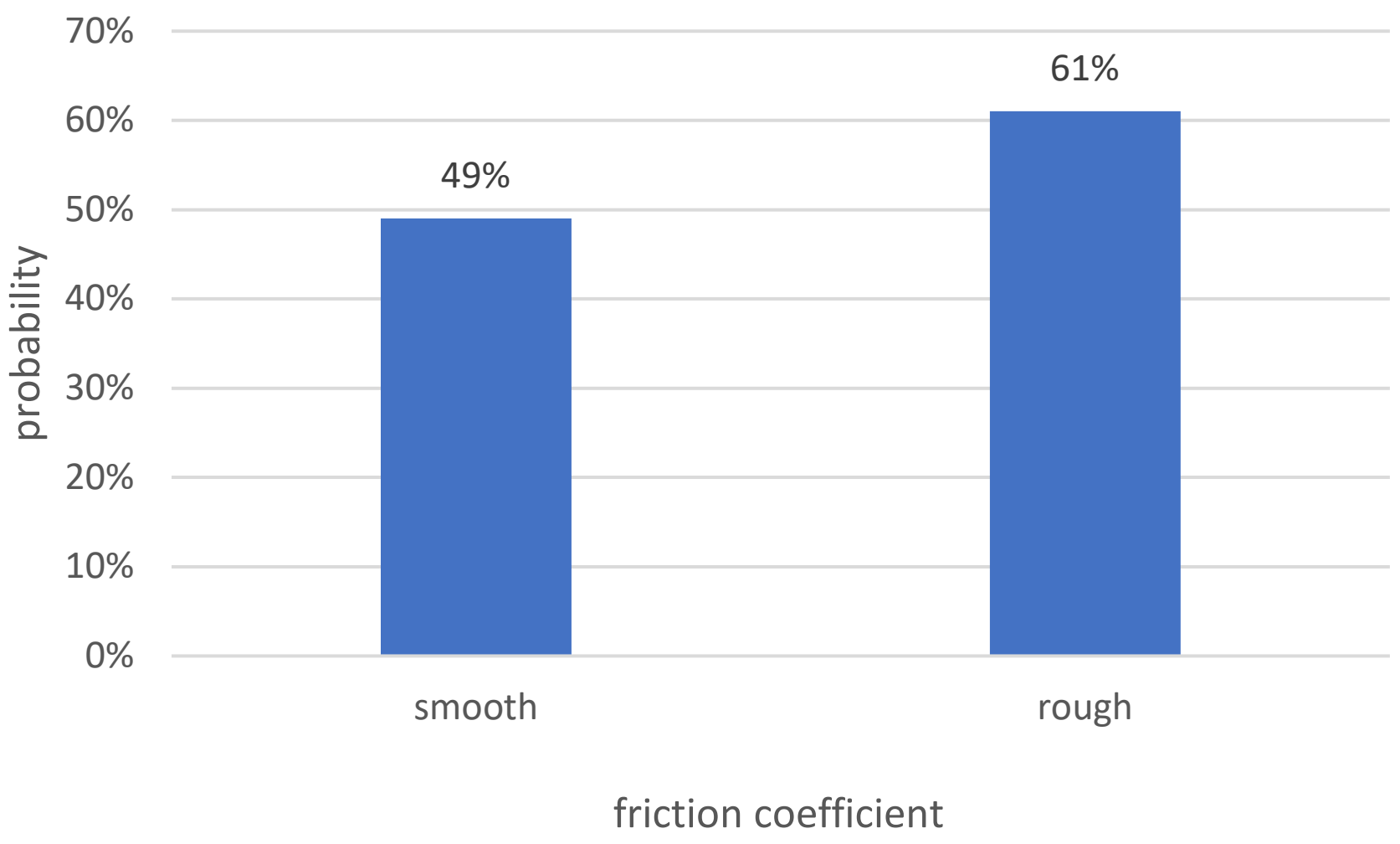

**Figure 2.13 Landed upright probability and ground friction coefficient**

Conclusion：Bottles are more likely to land upright on the ground with a larger friction coefficient, which means friction is conducive to the upright of the bottle.

Figure 2.14 shows that bottles containing 55% water and 55% non-newtonian fluid have almost the same probability to land upright on the rough ground after it is released from 180° rotating angle at a height of 86.4cm

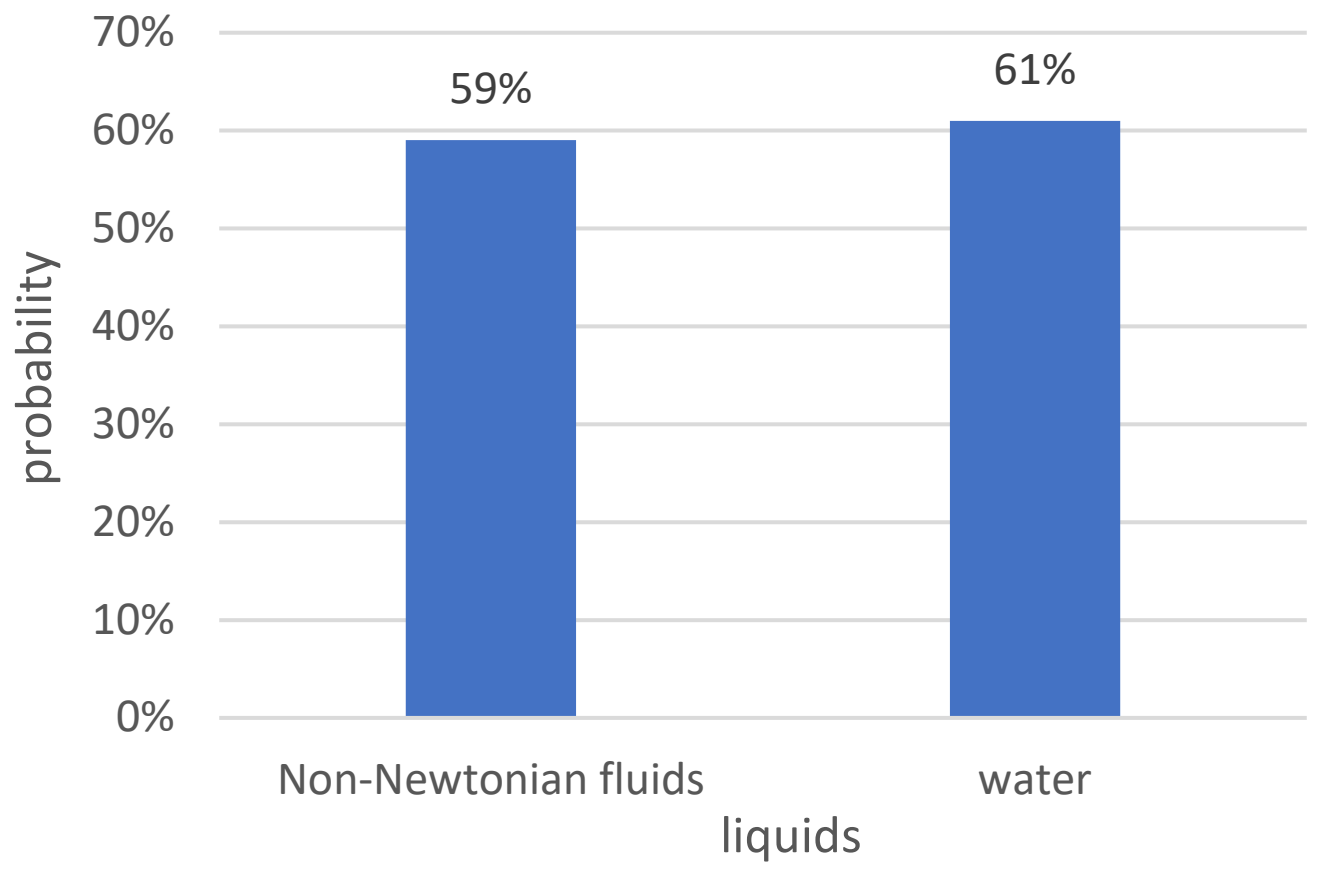

**Figure 2.14 Landed upright probability and liquids**

Conclusion：Non-Newtonian fluids have no significant stabilizing effect

### 2.1.4 Discussion: the impact of collision position

After controlling the amount of water, height, and rotation angle, we found that in some cases the probability the bottles landing upright reaches more than 50%, and among these, the rotation angle or angular momentum is the key to factor. Now we subdivide the first landing into left-first and right-first. As shown in Figure 2.15, left-first refers to the side closest to the platform landed first and the right-first refers to the side far from the platform landed first.

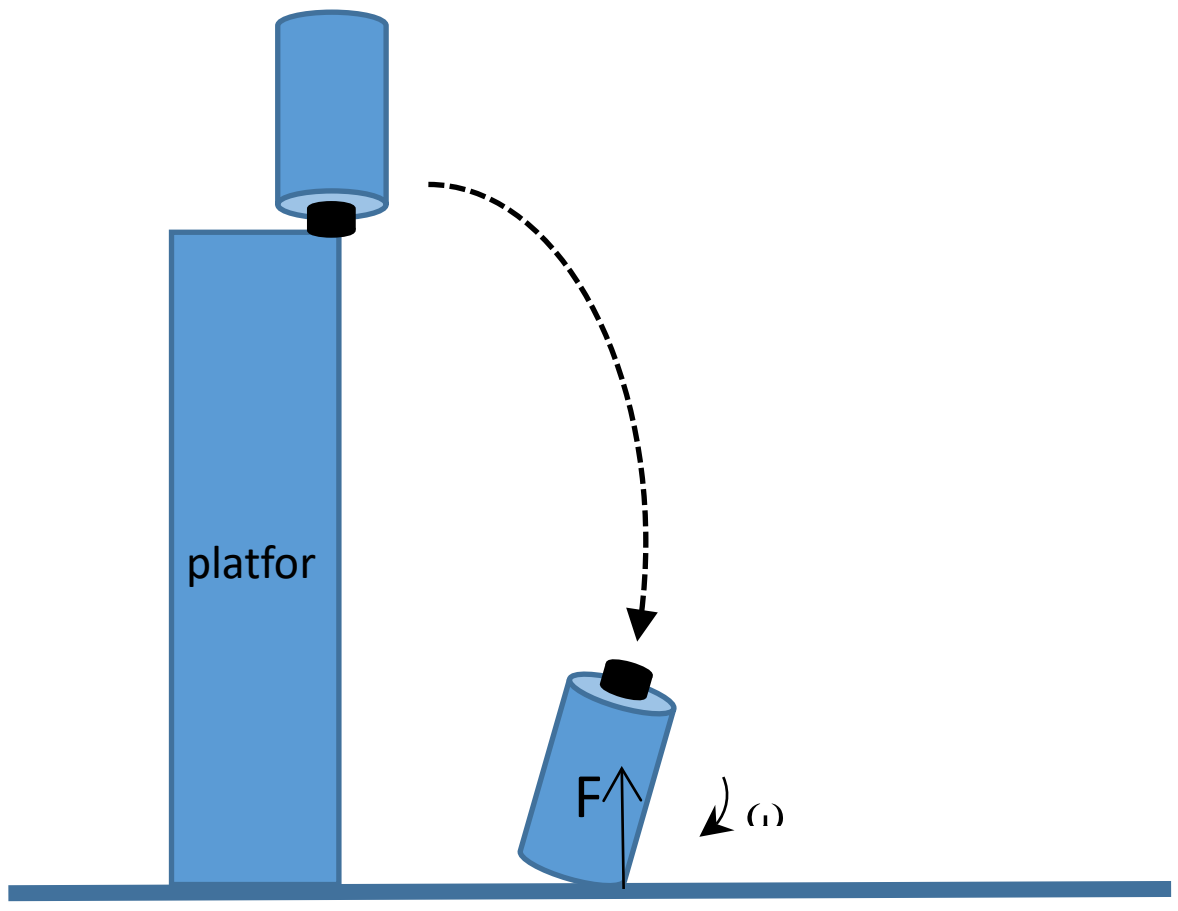

**Figure 2.15 Landing after rotation (right-first landing)**

In the experiment, we keep 55% water with a release height of 86.4cm while changing the rotation angle, and photographed by a camera (see Figure 2.16 and Figure 2.17) .

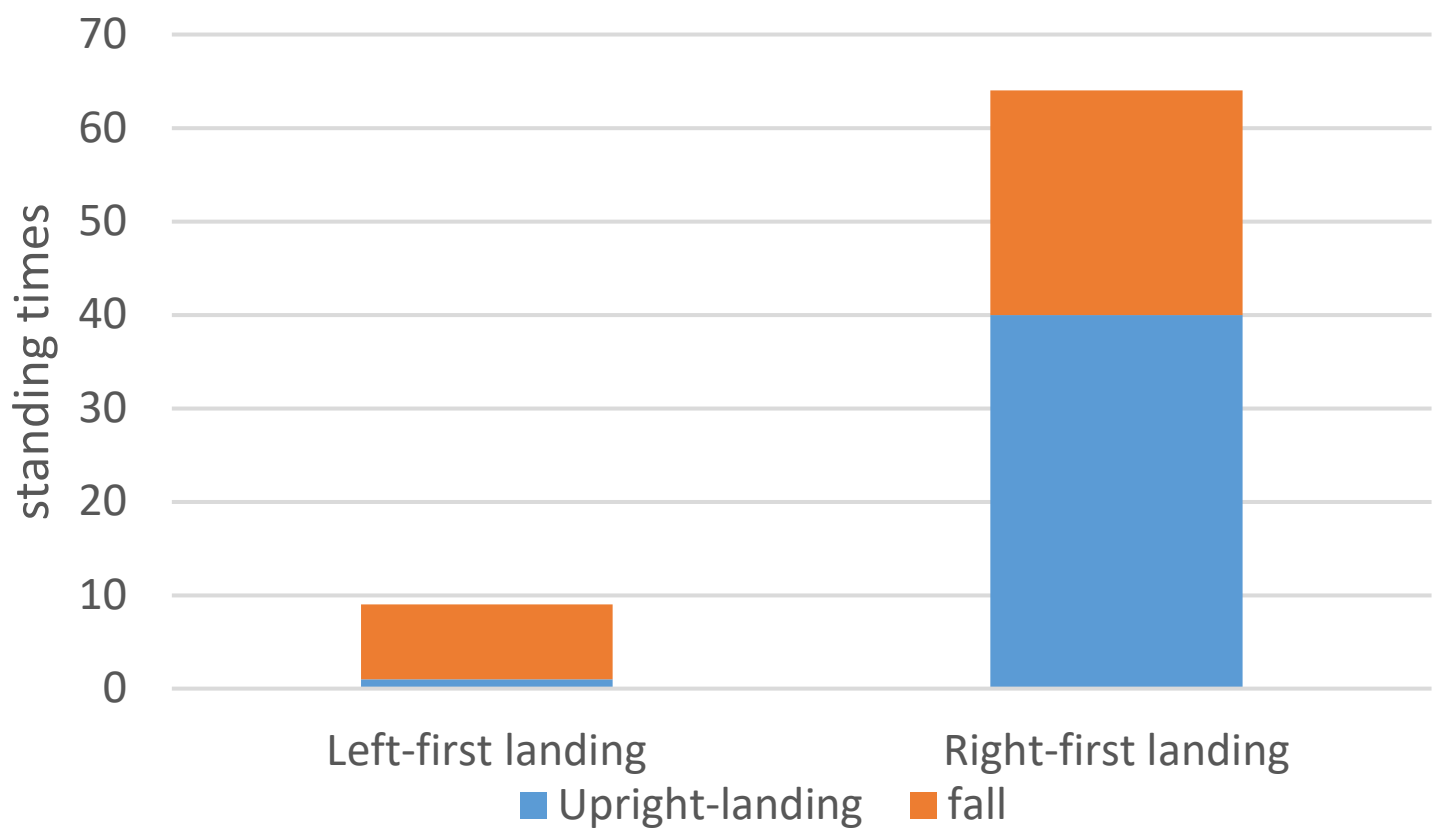

**Figure 2.16 Bottles landed times - 180° rotating angle**

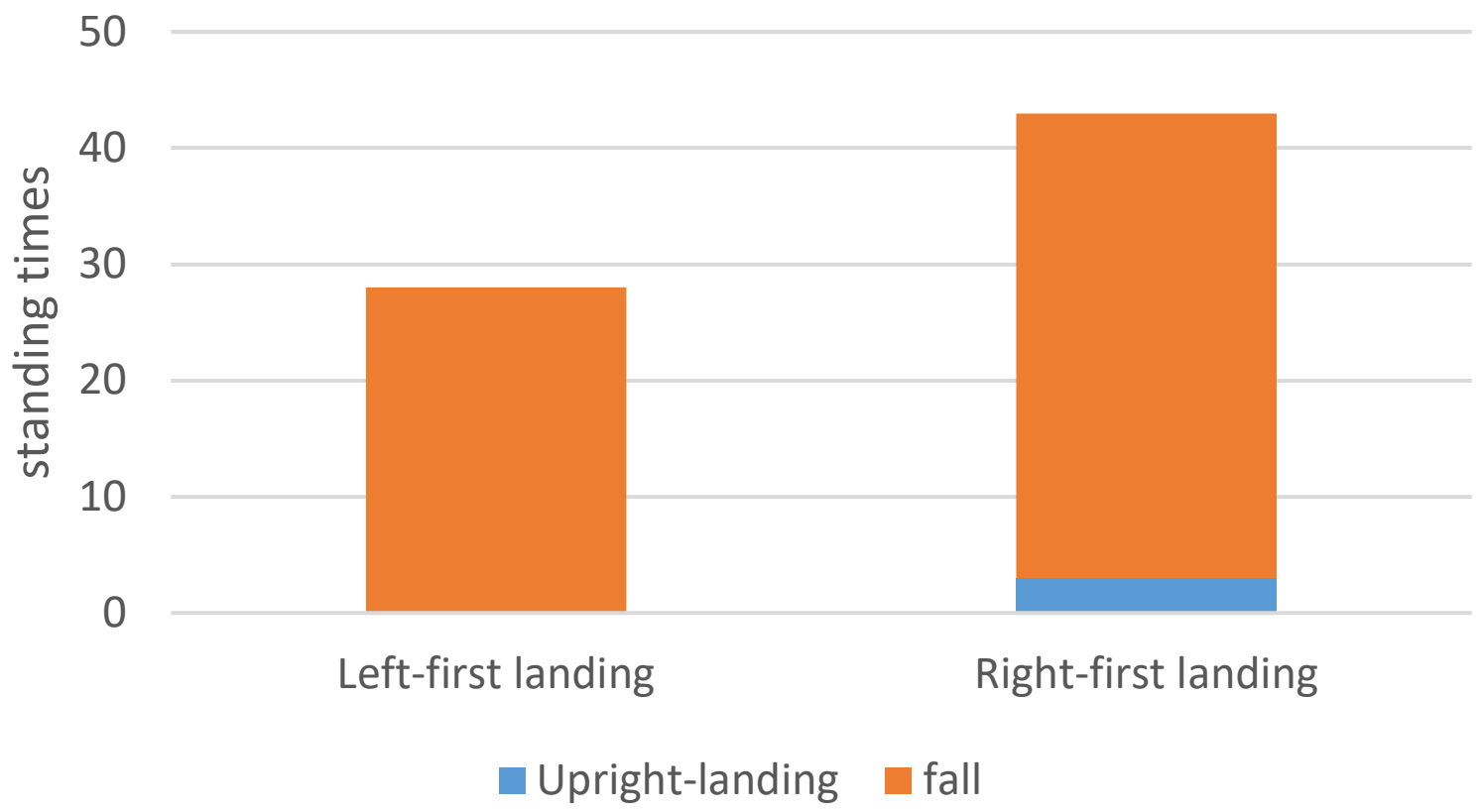

**Figure 2.17 Bottles landed - 90° rotating angle**

The results are shown in Figure 2.16 and figure 2.17, that is, when the bottle rotates 180° or 90°, the probability for the right-first upright-landing is significantly higher than that of the left-first. And the probability for 180° rotation is the highest.

When landing on the right, the impulsive moment from the ground is opposite to the angular momentum of the bottle, and the angular velocity is minimized after the collision. This explains why a certain amount of rotation matched with "good" collision position is essential for upright landing. In contrast, when landing on the left first, the impulsive moment from the ground is in the same direction as the angular momentum of the bottle, thus increasing the rotation of the bottle and causing the bottle to fly out.

## 2.2 Rolling

### 2.2.1 Verification of Theoretical Model

To verify our bottle-and-bead model, we conducted comparison experiment of rolling bottles with steel balls and bottle filled with water.

Experiment devices: two 500ml water bottle, two slopes (11°), ink, 5 steel balls, plane, ruler, weigher, tape, cell phone (with video recording parameters settled) (see Figure 2.18).

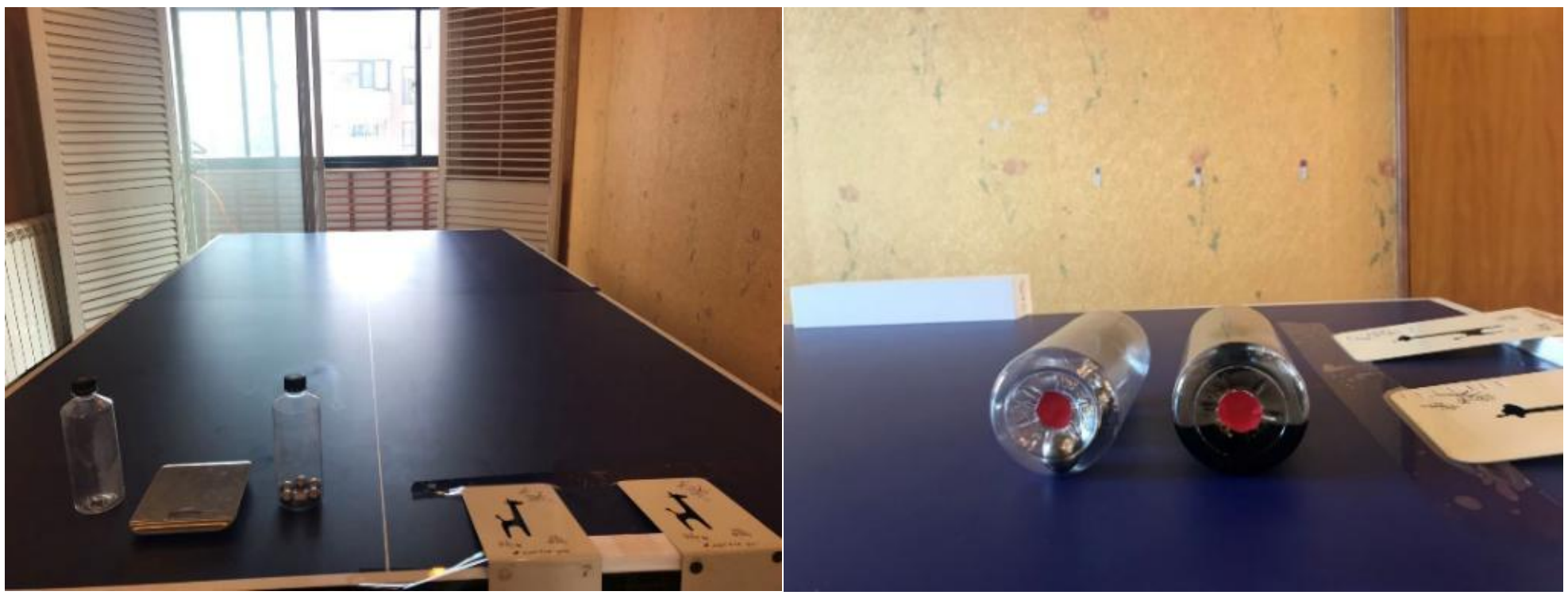

**Figure 2.18 Steel balls and water experiment devices**

Experimental procedures:

1. Fix the slope to the plane with tape and mark the height on the slope with a ruler.

2. Attach the phone to the rack.

3. Stick a bright dot in the bottom center of the water bottle to track as it moves.

Parameters:

$m_{\text{steel ball}}$=33g, $M_{\text{bottle}}$=35g, slope angle $\theta$=11°, $R_{\text{bottle}}$=3.25cm, $r_{\text{steel ball}}$=0.955cm

Variables:

X: the displacement of the bottom center point along the rolling direction;

V: the velocity of the bottom center point moving along the rolling direction;

a: the acceleration of the bottom center point moving along the rolling direction.

Two 35g bottles were filled with 165g dyed water and 5 steel balls (total mass 165g) respectively. Then they are released from rest and rolled down two identical slopes (11°) at 4cm above the plane at the same time.

Figure 2.19-21 show the displacement-time, velocity-time and accelaration-time graph (steel ball bottles vs. water bottles with the same mass and release height) of the center points of the bottom (the orange line refers to the data from water bottles while the blue line refers to that of steel ball bottles):

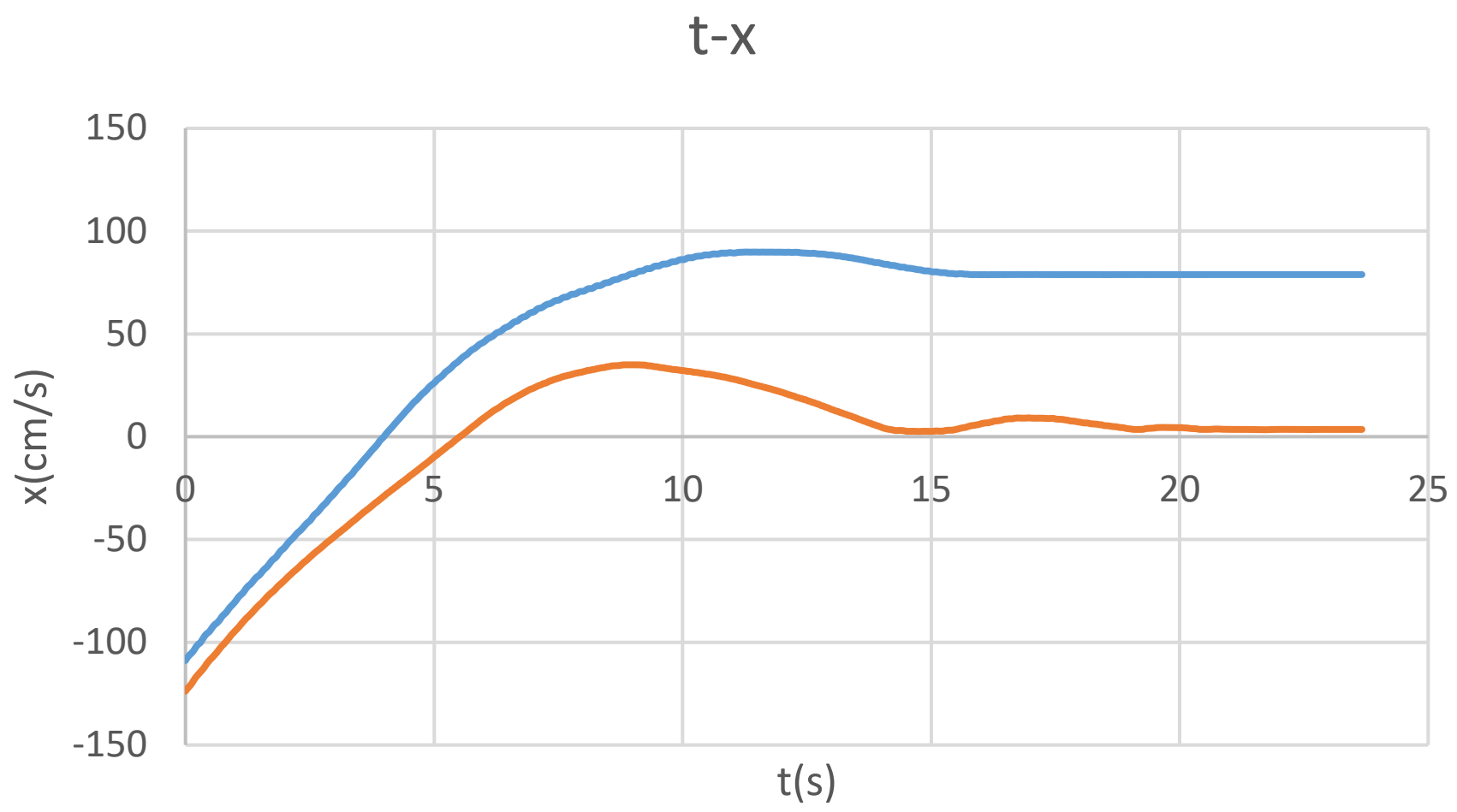

**Figure 2.19Displacement-time graph, steel balls (blue) vs. water (orange)**

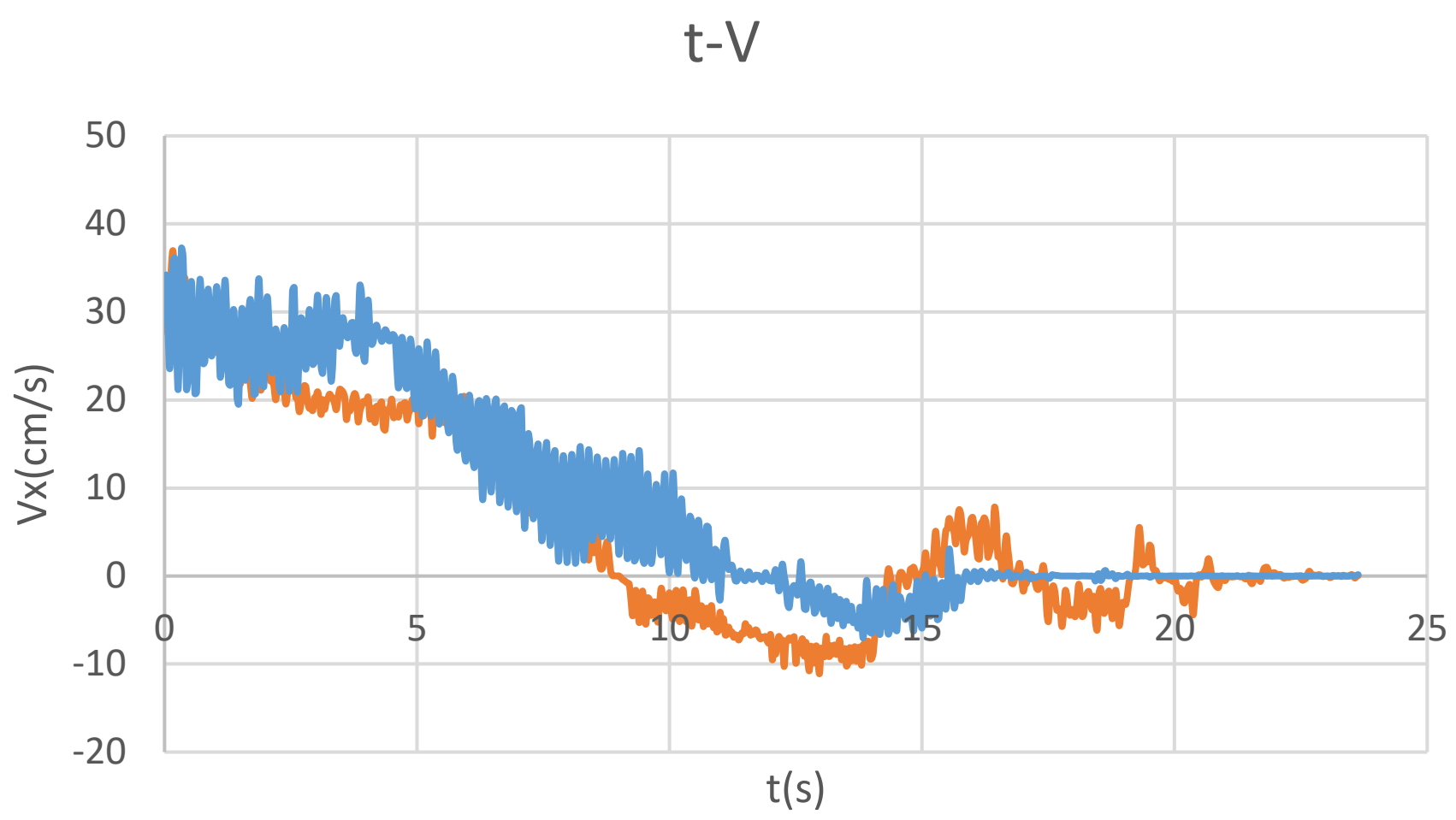

**Figure 2.20Velocity-time graph, steel balls (blue) vs. water (orange)**

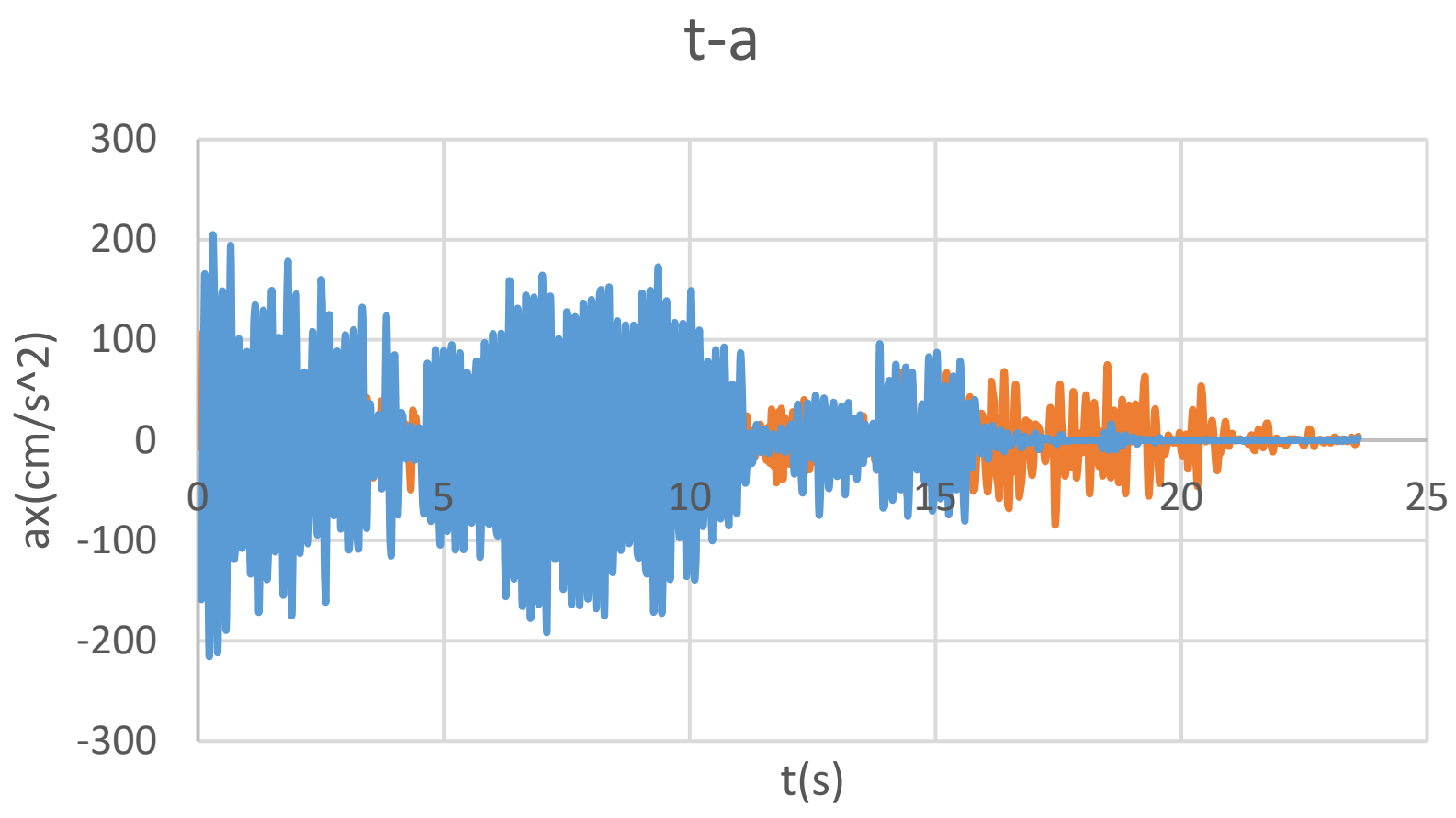

**Figure 2.21 Acceleration -time graph, steel balls (blue) vs. water (orange)**

By observing the control experiments of steel ball bottles and water bottles, as well as the analysis the diagrams of displacement, velocity and acceleration in both cases, we conclude that the motion patterns are almost the same in both cases: with a superposition of translation and vibration. Therefore, our theoretical bottle-and-bead model also applies to the water bottle system.

### 2.2.2 Experiment with Changing Water Amount

Experiment devices: slopes (16°), 520ml water bottles (cylinder), water (dyed yellow), injector, ruler, meter ruler, cell phone

Experiment method:

1. Fix the slope and mark the distance on it. The lower side of the slope is connected to a horizontal plane 3 meters in length. The initial condition of the bottle is considered as a translation-dominated motion when it reaches the horizontal plane.

2. Fix the phone and release the bottle from rest at 4cm on the slope. Change water amount. Fill the bottle with 15%, 25%, 35%, 45%, 55%, 65%, 75%, 85%, 95% of water respectively. Record the motion of the bottom center point of the bottle. Set the phone frames as 4k, 30fps.

**Experiment phenomena:**

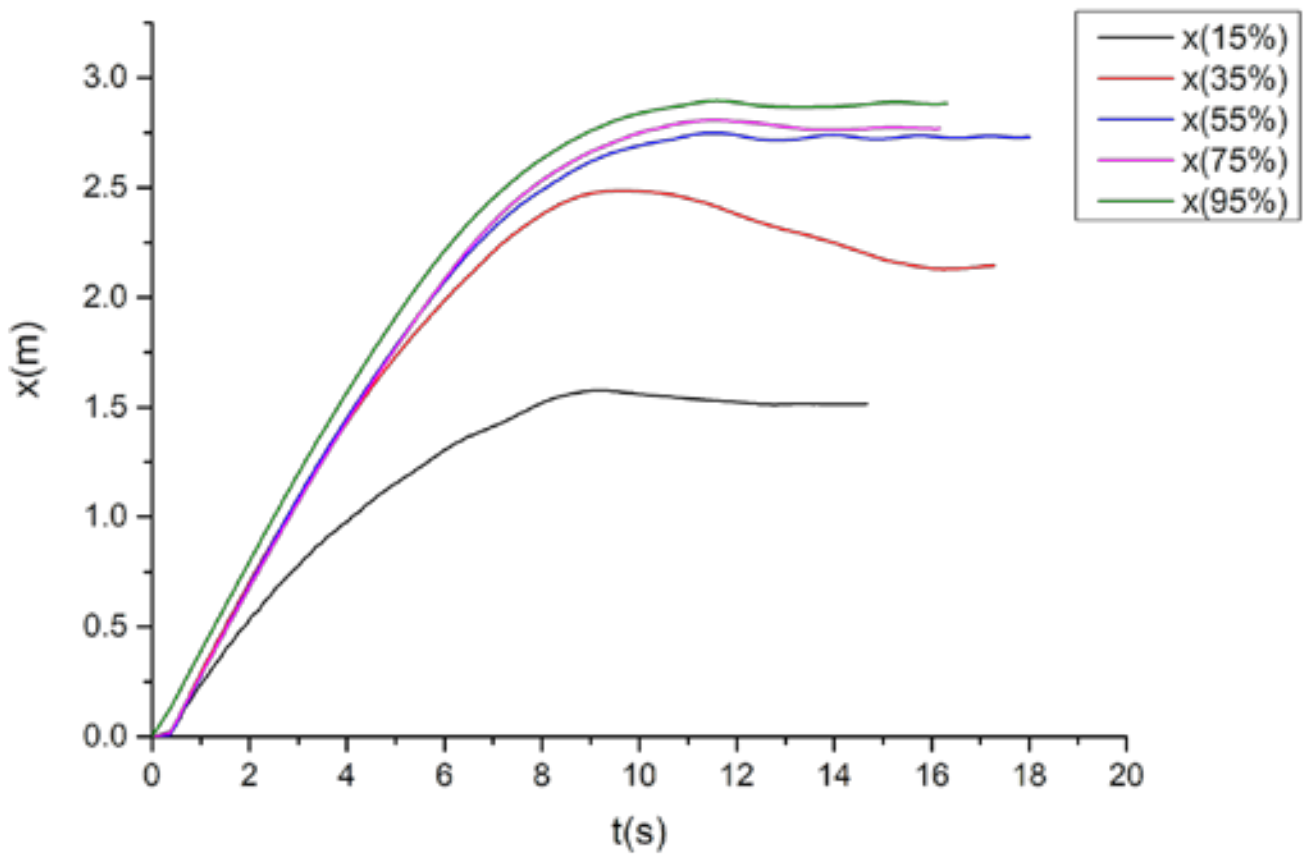

**Figure 2.22 Displacement-time graph with different water amounts**

Using Tracker and Origin, we obtain the x-t, v-t and a-t graphs under different water amounts. From Figures 2.22, we find that in the experiment, at all water amounts, the bottle exhibits a turn-back motion, that is, the final displacement is less than the maximum displacement achieved.

Figures 2.23 is the v-t graph of bottles with 15%, 35%, 55%, 75%, and 95% water.

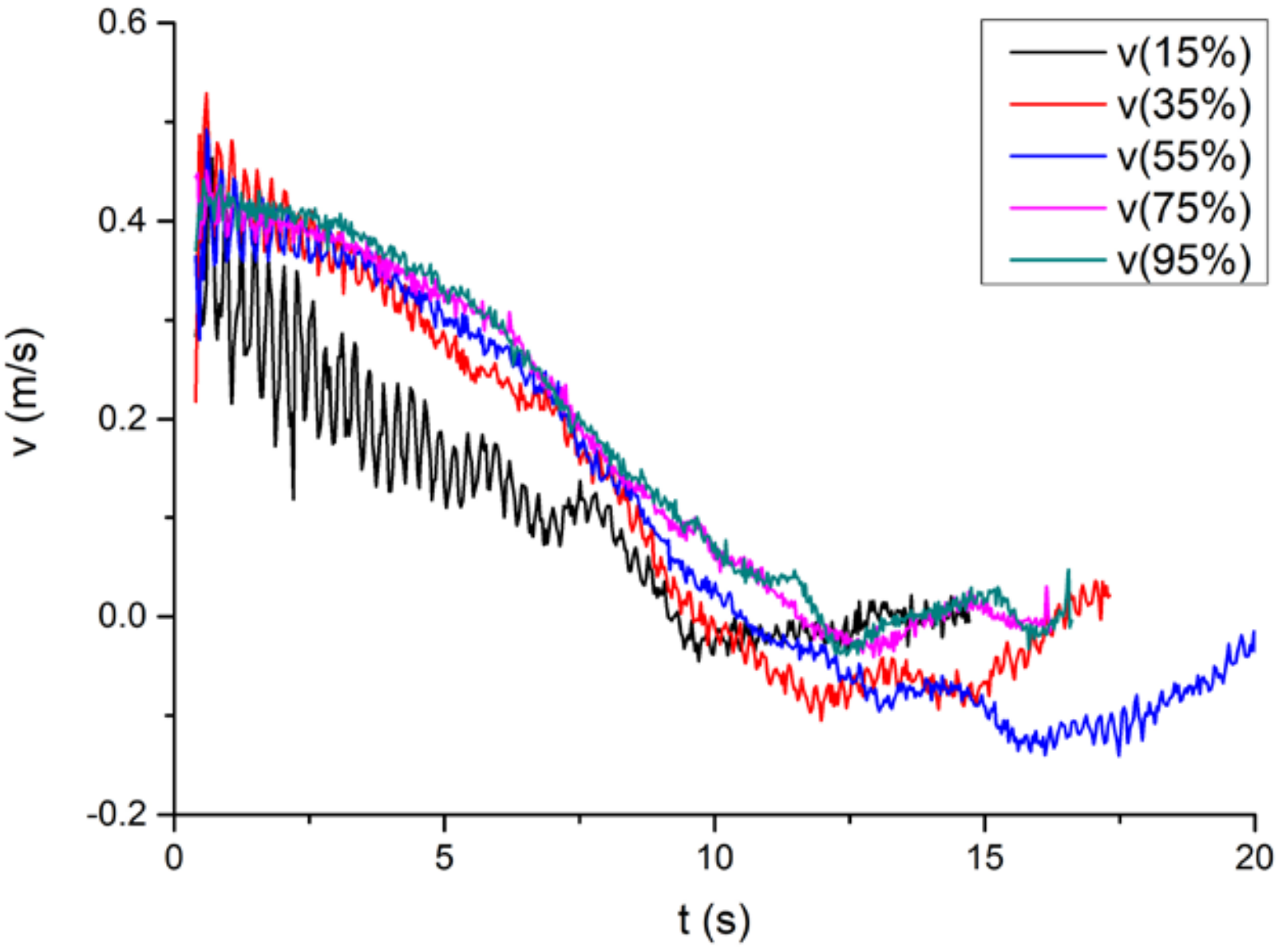

**Figure 2.23 Velocity - time, 15%, 35%, 55%, 75%, and 95% water, 4cm release height**

From Figures 2.19 to 2.23 above, we find that the short-term motion pattern of the water bottle is the same as described by our bottle-and-bead model: the superposition of oscillation and translational motion. However, the long-term pattern is more complicated. We see that as the bottle moves, both translational kinetic energy and oscillation energy are attenuating. The attenuation of translational kinetic energy and oscillation kinetic energy is represented by the decrease in bottle velocity and oscillation amplitude, respectively. The translational velocity will decrease to zero before it increases in the opposite direction and changes back and forth around zero. Finally, the translational velocity decreases to zero, the bottle retains oscillation until it becomes totally stationary, that is, both translational velocity and oscillation velocity become zero.

Both translational and oscillation motion contribute to the change in acceleration. The acceleration attenuation is obvious at the beginning of the motion, and then decreases to a certain range; the attenuation becomes slow.

Now we compare the motion with different amount of water. As the water amount increases, the oscillation amplitude is smaller; the velocity of translational motion is larger due to greater gravitational potential energy of the water and bottle, meanwhile the bottle takes longer time to come to rest. Only a relatively small amount of water (15-40%) can attenuate the motion effectively and stabilize the bottle in a shorter time.

Now we compare the oscillation frequency in theory and experiment.

When the bottle is released from the same height, the increase of water amount has little impact on the initial oscillation frequency of the bottle, which is around 4Hz. During the motion, the amplitude of the bottle vibration decreases gradually while the frequency increases.

According to our theoretical model, the oscillation frequency of the bottle could be expressed as $\frac{1}{2\pi}\sqrt{\frac{2M+m}{2M}\frac{g}{R}}$, and it is related to the mass of water. In experiment, we observed that when the bottle moves, the water inside does not move as a whole, because some of the water does not oscillate with the bottle due to inertia. Here, we define the amount of the water involved in oscillation as the effective mass, when $f = 4\text{Hz}, M = 0.035kg, R = 0.0325m$, the result is $m \approx 0.0765kg$.

### 2.2.3 Experiment with Changing Initial Conditions

In this part of experiment, we will change initial conditions and switch the motion pattern from the motion dominated by translation to that dominated by oscillation. Water amount and releasing height are also controlled.

Experiment design: we release the bottle from a certain height on the slope, wait until it reaches the horizontal plane and obtain a motion dominated by translation.

For the motion dominated by oscillation, we release the bottle and wait until it passes through a horizontal plane before colliding with a vertical wall. After the collision, the bottle will switch to a motion dominated by switch oscillation.

Experiment devices: slope (16°), 520mL water bottle (cylinder), water (dyed-yellow), desk and chair (used as a wall), injector, ruler, meter ruler, cell phone

Experiment procedures:

1. Set the slope and mark the distance on it. The lower side of the slope is connected to a plane of 2.11 meters in length.

2. The wall is placed perpendicular to the bottle movement track at 2.11 meters away from the releasing point. The initial condition of the bottle is considered as translational movement when it reaches the plane.

3. Fix the wall to prevent it from moving.

4. Release the bottle with a certain amount of water from rest from a certain height on the slope. Record the movement of the bottom center point of the bottle. Set the phone frames as 4k and 30fps.

Using Tracker, we obtain the velocity-time and acceleration-time graphs of the rolling bottles with different initial conditions.

Figure 2.24 shows the velocity-time and acceleration-time graphs of a bottle with 55% water, rolling down from various release heights and hitting the wall.

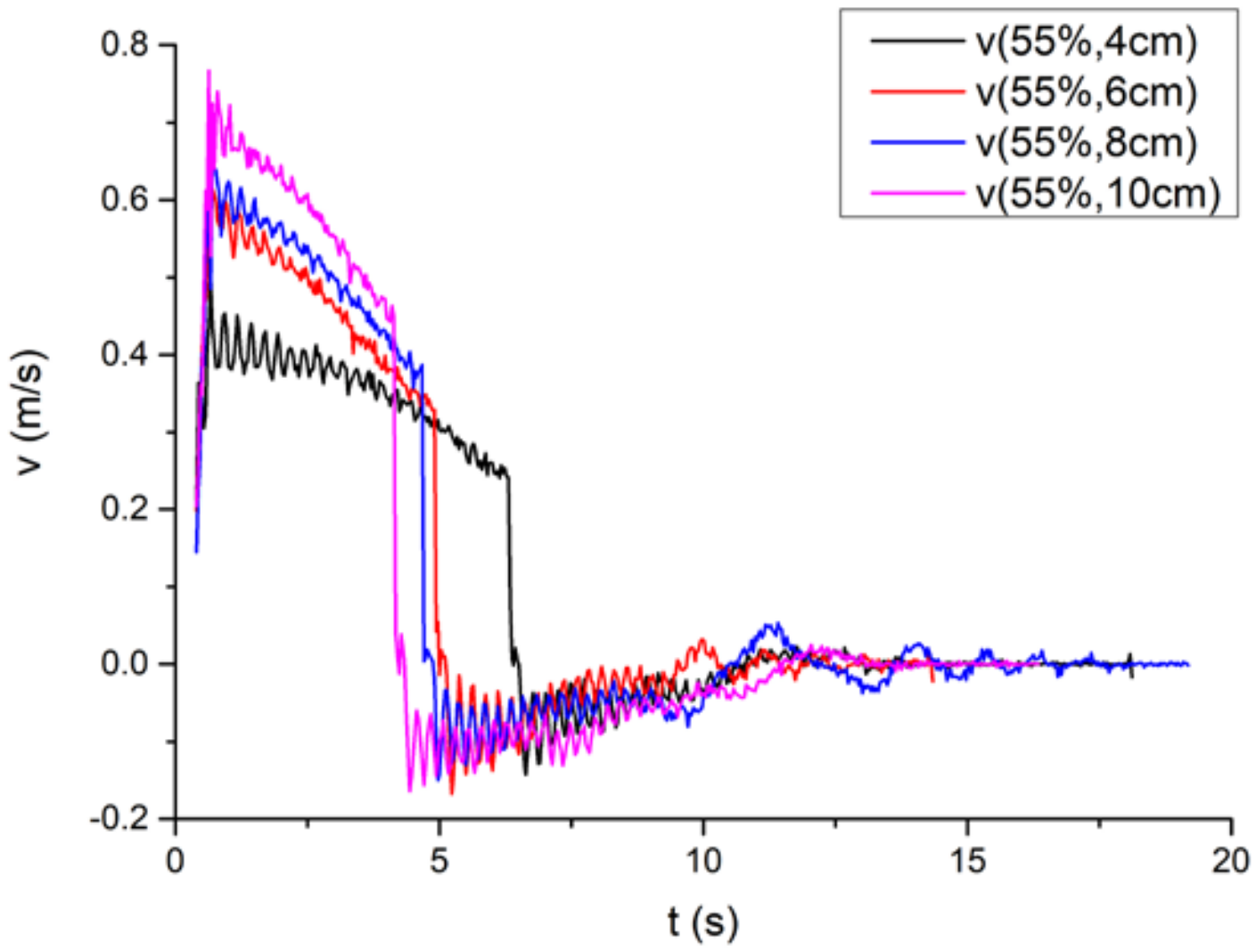

**Figure 2.24 Velocity- time, 55% water, 4, 6, 8, 10cm release height**

The motion can be divided into two phases:

(1) After being released from the slope, the bottle mainly moves translationally. In this process, both translation and vibration are attenuating. The translational velocity decreases, so does the vibration amplitude.

(2) After collision, the oscillation suddenly increases while the translational velocity is suddenly decreased and the oscillation becomes the dominate motion. The translational velocity will gradually change around zero, and the bottle rolls back and forth until it completely stops.

Figure 2.25 shows the velocity-time and acceleration-time graphs of a bottle with 100% water, rolling down from various release heights and hitting the wall.

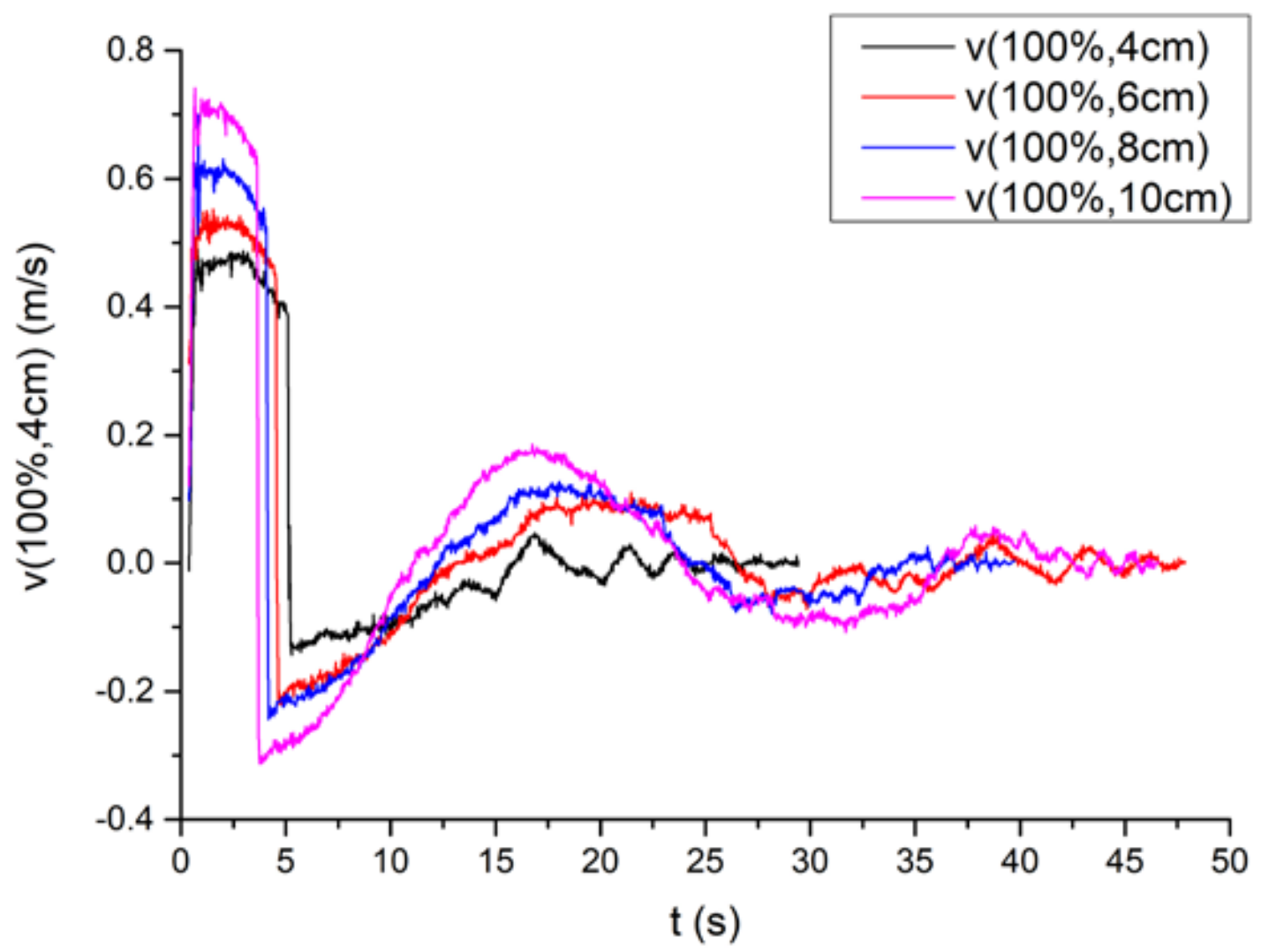

**Figure 2.25 Velocity- time, 100% water, 4, 6, 8, 10cm release height**

The higher the release height is, the faster the linear velocity of the bottle is when it reaches the plane. When oscillation is ignored, the translational motion of the bottle is nearly uniform deceleration motion, and the deceleration is greater for higher the release height and faster velocity. The initial angular velocity of water and its amplitude are relatively high in the beginning and decrease gradually later.

A bottle with 100% water moves in a similar pattern, though with significant smaller amplitude. It will take longer for the bottle to reach stationary. During the experiment dominated by translational motion, we found that water in the bottle helps the bottle to slow down, and the oscillation of the water helps to absorb energy from the collision, both contributing to make the bottle quickly come to rest. Therefore, it can be deduced that water in a bottle helps to restore stability during rolling motion.

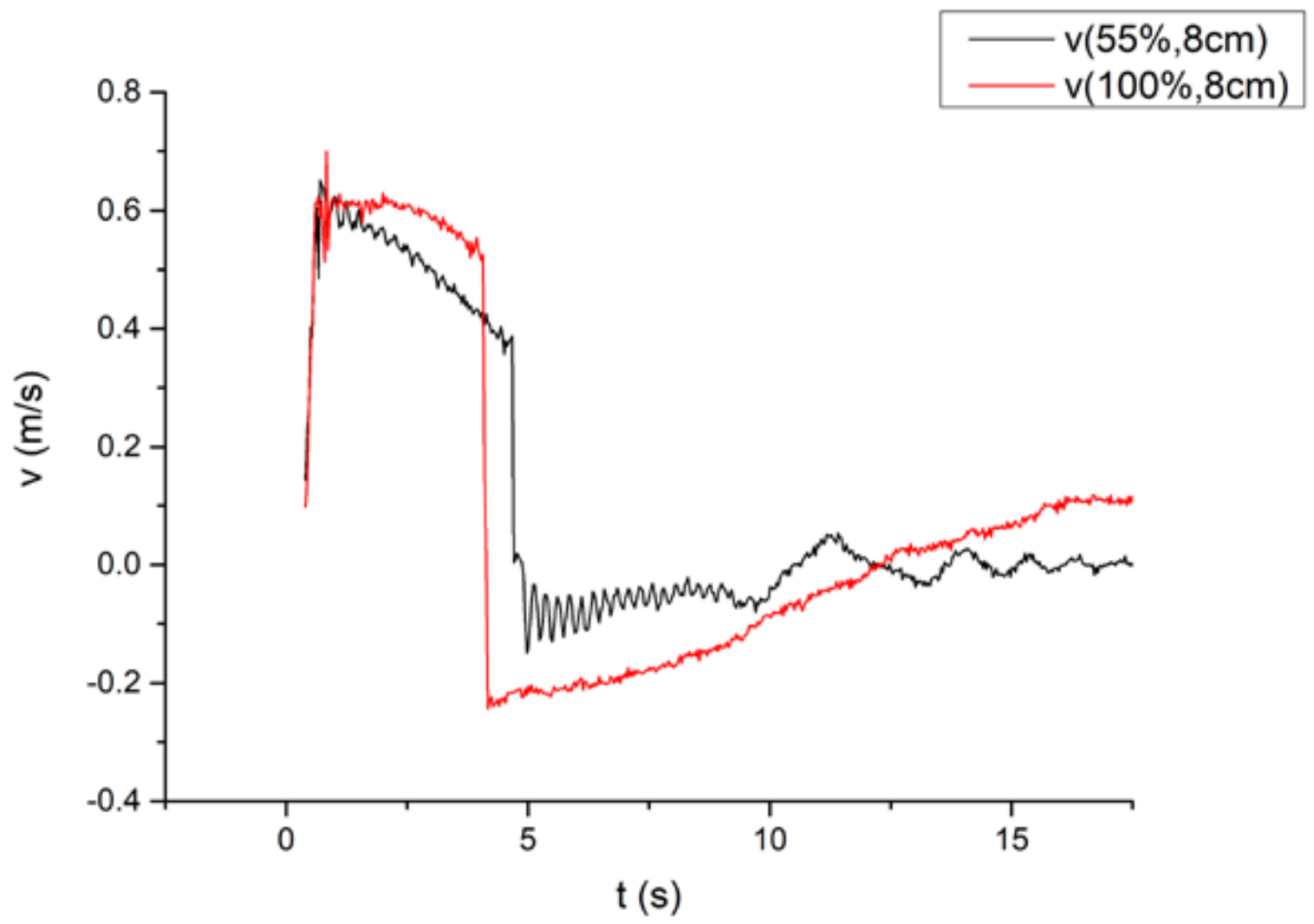

**Figure 2.26 Velocity- time, 55%, 100% water, 8cm release height**

In figure 2.26, when we compare the movement of a 55% water bottle with that of a 100% water bottle, it is found that former one has a faster attenuation of vibration amplitude and a more obvious frequency change. For a 100% water bottle, the vibration frequency and amplitude stay basically unchanged. This verifies the assumption about effective water quantity. When a water bottle is filled 55% with water, some of the water inside will not vibrate when the horizontal velocity is too fast. As the translational velocity gradually decreases over time, the water amount involved in vibration increases, thus the vibration frequency increases too. And for a water bottle all filled with water, the water inside has no space to move, so the bottle and the water inside can be seen as a whole when rolling. This explains why the frequency is much faster in 100% water bottle and it does not change significantly over time.

**Potential sources of error:**

(1) The bottle is not a perfect cylinder (uneven underside observed). The bottom is uneven. This may cause irregular motion.

(2)The camera lens is too close to the bottle, resulting in large deformation at both ends of the frame, which affects the accuracy of the data collected by the tracker from both ends.

### 2.2.4 Experiment with Different Liquid

Except for water amount, releasing height, and the initial condition of the bottle, we wonder what impacts does coefficient of viscosity has on the dynamics of a rolling liquid bottle. We use corn starch to make up a thick solution for comparison.

Experiment devices: slope (16°), 520mL water bottle (cylinder), starch solution ( 1:10 mixed with water, density 1.1g/cm$^3$), injector, ruler, meter ruler, phone,

Experiment procedure:

1. Set the slope and mark the distance on it. The lower side of the slope is connected to a plane of 2.11 meters in length.
2. Release the bottle with a 15% of starch solution from rest from a certain height on the slope. Record the movement of the bottom center point of the bottle. Set the phone frames as 4k and 30fps.

We experimented under the temperature of 10℃. The starch solution is a kind of non-Newtonian fluid. Using Tracker, we obtain the graphs of the velocity and acceleration of the bottles containing the starch solution.

Figure 2.27 and 2.28 show the displacement and velocity graphs in a bottle partially filled (15%) with starch solution rolling down from 4cm, 6cm, 8cm, 10cm, and 12cm height. We find that the oscillation frequency is higher and the amplitude is smaller for the starch solution compared to water filled bottle.

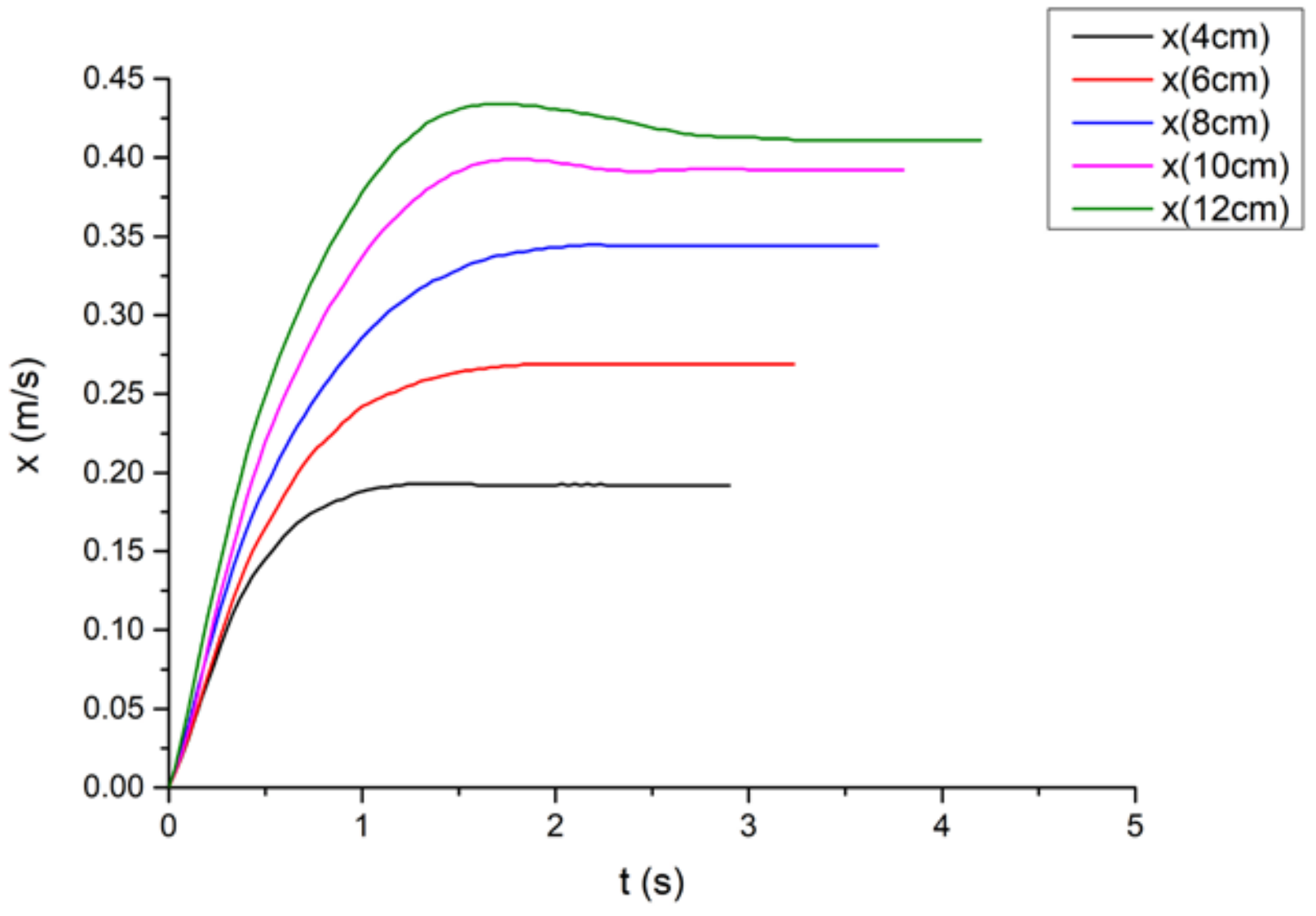

**Figure 2.27 Displacement- time graph of a bottle partially filled (15%) with starch solution**

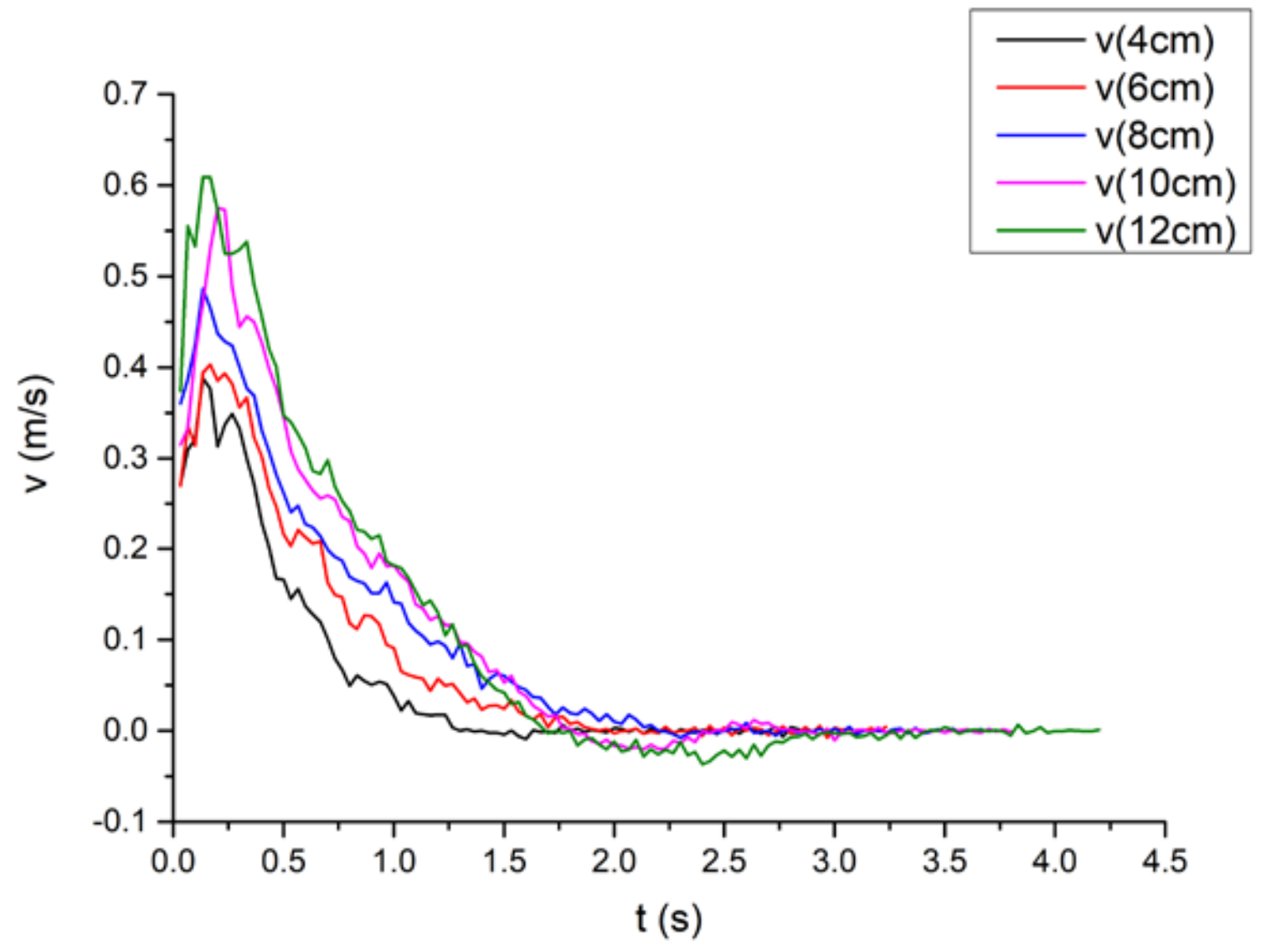

**Figure 2.28 Velocity- time graph of a bottle partially filled (15%) with starch solution**

From figures 2.27 and 2.28, we can see that a rolling bottle with starch solution can be stabilized within 2 seconds. From Fig2.20 in Sec.2.2.2 we know that the time to stabilized a water bottle is 10~20 seconds. We come to the conclusion that non-Newtonian fluid is more favorable to stabilize a rolling bottle.

# 3. Practical Applications

Such problems can also be examined in practical applications, such as the movement of a tanker. We can simplify the physical phenomena of a stationary tanker as follows (see Figure 3.1):

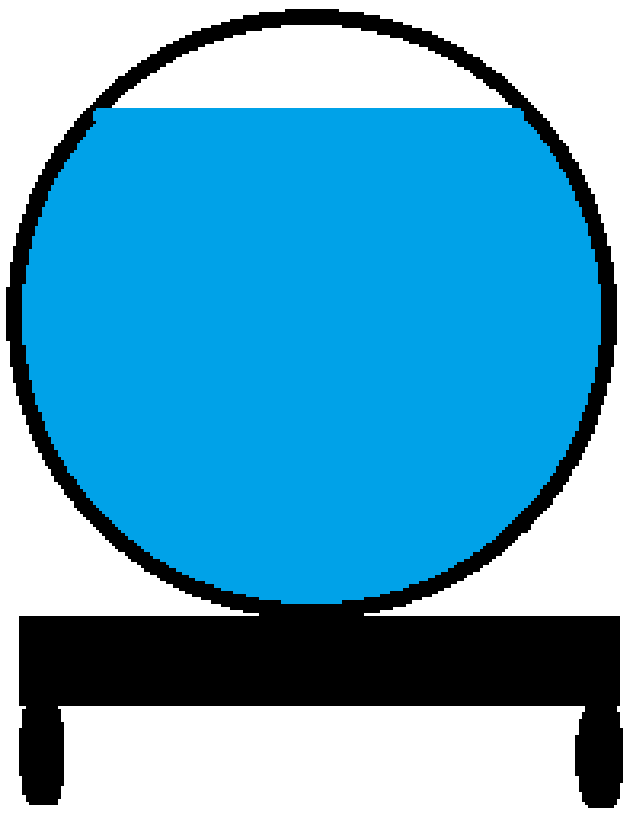

**Figure 3.1 Section of a stationary tanker**

Suppose the reservoir on a tanker is cylindrical in shape and its wall thickness is ignored, the chassis and wheels at the bottom can be seen as a combination of a cuboid plus cylinders. When the tanker is stationary, the symmetry axis of the liquid inside is in the vertical direction.

When the tanker rotates, assuming the liquid inside has reached equilibrium in the non-inertial system of the tanker, its basic shape remains unchanged, and there is a certain angle between the symmetry axis and the vertical direction (see Figure 3.2):

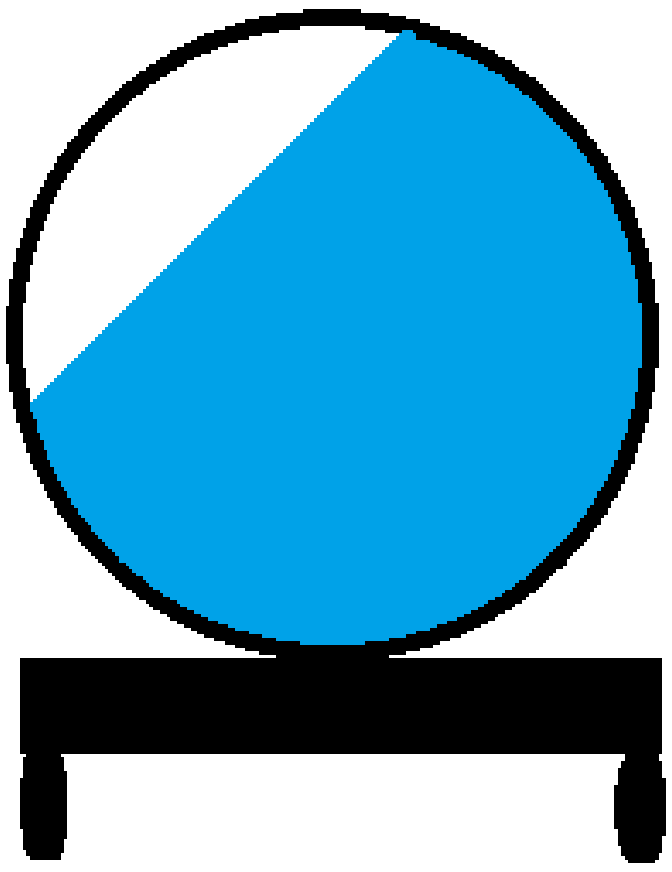

**Figure 3.2 Section of a turning tanker**

Assuming the volume of the liquid inside is $V_L$, the mass is $m_L$ and the density is $\rho$; The mass of the reservoir is $m_c$, and the radius is $r$; The total mass of the chassis and the wheels is $m_0$, and the width is $2L$; The height of the center of gravity is half the height of the chassis, which is $h$.

Taking a turning tanker as the reference frame, its liquid and the whole body are affected by inertia forces, namely $m_L \frac{V^2}{R}$ and $(m_0 + m_c)\frac{V^2}{R}$ respectively. The applying points are located on the centroid of the liquid and the body respectively.

Therefore, we must calculate the centroid of the liquid inside at first. The shape of the liquid can be seen as a notched sphere. We assume the angle between its symmetry axis and the vertical direction is $\theta$ (see Figure 3.3).

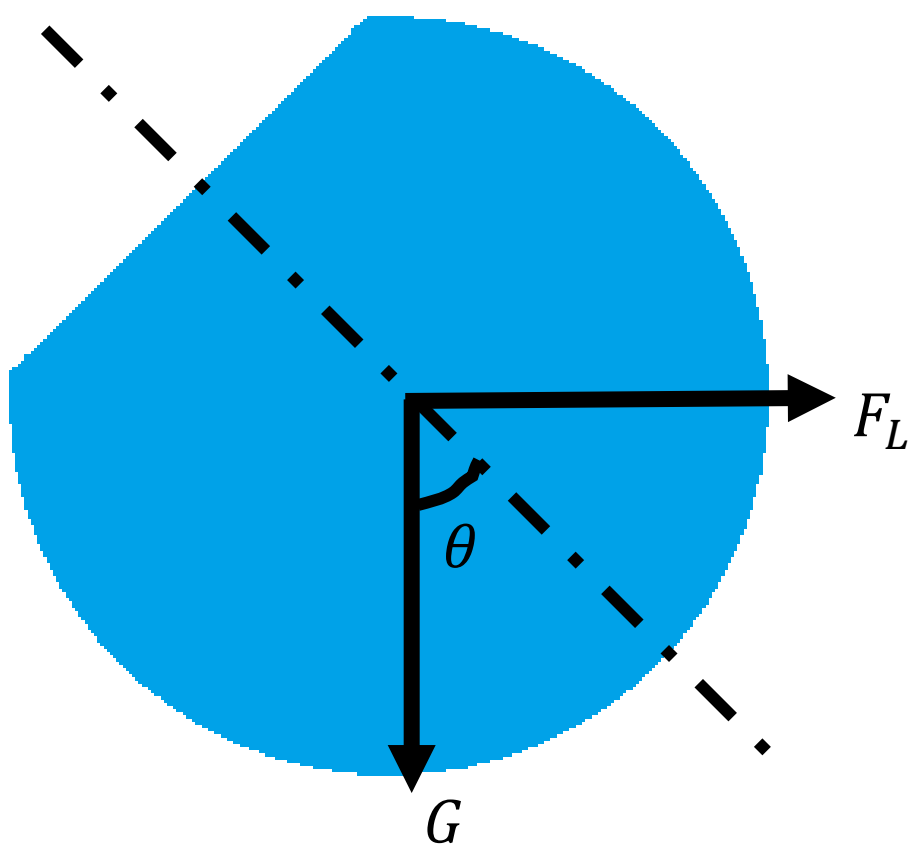

**Figure 3.3 Diagram of the shape of the liquid inside**

Establish a cartesian coordinate system, with the center of the notched sphere as the origin point, the symmetry axis as axis $x$, and any ray perpendicular to axis $x$ as the positive direction of axis $y$. Then use the infinitesimal element method to the notched sphere along axis $x$ (see Figure 3.4).

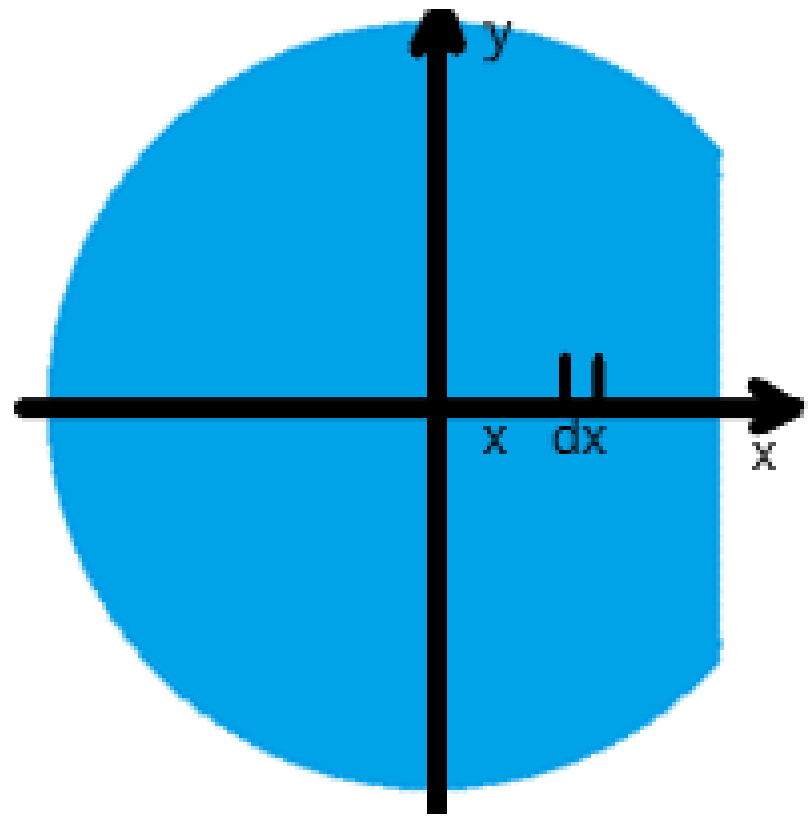

**Figure 3.4 Mathematical model**

According to the principle of symmetry, the centroid of the notched sphere will only be on axis $x$. Then based on the definition of centroid $r = \frac{\sum m_i r_i}{M}$, the horizontal axis of the centroid of the notched sphere is calculated as follows:

$$X = \frac{\int_{-r}^{h-r} \pi\rho(r^2 - x^2) \cdot xdx}{m_L} = \frac{\pi\rho r^2}{2m_L}[(h-r)^2 - r^2] - \frac{\pi\rho}{4}[(h-r)^4 - r^4]$$

Obviously, a liquid notched sphere has the following relationship with angel $\theta$:

$$tan\theta = \frac{F_L}{m_L g} = \frac{v^2}{Rg}$$

In the next step, the forces applied on the tanker are examined. When the tanker turns, assume the supporting force on the inner wheel from the ground is $N_1$, that on the outer wheel is $N_2$; the centrifugal force on the internal liquid is $F_L$, that on the reservoir is $F_c$, and that on the chassis and wheels is $F_0$; the turning radius is $R$.

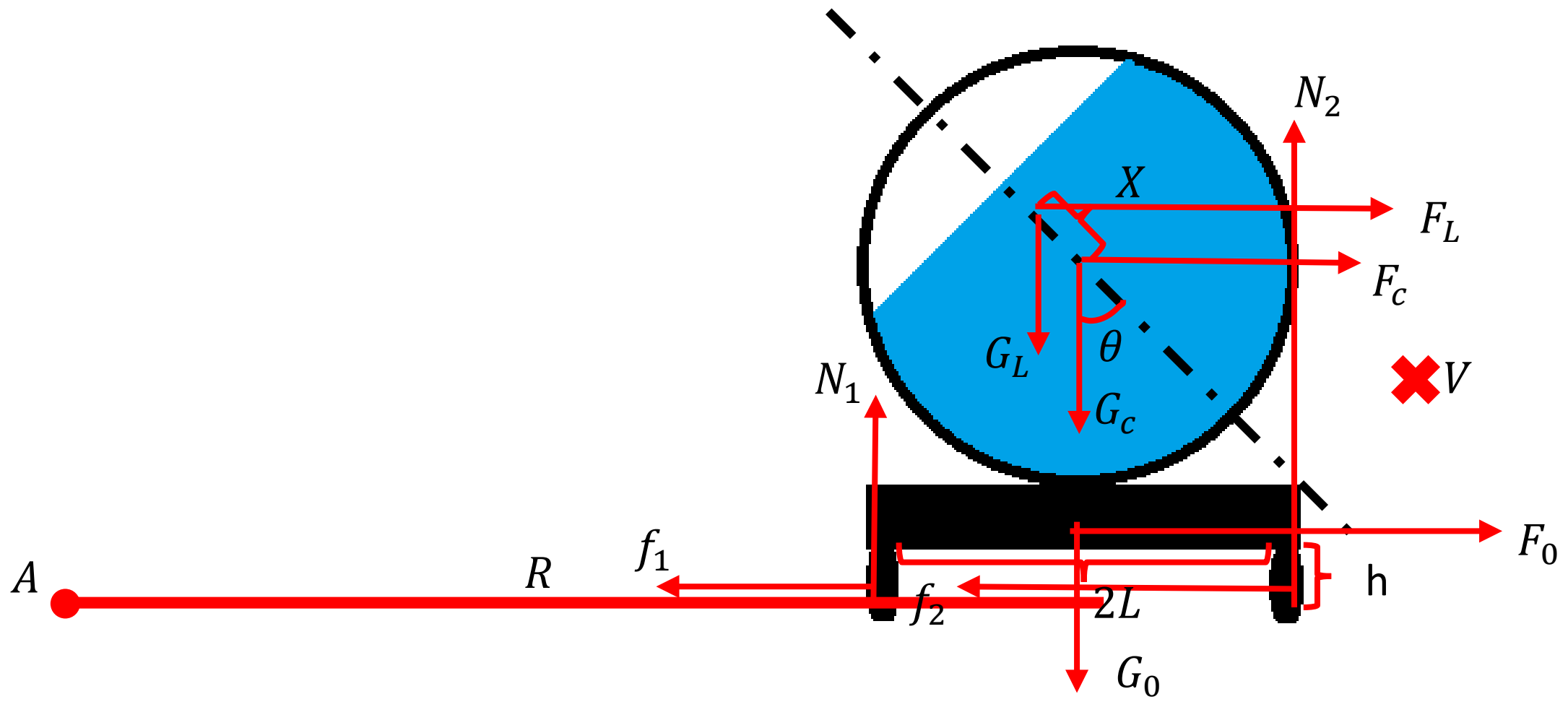

**Figure 5 Diagram of a turning tanker**

According to the principle of force balance, the following expression can be obtained:

$$N_1 + N_2 = (m_0 + m_c + m_L)g \ldots\ldots ①$$

$$f_1 + f_2 = F_L + F_c + F_0 \ldots\ldots ②$$

Taking $A$ as the rotation axis, the following expression can be obtained according to the torque balance principle:

$$N_1(R-L) + N_2(R+L) = m_L g(R - Xsin\theta) + (m_0 + m_c)gR + F_0 h + F_c(2h + r)$$

$$+F_L(2h + r + Xcos\theta) \ldots\ldots ③$$

It can be inferred that the faster the tanker is, the greater the centrifugal force applied on it is, that is, $F_0, F_L, F_c$ will increase with the velocity. By observing the two expressions above, we found that when $F_0, F_L, F_c$ change, only $N_1$ and $N_2$ change with when, yet the sum of $N_1$ and $N_2$ remains unchanged. Therefore, when the torque on the right side of expression ② increases, the supporting torque on the left side can only be increased by increasing $N_2$ and decreasing $N_1$. However, when $N_1$ decreases to zero, the tanker is likely to roll outwards. Or, when the friction of the ground is not enough to provide the centripetal force needed, the tanker will also slide outward.

Assume the friction coefficient of road is $\mu$ and take the contact point between the outside of the tanker and the road surface as the rotation axis. The force balance expression under critical conditions is as follows:

$$N_2 = (m_0 + m_c + m_L)g \ldots\ldots ④$$

$$m_L g(L + Xsin\theta) + (m_0 + m_c)gL = m_c(2h + r)\frac{V^2}{R} + m_L(2h + r + Xcos\theta)\frac{V^2}{R}$$

$$+ m_0 h\frac{V^2}{R} \ldots\ldots ⑤$$

Or:

$$\mu g(m_0 + m_c + m_L) = (m_0 + m_c + m_L)\ \frac{\mathrm{V}^2}{\mathrm{R}} \ldots\ldots ⑥$$

When one of these two critical conditions is met, the tanker is in danger of turning over.

# 4. Conclusions

In throwing, we define two kinds of falling conditions. The first is free falling. The probability of upright landing under this condition is very low, mostly under 2%. The theory provides an explanation for this phenomenon that the velocity and angular velocity of falling don’t match each other. The other kind of falling is flipping. After conducting the experiments of different relevant parameters over a thousand times, we find the most favorable condition for an upright landing. The water amount should be between 35% to 45%; the falling height should be around 80cm; the initial angle should be 180°; the landing position should be the farther end to the falling platform. In theory, we consider the impact of water amount in statics and dynamics. In statics, the more favorable condition for upright landing is a lower center of mass and larger collision angle. In dynamics, Considering other parameters, an accurate matching between the angular momentum and the moment of impulse for an upright landing. Thus we need a matched velocity, angular velocity, and direction of rotation. In experiments, we further prove that friction is favorable to an upright landing.

In rolling, we develop a bottle-and-bead model, which concludes that the motion of a rolling bottle with water is a combination of vibration motion and translational motion. We use an experiment with a bottle filled with steel beads to verify the theoretical model. It is found that the bottle vibrates when there is no linear velocity, and it becomes a superposition of translational motion and vibration when linear velocity exists. The limits of the bottle-and-bead model are that damping impact is not considered, thus the energy does not attenuate, water cannot bring the bottle to stability, and that the change between vibration and translation motion is not considered. However, in the rolling experiments, we observe the attenuation of translation velocity and vibration amplitude. An obvious switch between vibration and translation can be seen when the bottle hits an obstacle. Also, experimental data prove that water can help stabilize the bottle. In analyzing the experiments with corn starch solution, we come to the conclusion that starch solution, compared to pure water, is more favorable to the stabilization of a rolling bottle.

# Acknowledgments

The authors would like to thank Professor Wang Sihui of Nanjing University, Professor Pan Zhimin of Nanjing Foreign Language School, He Zhaoqin of Metatest Corporation, and Jin Kaiwen and Xiao Lintao for their kind support. More specifically, the authors are grateful to Professor Wang Sihui for her guide in topic selection, theoretical model building, experiment design, group discussion and paper revision. Our thanks go to Mr. Pan Zhimin for his help in team building, topic selection, communication, psychological counseling and final paper revision. We also want to thank He Zhaoqin and Jin Kaiwen for providing some experiment sites and devices, and Xiao Lintao for advice in the group discussion, sharing physics knowledge and providing valuable suggestions in theories and experiments.

The three authors worked closely through this paper and completed the researches and examinations all together. Among them, Gu Yanwen was responsible for finding solutions in theory and writing the “Theoretical Analysis”, the experiment design, and participated in some of the experiments about “Throwing” in the first part; Bai Yunzhou and Xin Yuxi were responsible for conducting most of the “throwing” experiments and all of the “rolling” experiments. In the writing process, the three authors were respectively responsible for their own parts, and Bai Yunzhou also took on the summary and draft work of this paper.

# Appendix 1: A Bottle Containing Steel Balls, Experimental Phenomenon Probe

In order to facilitate further research on this subject, we expand the experiment content. We roll down a bottle containing the same mass of steel balls from different heights, and a bottle containing different mass of steel balls from the same height respectively. During this time, we observe the phenomena and analyze the data collected.

**Experiment 1. Roll down a bottle containing the same mass of steel balls from different heights, compare**

Take a bottle with 4 steel balls, roll it down without initial velocity from the same slope at different heights from the bottom, namely 1cm, 2cm, 3cm, 4cm, 5cm and 7cm.

The image of the tracking data (with the same mass but at different heights) is as follows:

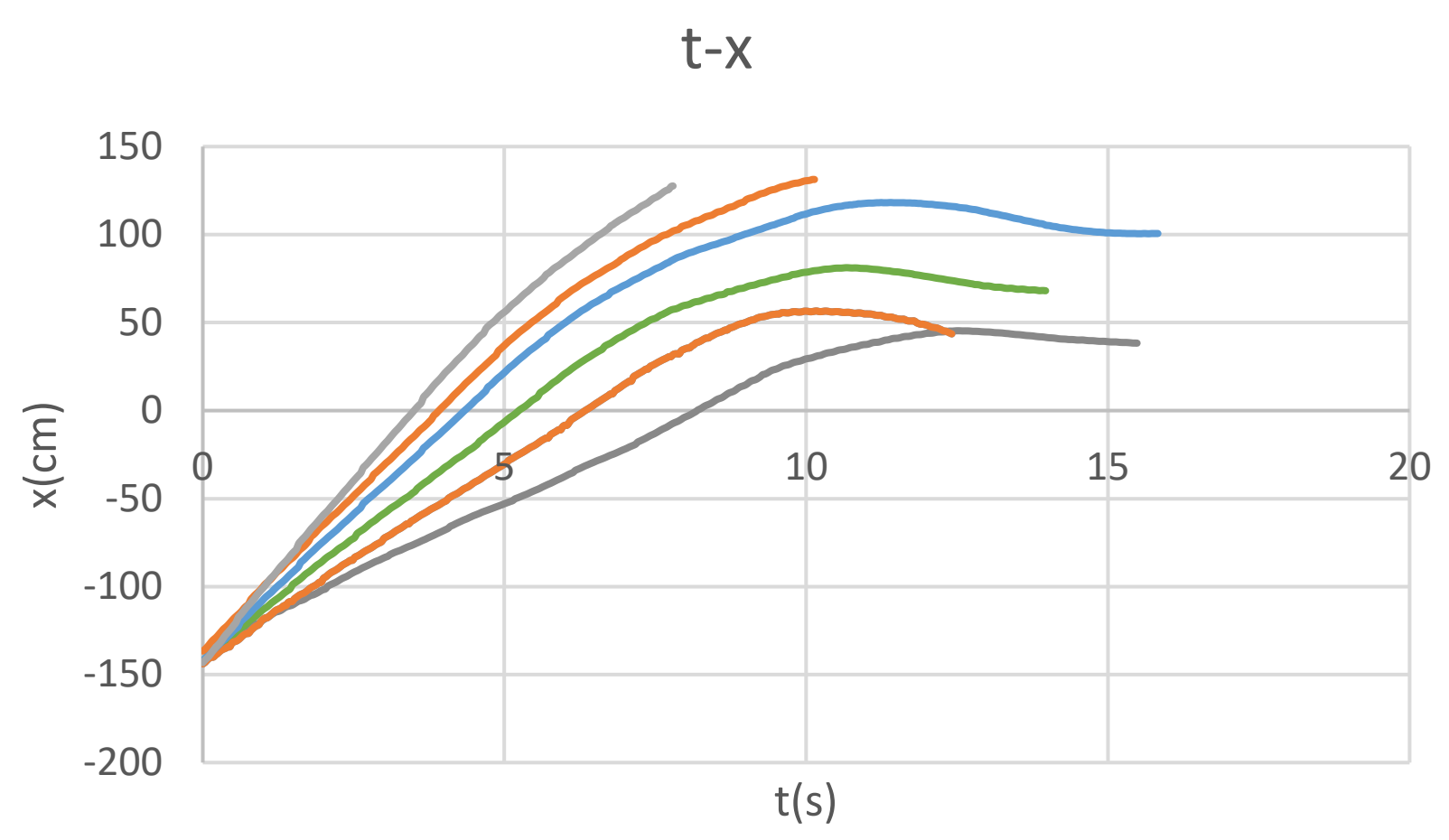

**Figure 1 Displacement-time, a bottle containing four steel balls rolling down from different heights, compare**

Note: From top to bottom, the heights are 7cm, 5cm, 4cm, 3cm, 2cm and 1cm respectively.

It was found that the motion patterns and tracks of the bottle are almost the same at all heights, which can be seen from the velocity and acceleration diagrams below: (a bottle containing four steel balls rolling down without initial velocity from a height of 4cm from the bottom)

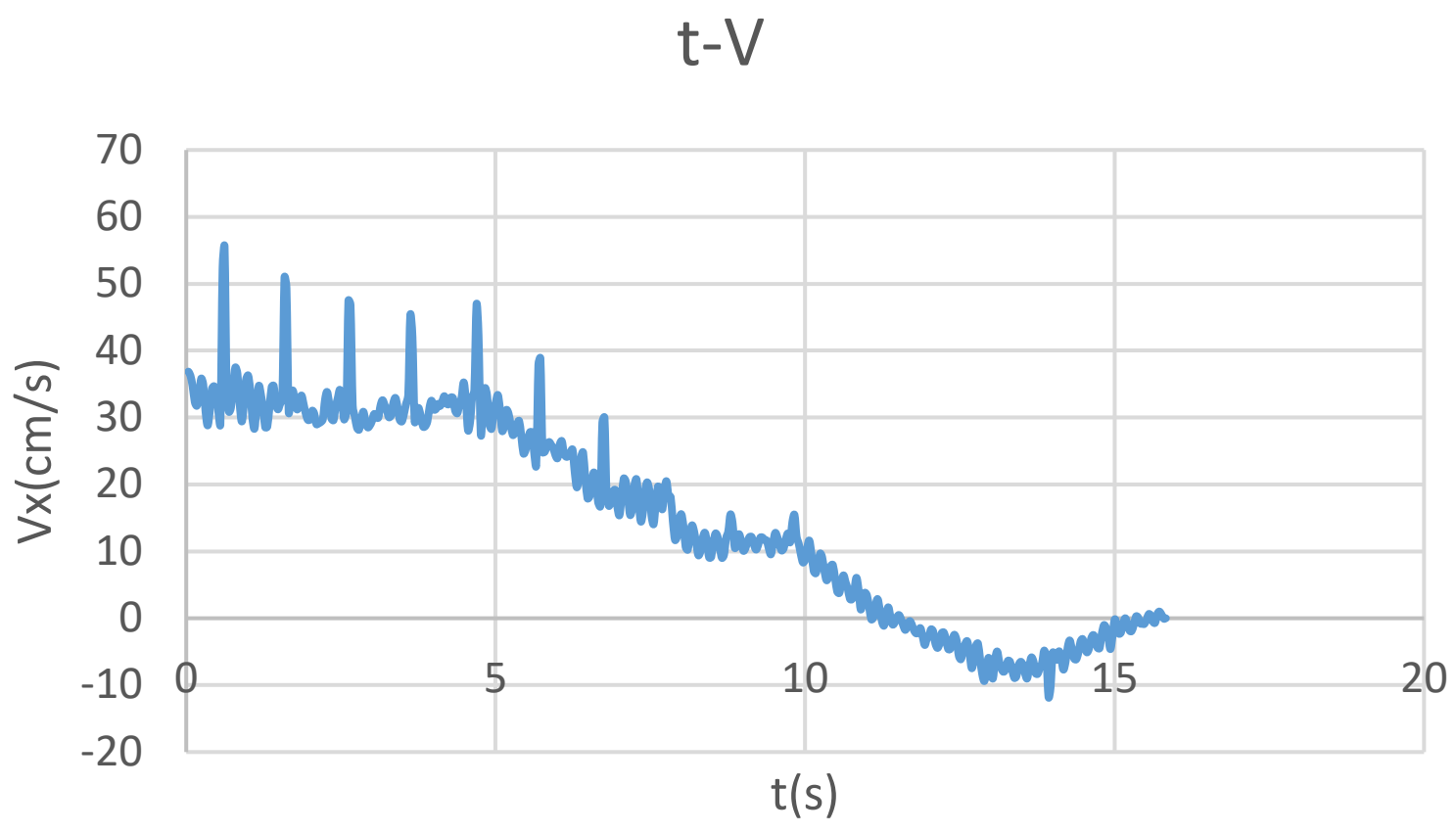

**Figure 2 Velocity-time, four steel balls, without initial velocity, 4cm release height**

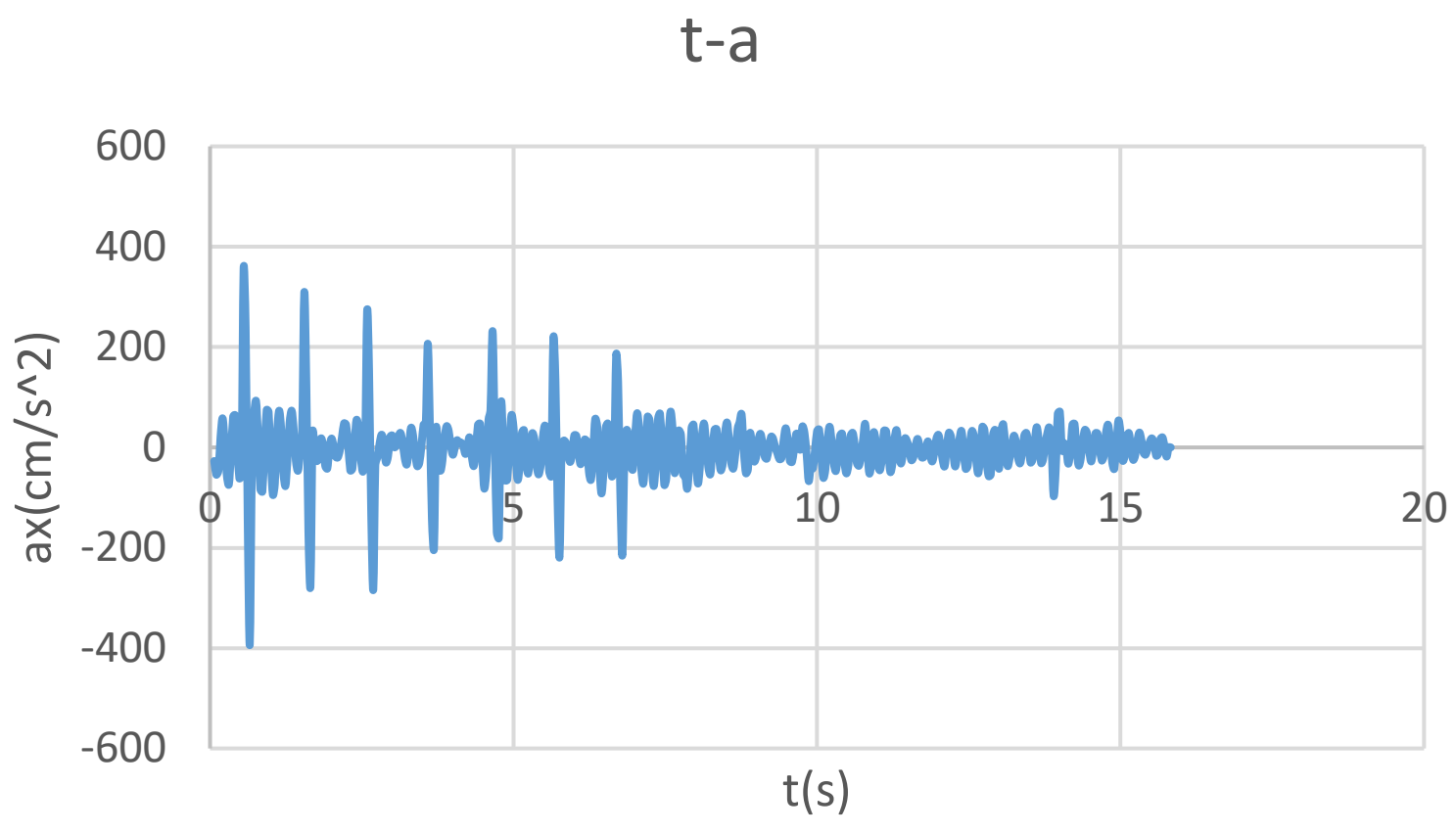

**Figure 3 Acceleration-time, four steel balls, without initial velocity, 4cm release height**

Velocity-time diagrams, at different heights:

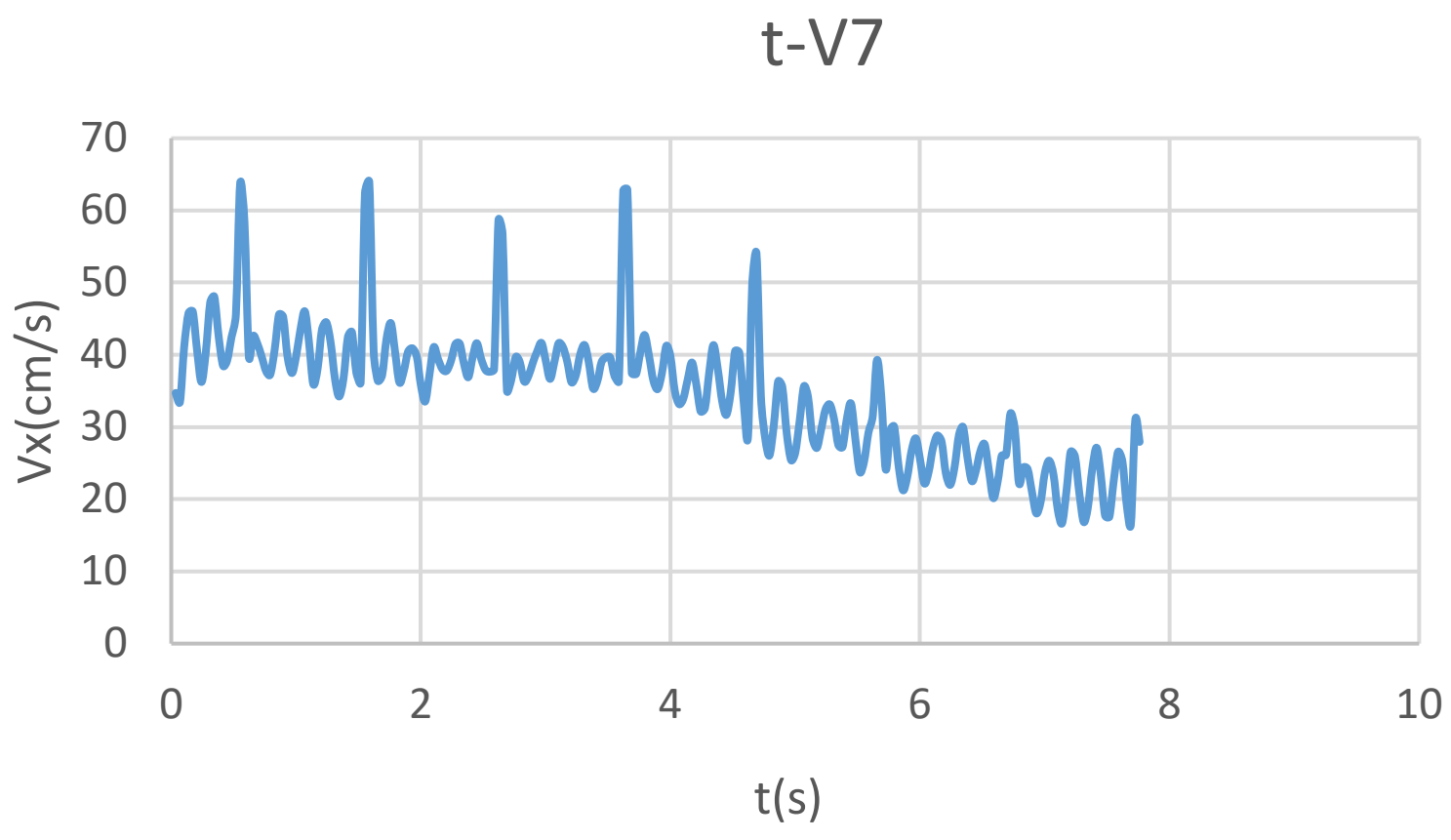

**Figure 4 Velocity-time, four steel balls, without initial velocity, 7cm release height**

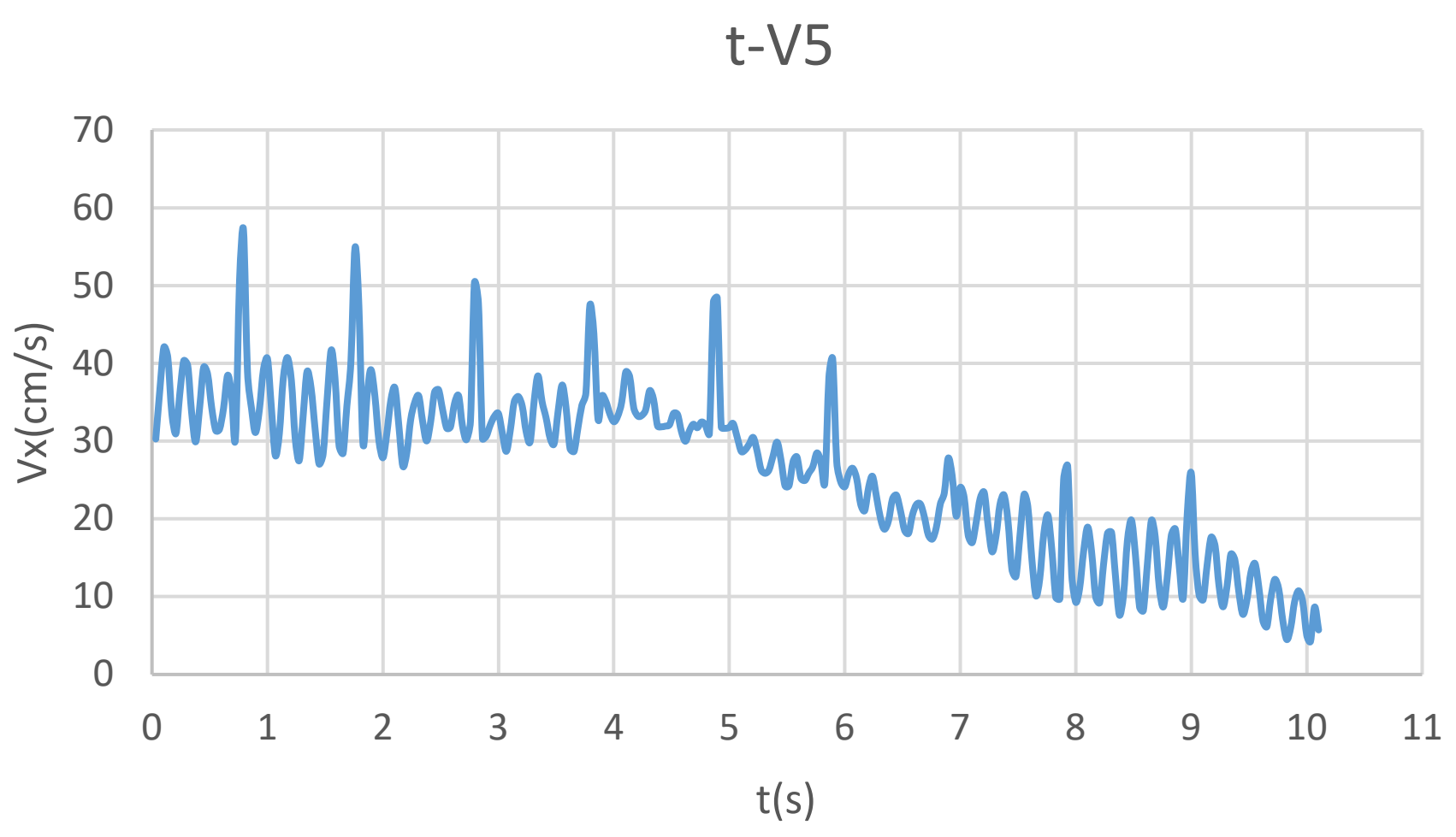

**Figure 5 Velocity-time, four steel balls, without initial velocity, 5cm release height**

For a bottle containing four steel balls rolling down without initial velocity from a height of 4cm from the bottom, see Figure 2

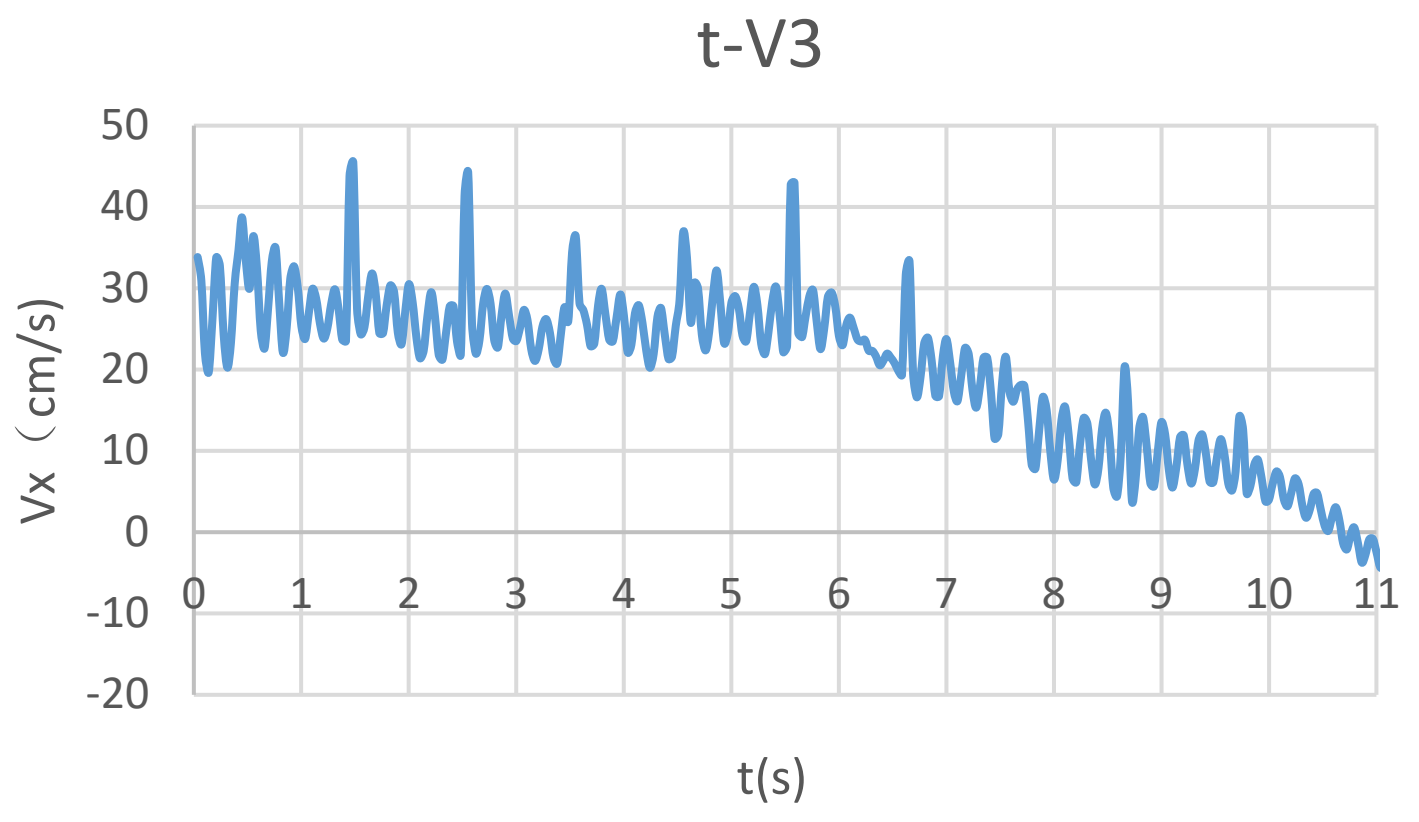

**Figure 6 Velocity-time, four steel balls, without initial velocity, 3cm release height**

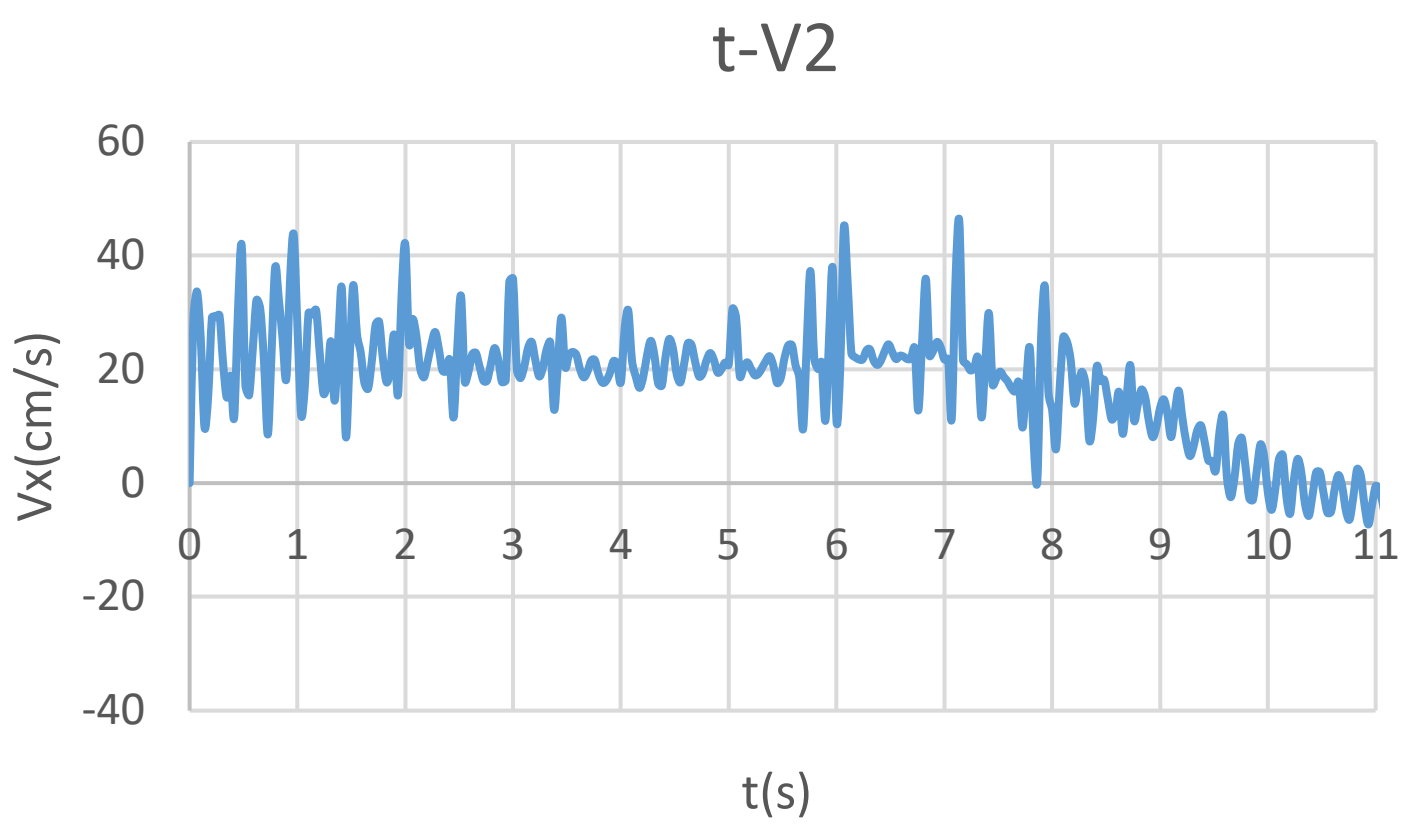

**Figure 7 Velocity-time, four steel balls, without initial velocity, 2cm release height**

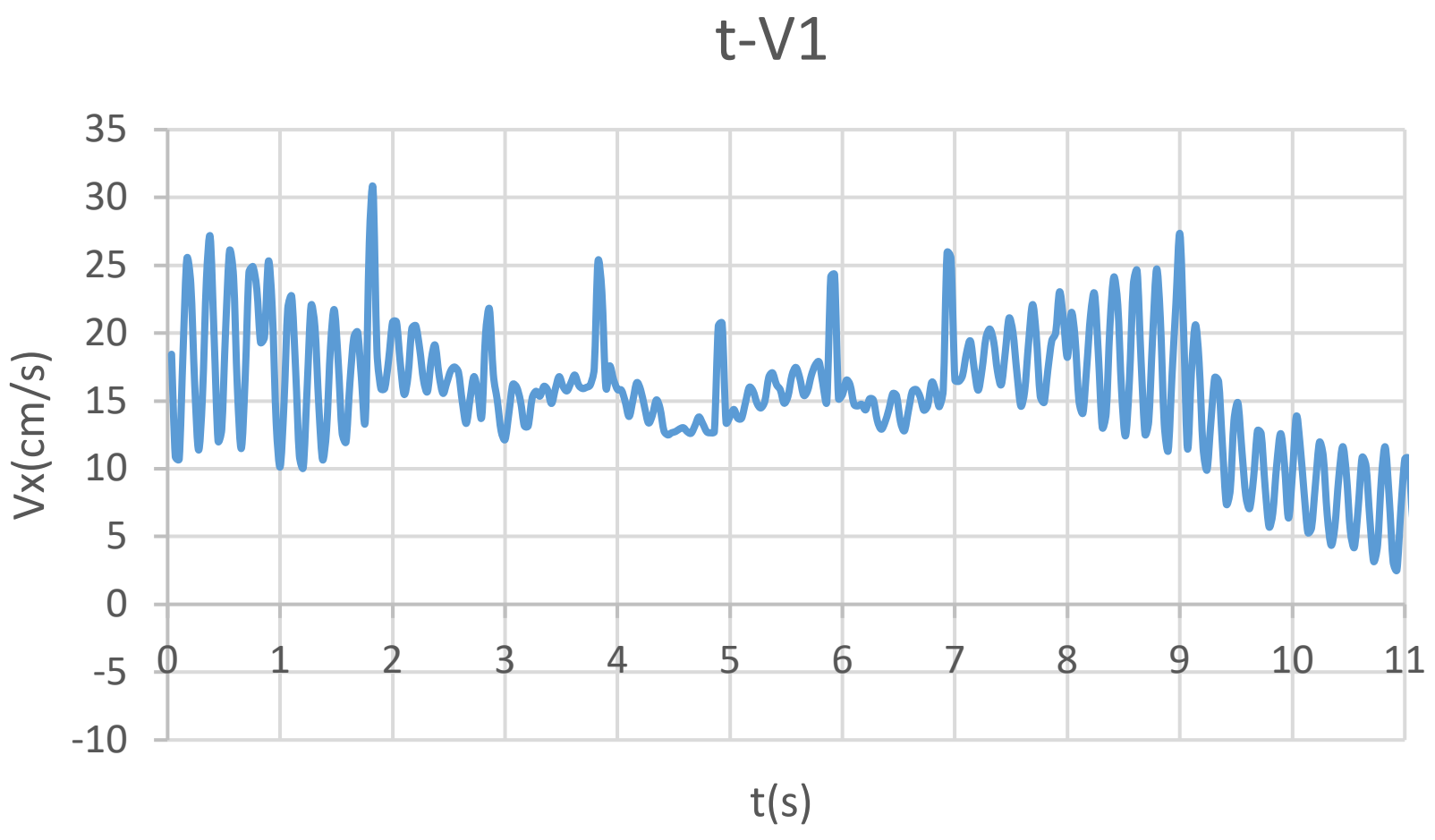

**Figure 8 Velocity-time, four steel balls, without initial velocity, 1cm release height**

In the six cases above, the velocity change periods are all around 0.173s, and the frequencies are about 5.78Hz.

Conclusion of experiment 1:

1. With the same mass, the higher the bottle is released, the faster it reaches the plane and the greater the horizontal displacement before stationary.

2. The centroid of the bottle and the steel balls, although released at different heights, vibrate at the same frequency.

**Experiment 2. Roll down bottles containing different numbers of steel balls from the same height.**

Take the same bottle, place one steel ball and four steel balls respectively, and roll it down without initial velocity at the same height of 7cm from the bottom.

The image of the tracking data (with different masses but at the same height) is as follows (blue line refers to the bottle with one steel ball; orange line refers to the bottle with four steel balls):

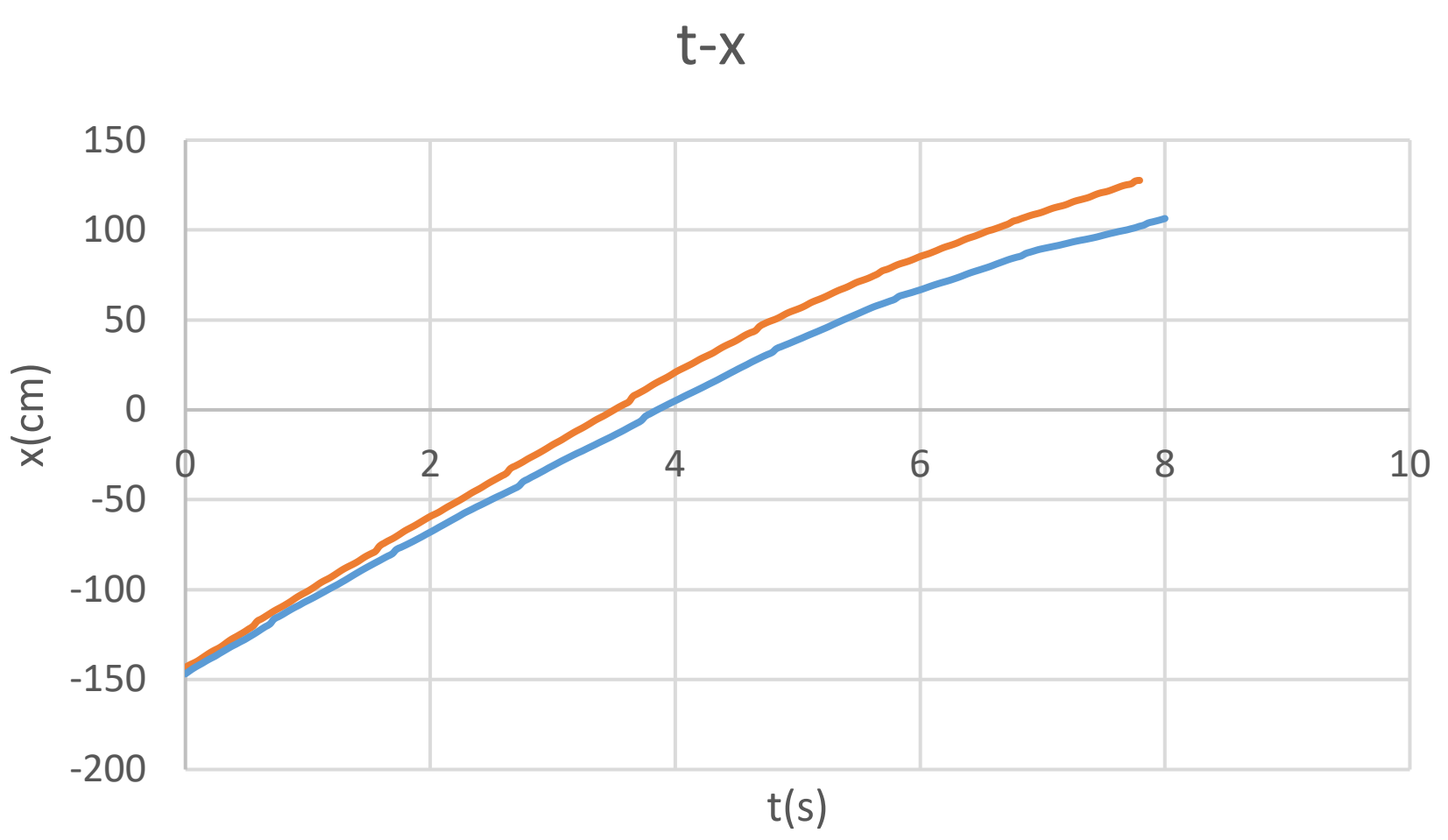

**Figure 9 Displacement-time, a bottle containing one steel ball and four steel balls respectively rolling down without initial velocity from 7cm**

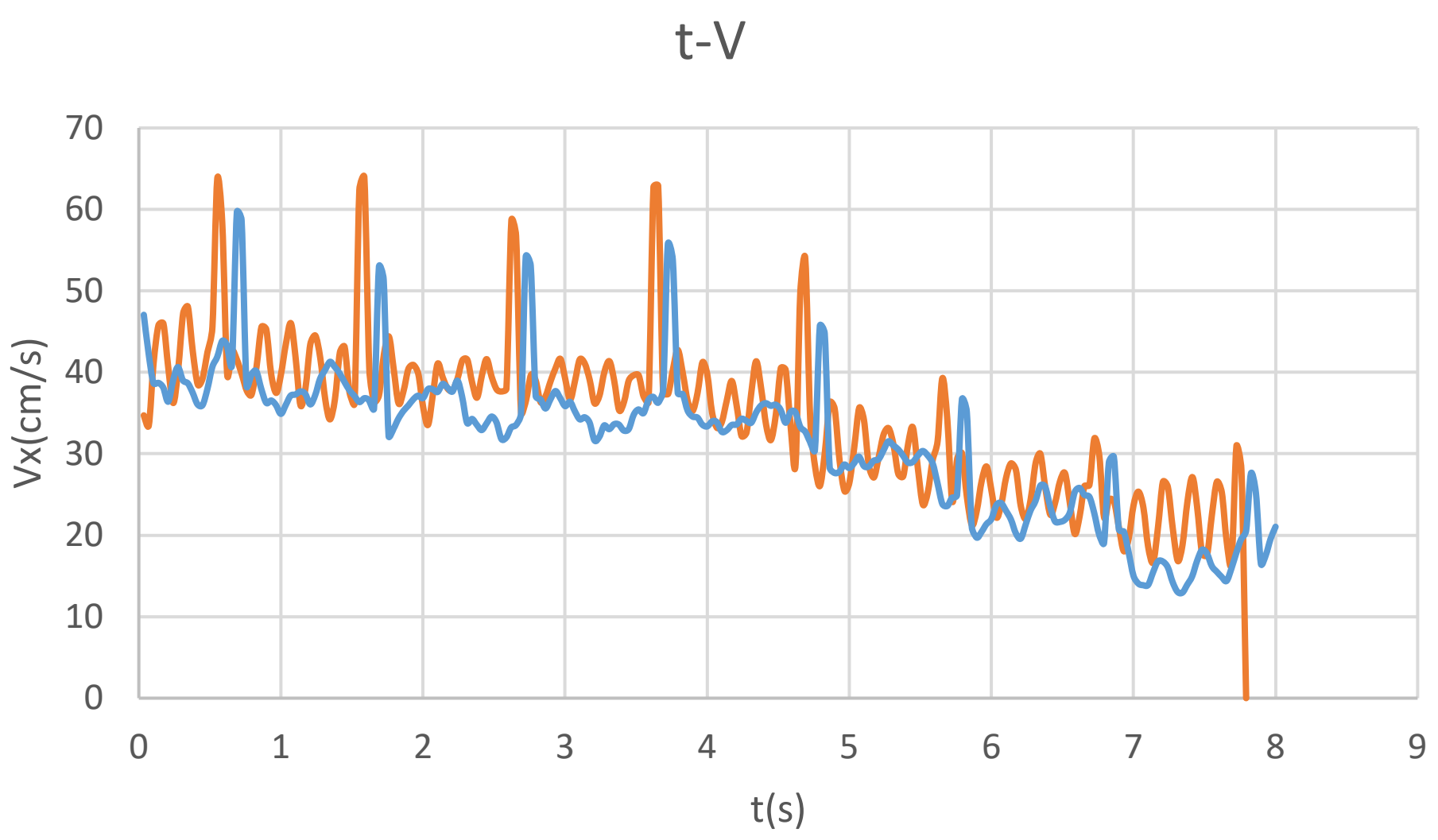

**Figure 10 Velocity-time, one steel ball/four steel balls, without initial velocity, 7cm release height**

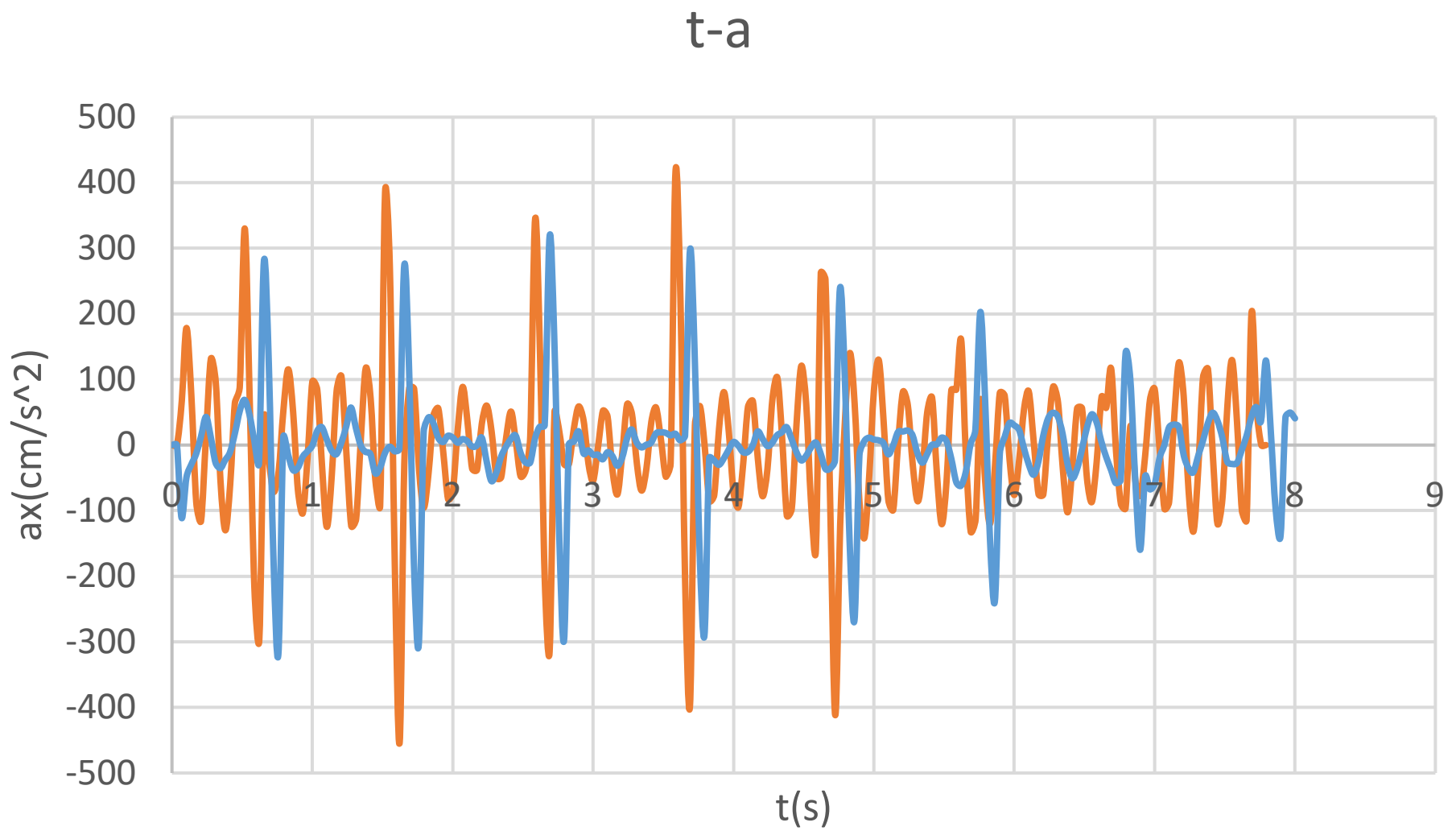

**Figure 11 Acceleration-time, one steel ball/four steel balls, without initial velocity, 7cm release height**

Conclusion of experiment 2:

With the same release height, bottles reach the plane almost at the same velocity, with little impacted by the change of mass. The centroid of the bottle and the steel balls vibrate at the same frequency. The heavier the steel balls, the lower the vibration frequency (In this experiment, the frequency for one steel ball is 32.25HZ, that for four steel balls is 5.84HZ).

**Error analysis:**

1. The vibration of the steel balls and the bottle system is not completely simple harmonic vibration due to the instability of steel balls.

2. The platform used in the experiment is not completely flat, resulting in large damping, which impacts the experiment results.

3. Not enough accuracy is achieved when using Tracker to track function points.

4. For a bottle containing only a few steel balls, it is not completely accurate to take the bottom center as the centroid.

5. Visual errors exist in the video.